# Extension of Hall-symbols of crystallographic space groups to magnetic space groups

Javier González-Platas[1], Nebil A. Katcho[2] and Juan Rodríguez-Carvajal[2]

[1]Departamento de Física. Instituto Universitario de Estudios Avanzados en Física Atómica, Molecular y Fotónica (IUDEA). MALTA Consolider Team. Universidad de La Laguna, Avda. Astrofísico Fco. Sánchez s/n, La Laguna, Tenerife, E-38204, Spain.

[2]Institut Laue-Langevin, Diffraction Group. 71 Avenue des Martyrs, CS 20156, 38042, Grenoble, Cedex 9, France

**Abstract**

The Hall-symbols for describing unambiguously the generators of space groups have been extended to describe whatever setting of the 1651 types of magnetic space groups (Shubnikov groups). A computer program called MHALL has been developed for parsing the Hall symbols, generate the full list of symmetry operators and identify the transformation to the standard setting.

**Introduction**

The international symbols for crystallographic space groups (CSG) are based in the Hermann-Mauguin (HM) symbols containing information about the Bravais lattice of the translation subgroup and the nature and orientation of the symmetry elements. However, as pointed out by S.R. Hall (Hall, 1981), there is no explicit information about the position of the origin determining partially the translational part of the symmetry operators. This has, consequently, an important drawback: the impossibility of generating the full set of symmetry operators from the symmetry elements appearing in the symbol, in the general case, coinciding with the lists of the *International Tables for Crystallography* (ITC). The computer programs, not using a database, doing that task have always some special code to handle special cases in order to generate strictly the same set of operators. The proposal by S.R. Hall (Hall, 1981) solves this problem in an elegant way allowing, at the same time, the possibility of describing reasonable arbitrary settings of CSG. The word "reasonable" is used here to emphasize the fact that the symmetry operators (with rotational part described by an integer matrix) may be very complicated (with rotational part described by a rational matrix) if a unit cell setting, for instance, is selected in which a three-fold axis is along an arbitrary [uvw] direction. However, even in these cases, the Hall-symbol, as treated in (S. R. Hall and R. W. Grosse-Kunstleve, 2010, hereafter referred as HG-K) can be written by using the standard symbols described below followed by a setting change included at the end of the symbol.
In this document, we propose to modify slightly the original Hall-symbols in order to extend them for describing magnetic space groups (MSG). For that, we had to change the original symbol " ' " to " ^ ". The reason is that the prime symbol " ' " is reserved for the time reversal operator to be consistent with the currently used symbols in Belov-Neronova-Smirnova (BNS) and Opechowski-Guccione (OG) notations for MSG. Here we will not discuss the derivation of MSG or their relation with families of CSG that may be found in literature (see, for instance, D. Litvin, 2001); we shall be concentrated in the new proposed nomenclature aspects.

The aim of the Hall-symbols is to have an unambiguous descriptor of a CSG or MSG and not to substitute the HM, BNS or OG (or whatever new proposal that may be adopted) symbols used for a standard setting. Once we have a correspondence with the standard setting of CSG or MSG, an appropriate computer program can parse the symbols and generate the full set of symmetry operators. We have to keep in mind that two different Hall-symbols may represent the same space group type in a particular setting. This is obvious by the fact that we can use different generators to obtain the same full representative operators of the group type.

**New proposed Hall-symbols**

This section summarises the notation described in HG-K, and the modifications we propose.
The Hall-symbols try to describe a CSG or a MSG using a short set of generators. Each generator corresponds to a Seitz operator acting on atom positions or on magnetic moments. The action of the operator $S=\{R, \theta|t\}$ (where R is a proper or improper rotation matrix, $t$ is a translation vector and $\theta$ is an integer representing the time reversal operation) on the atom position $r$ having a magnetic moment $m$ is given by:

$$r' = \{R, \theta|t\}\, r = R\, r + t \quad (1)$$

The corresponding magnetic moment (axial vector) is obtained as:

$$m' = \{R, \theta|t\}\, m = \theta\, det(R)\, R\, m \quad (2)$$

The integer $\theta$ is equal to -1 if the operator is "primed" (combined with time reversal operator 1') and equal to 1 in the "non-primed" case; $det(R)$ stands for determinant of matrix R. The Hall-symbol is the combination of a series of Seitz operators with a lattice symbol and the presence or not of a centre of symmetry at the origin. Each operator is constituted by a particular notation for the operator S that describes unambiguously the matrix R, the value of $\theta$ and the translation vector $t$. The operator S may by represented as a Seitz matrix or in the extended form (see below) of Jones faithful notation. For instance, a Seitz matrix representing a primed two-fold screw rotation along the b-axis is given by:

$$2_1^{y'} = \{2'_{010}\,|\,0,\tfrac{1}{2},0\} = \begin{pmatrix} R & t \\ 0 & 1 \end{pmatrix}^\theta = \begin{pmatrix} -1 & 0 & 0 & 0 \\ 0 & 1 & 0 & 1/2 \\ 0 & 0 & -1 & 0 \\ 0 & 0 & 0 & 1 \end{pmatrix}'$$

Notice that we have no way to explicitly writing the time reversal operator within the matrix; only the knowledge of the presence of time reversal is used in applying effectively the equations (1) and (2). The same operator in Jones faithful notation (extended with 1 or -1 in the last position representing $\theta$) can be written as: $-x, y+1/2, -z, -1$, which is more compact than the full matrix. We will use this last representation instead of matrices hereafter.

The newly proposed Hall-symbols have the general form (compare to equation A1.4.2.4 in HG-K):

$$L\,[N_t^{A\theta}]_1 [N_t^{A\theta}]_2 ... [N_t^{A\theta}]_p\, V \quad (3)$$

The symbol $L$ is the lattice symbol, $L=$ P, A, B, C, I, R, H, F, X. The symbol may include a preceding minus sign for indicating that a centre of symmetry exists at the origin. The X symbol is used for non-conventional lattice centrings, in such a case the explicit generators of the lattice must be provided within the *p* provided operators. The operators related with the **L** symbols are described in Table 1.

---

**Table 1: Lattice symbols L** (similar to table A1.4.2.2. in HG-K)

The lattice symbol **L** implies the following generators in Jones' faithful notation. The identity is always implied even if not provided, the time reversal operator is provided as the last item, the different generators are separated by ";". In the general case, X, we have used the notation $1\mathbf{t} \equiv x+t_1, y+t_2, z+t_3, 1$ where $\mathbf{t} = (t_1, t_2, t_3)$, with $t_i$ rational numbers.

| | | | |
|---|---|---|---|
| P: | $x, y, z, 1$ | -P: | $-x, -y, -z, 1$ |
| A: | $x, y+1/2, z+1/2, 1$ | -A: | $-x, -y, -z, 1; x, y+1/2, z+1/2, 1$ |
| B: | $x+1/2, y, z+1/2, 1$ | -B: | $-x, -y, -z, 1; x+1/2, y, z+1/2, 1$ |
| C: | $x+1/2, y+1/2, z, 1$ | -C: | $-x, -y, -z, 1; x+1/2, y+1/2, z, 1$ |
| I: | $x+1/2, y+1/2, z+1/2, 1$ | -I: | $-x, -y, -z, 1; x+1/2, y+1/2, z+1/2, 1$ |
| R: | $x+2/3, y+1/3, z+1/3, 1; x+1/3, y+2/3, z+2/3, 1$ | -R: | $-x, -y, -z, 1; x+2/3, y+1/3, z+1/3, 1; x+1/3, y+2/3, z+2/3, 1$ |
| H: | $x+2/3, y+1/3, z, 1; x+1/3, y+2/3, z, 1$ | -H: | $-x, -y, -z, 1; x+2/3, y+1/3, z, 1; x+1/3, y+2/3, z, 1$ |
| F: | $x+1/2, y+1/2, z, 1; x, y+1/2, z+1/2, 1; x+1/2, y, z+1/2, 1$ | -F: | $-x, -y, -z, 1; x+1/2, y+1/2, z, 1; x, y+1/2, z+1/2, 1; x+1/2, y, z+1/2, 1$ |
| X: | $1\mathbf{t}_1 ; 1\mathbf{t}_{2;} \ldots 1\mathbf{t}_p$ | -X: | $-x, -y, -z, 1; 1\mathbf{t}_1 ; 1\mathbf{t}_{2;} \ldots 1\mathbf{t}_p$ |

---

The symbol $N$ is 1, 2, 3, 4 or 6 for proper rotations and $\bar{1}, \bar{2}, \bar{3}, \bar{4}$ or $\bar{6}$ (or in text-only characters -1, -2, -3, -4 or -6) for improper rotations (the symbol -2 corresponds to mirror or glide planes depending on the associated translation). The symbol $\theta$ is " ' " when the operator is associated with time reversal or it is absent otherwise. The symbol $A$ indicates the direction of the rotation axis. The possible values of $A$ are $x$, $y$, $z$, ^, ″ and *, for rotation axis along **a**, **b**, **c**, **a-b** (or alternatively **b - c** or **c - a**), **a+b** (or alternatively **b + c** or **c + a**) and **a+b+c**, respectively. The **t** translation symbols are 1, 2, 3, 4, 5, 6, *a*, *b*, *c*, *n*, *u*, *v*, *w*, *d*. They are described in Table A1.4.2.3 of HG-K. This table is reproduced below as Table 2. Even if not explicit in Table 2, the symbols of screw axes $4_2$ and $6_3$ (with translation of 1/2 along the axis $A$) are legal Hall symbols and may be written as $4_c$ and $6_c$ if they are oriented along the **c**-axis. These translations apply additively [e.g. *ad* signifies a (3/4, 1/4, 1/4) translation]. Notice that we can use, for anti-translations, a lattice symbol instead of two translation symbols. For instance, the following symbols are equivalent: $1'_{bc} = 1'_A, 1'_{ac} = 1'_B, 1'_{ab} = 1'_C, 1'_n = 1'_I$

Examples of valid symbols for the $[N_\mathbf{t}^{A\theta}]$ generators are: $4_a^{x'}, -2_n^z, 3_1^*, \bar{1}'_{avw}, 1'_{ad}$ ; representing respectively the operators (in Seitz notation, see Glazer *et al*, 2014):
$\{4_{100}^+ '|\frac{1}{2}, 0, 0\}, \{m_{001}|\frac{1}{2}, \frac{1}{2}, \frac{1}{2}\}, \{3_{111}^+|\frac{1}{2}, \frac{1}{2}, \frac{1}{2}\}, \{\bar{1}'|\frac{1}{2}, \frac{1}{4}, \frac{1}{4}\}, \{1|\frac{3}{4}, \frac{1}{4}, \frac{1}{4}\}$

The change of basis operator **V** is optional and may be used as described in HG-K; however, we propose an alternative change of notation that is more clearly related to its purpose: a change of reference frame,

---

**Table 2: Translation symbols t** (same as table A1.4.2.3. in HG-K)

Alphabetical symbols (given in the first column) specify translations along a fixed direction. Numerical symbols (given in the third column) specify translations as a fraction of the rotation order $|N|$ and in the direction of the implied or explicitly defined axis. Putting several symbols together means the addition of the translation vectors corresponding to the given symbols.

| Translation Symbol | Translation Vector | Subscript Symbol | Fractional Translation |
|---|---|---|---|
| a | 1/2, 0, 0 | 1 in $3_1$ | 1/3 |
| b | 0, 1/2, 0 | 2 in $3_2$ | 2/3 |
| c | 0, 0, 1/2 | 1 in $4_1$ | 1/4 |
| n | 1/2, 1/2, 1/2 | 3 in $4_3$ | 3/4 |
| u | 1/4, 0, 0 | 1 in $6_1$ | 1/6 |
| v | 0, 1/4, 0 | 2 in $6_2$ | 1/3 |
| w | 0, 0, 1/4 | 4 in $6_4$ | 2/3 |
| d | 1/4, 1/4, 1/4 | 5 in $6_5$ | 5/6 |

---

**Table 3: Operators $[N^A]$** (similar to table A1.4.2.4. in HG-K)

Jones faithful operators for proper rotations along the three principal unit-cell directions are given below. For improper rotations (-2, -3, -4 and -6) the symbols are identical except that all signs are reversed. The identity 1 ($x,y,z$) and inversion centre -1 ($-x,-y,-z$) are omitted because they do not depend on the axis. The same holds for time reversal.

| Rotation Axis | A-symbol | Rotation order and Jones faithful symbol | | | |
|---|---|---|---|---|---|
| | | 2 | 3 | 4 | 6 |
| **a** | $x$ | $x, -y, -z$ | $x, -z, y-z$ | $x, -z, y$ | $x, y-z, y$ |
| **b** | $y$ | $-x, y, -z$ | $z-x, y, -x$ | $z, y, -x$ | $z, y, z-x$ |
| **c** | $z$ | $-x, -y, z$ | $-y, x-y, z$ | $-y, x, z$ | $x-y, x, z$ |

---

involving the basis vectors and origin, to which refer the symmetry operators.

**Default axes from the positional order of operators**

The superscripts indicating the direction of rotation axes may be suppressed by adopting the following rules (see section A1.4.2.3.1. in HG-K) for the symbol given in expression (3):

(*i*) The first rotation (or roto-inversion) axis direction is always along **c**

(*ii*) The second rotation (if |N| is 2) axis direction is along **a** if preceded by an |N| of 2 or 4, it is along **a-b** if preceded by an |N| of 3 or 6

(*iii*) The third rotation (if |N| is 3) has an axis direction of **a** + **b** + **c**

This implies that in the major part of cases the symbol "^" is not needed. If we ignore these rules, the superscript of the provided operators should be explicitly written. The rotational part of $[N_t^{A\theta}]$ operators ( $[N^A]$) are given explicitly in Tables 3, 4 and 5.

Another additional convention is that anti-translations ($1_t'$) should appear after the rotational operators and non-conventional lattice translations ($1_x$).

---

**Table 4: Operators** $[N^A]$ (similar to table A1.4.2.5. in HG-K)

Jones faithful operators for proper two-fold rotations along the diagonal unit-cell directions.

| Preceding rotation | Rotation symbol | Axis | Jones faithful symbol |
|---|---|---|---|
| $N^x$ | 2^ | **b-c** | -x, -z, -y |
|  | 2″ | **b+c** | -x, z, y |
| $N^y$ | 2^ | **a-c** | -z, -y, -x |
|  | 2″ | **a+c** | z, -y, x |
| $N^z$ | 2^ | **a-b** | -y, -x, -z |
|  | 2″ | **a+b** | y, x, -z |

---

**Table 5: Operator** 3* (similar to table A1.4.2.6. in HG-K)

Operator for proper three-fold rotation along the body diagonal unit-cell direction.

| Rotation symbol | Axis | Jones faithful symbol |
|---|---|---|
| 3* | **a+b+c** | z, x, y |

---

**New change of basis operator V**

In HG-K, the authors extend the original shift of origin symbol to a complete change of basis; however, they use a vector form in which a symbol close to Jones faithful representation of symmetry operators is used to represent this time a change of basis. Symmetry operators are implicitly referred to a reference frame {O, **a**, **b**, **c**}, where O represents a point in 3D space and **a**, **b**, **c** are the basis vectors. In previous forms of the Hall-symbol (see HG-K), the operator **V** is represented in vectorial form **V**= ($v_1$, $v_2$, $v_3$) with $v_i = r_{i1}X + r_{i2}Y + r_{i3}Z + t_i$, where $r_{ij}$ and $t_i$ are rational numbers. The operator can be written as a Seitz matrix as given in HG-K:

$$\mathbf{V} = \begin{pmatrix} r_{11} & r_{12} & r_{13} & t_1 \\ r_{21} & r_{22} & r_{23} & t_2 \\ r_{31} & r_{32} & r_{33} & t_3 \\ 0 & 0 & 0 & 1 \end{pmatrix} \quad (4)$$

In case of a simple change of origin, $r_{ij}=0$ if $i \neq j$ and $r_{ii}=1$, we conserve the same notation as proposed in (Hall, 1981) and HG-K, where the Hall-symbol finishes with an integer vector in parenthesis of the form ($n_1$, $n_2$, $n_3$), meaning $t_1 = n_1/12$, $t_2 = n_2/12$ and $t_3 = n_3/12$.

The meaning of the change of basis operator **V** is the following: the new basis vectors **a'**, **b'**, **c'** are obtained as:

$$(\mathbf{a'} \quad \mathbf{b'} \quad \mathbf{c'}) = (\mathbf{a} \quad \mathbf{b} \quad \mathbf{c}) \begin{pmatrix} r_{11} & r_{21} & r_{31} \\ r_{12} & r_{22} & r_{32} \\ r_{13} & r_{23} & r_{33} \end{pmatrix} \quad (5)$$

Notice that the 3×3 submatrix is transposed to that of **V**. The new origin O' have coordinates ($t_1$, $t_2$, $t_3$) with respect to the original {O, **a**, **b**, **c**} basis.

The symmetry operators provided in the Hall-symbol (generators) should be transformed, if the operator **V** is provided, according to the equation:

$$\mathbf{S}'_n = \mathbf{V}\mathbf{S}_n\mathbf{V}^{-1} \quad (6)$$

Where the new operators $\mathbf{S}'_n$ are now implicitly referred to the basis {O', **a'**, **b'**, **c'**}.

We maintain the original convention for change of origin but, in order to be compatible with this form, we propose to use the colon ":" as separator between the first part of the symbol and **V** to tell the parsing program that we use the new symbolic form for the **V** operator. So the expression (3), in the general case of basis change, is written as:

$$\mathbf{L}\,[N^{A\theta}_{\mathbf{t}}]_1[N^{A\theta}_{\mathbf{t}}]_2...[N^{A\theta}_{\mathbf{t}}]_p : r_{11}a + r_{12}b + r_{13}c, r_{21}a + r_{22}b + r_{23}c, r_{31}a + r_{32}b + r_{33}c; t_1, t_2, t_3 \quad (7)$$

Notice the semicolon separating the "rotational" part of the matrix (4) from the coordinates of the new origin ($t_1$, $t_2$, $t_3$).

**Detailed examples**

The header "ASCII" given in the examples below means that we use characters from the ASCII character set available in computers and coded with 1 byte (8 bits), so subscripts and superscripts are not used. Instead, the translation symbols are just juxtaposed to the rotational symbols. When using the text (ASCII) form of the Hall-symbols we use one or more spaces between symbols as in:

$$\mathbf{L} \quad [N_{\mathbf{t}}^{A\theta}]_1 \quad [N_{\mathbf{t}}^{A\theta}]_2 \quad ... \quad [N_{\mathbf{t}}^{A\theta}]_p \quad \mathbf{V}$$

When writing Hall symbol in text-only mode the operator $[N_{\mathbf{t}}^{A\theta}]$ should be written in the form:

*N'At*, *NA'***t** or *NA***t'** with the axis symbol preceding the translation part. Notice that the prime symbol commutes with the axis and translation symbols.

An example of legal Hall-symbol for a magnetic group is $P\,6_c\,\bar{2}\,1'_c$ that in ASCII form is represented as

```
P 6c -2 1'c
```

or

```
P 6c -2 1c'
```

The above Hall-symbol corresponds to the MSG $P_c\,6_3cm$ (in BNS notation) in the standard setting. We can also use the symbol (see comment above about Table 2): `P 63 -2 1'c` in which the traditional form of a screw axis $6_3$ is explicitly given.

One can create a Hall-symbol without knowing to which MSG standard type corresponds. A computer program can identify the standard MSG type and calculate the change of basis to obtain the standard setting. For instance, a symbol like `-P 4n' 1u'` ($\bar{P}\,4'_n\,1'_u$) corresponds to the group P_C4_2/m ($P_C\,4_2/m$) in a particular setting. To obtain the standard setting one has to apply the following basis change: **-a**/4-**b**/4, **a**/4-**b**/4, **c**;-1/8,-1/8,0.

Let us consider conventional settings of MSG as examples. MSG are usually divided in four types:

**Type-1** MSG, also called colourless groups, are those having θ=1 for all operators. They are isomorphous to the 230 CSG. The BNS and OG symbols coincide in this case with the HM symbols of CSG. The Hall-symbols are the same as those of the CSG. Examples of the symbols can be found in Table A1.4.2.7 of HG-K.

**Type-2** MSG, also called paramagnetic groups, are those having two set of operators. The first set is the same as those of Type-1 groups and in the second set the operators have θ = -1 combined with all operators of the first set. Their number is also 230 and the BNS and OG symbols are those of the corresponding CSG followed by the time reversal operator. The Hall-symbols are obtained from the Hall-symbols of Type-1 groups by putting the time reversal operator as the last generator.

Examples:

| BNS/OG symbol | BNS/OG ASCII | Compact ASCII | Hall symbol | Hall ASCII |
|---|---|---|---|---|
| $P4_12_12\,1'$ | `P 41 21 2 1'` | `P4_12_121'` | $P\,4_{abw}\,2_{nw}\,1'$ | `P 4abw 2nw 1'` |
| $P4_2/mmc1'$ | `P 42/m m c 1'` | `P4_2/mmc1'` | $\bar{P}\,4_c\,2\,1'$ | `-P 4c 2 1'` |

The full set of operators for the first group ($P4_12_12\,1'$) can be obtained by the generators: -y+1/2, x+1/2, z+1/4, 1 ($4_{abw}$); x+1/2, -y+1/2, -z+3/4, 1 ($2_{nw}$); x, y, z, -1 (1').

The full set of operators for the second group ($P4_2/mmc1'$) can be obtained by the generators: -x, -y, -z, 1 (-1); -y, x, z+1/2, 1 ($4_c$); x, -y, -z, 1 (2); x, y, z, -1 (1')

**Type-3** MSG, also called type-1 black and white (BW1) groups, are those having the same lattice as that of the paramagnetic group, but half of operators are associated with time reversal. Their number is 674 and the BNS and OG symbols are the same.

Examples:

| BNS/OG symbol | BNS/OG ASCII | Compact ASCII | Hall symbol | Hall ASCII |
|---|---|---|---|---|
| $Pc'cn'$ | P c' c n' | Pc'cn' | $\bar{P}2'_{ab}2'_{ac}$ | -P 2ab' 2ac' |
| $P6'_222'$ | P 62' 2 2' | P6_2'22' | $P6'_22'_c(001)$ | P 62' 2c' (0 0 1) |

The full set of operators for the first group ($Pc'cn'$) can be obtained by the generators: -x+1/2,-y+1/2, z,-1 ($2_{ab}'$); x+1/2,-y,-z+1/2,-1 ($2_{ac}'$); -x,-y,-z, 1 (-1)

The full set of operators for the second group ($P6'_222'$) can be obtained by the generators: x-y, x, z+1/3,-1 ($6_2'$); -y,-x,-z+2/3,-1 (2c'). Notice that the second operator without taking into account the change of origin would have the form: -y,-x,-z+1/2,-1 according to the rule (*ii*) and Table 4.

**Type-4** MSG, also called type-2 black and white (BW2) groups, are those containing anti-translations: lattice translations associated with time reversal. Of course, these anti-translations are not lattice vectors. Their number is 517 and the BNS and OG symbols are different. Here we have conserved the columns concerning the ASCII representations of the BNS and OG in compact form.

Examples:

| BNS symbol | OG symbol | BNS ASCII | OG ASCII | Hall symbol | Hall ASCII |
|---|---|---|---|---|---|
| $F_Sdd2$ | $P_Inn2$ | F_Sdd2 | P_Inn2 | $F2\bar{2}_d1'_n$ | F 2 -2d 1'n |
| $P_Cbcn$ | $C_Pm'cm$ | P_Cbcn | C_Pm'cm | $\bar{P}2_n2_{ab}1'_C$ | -P 2n 2ab 1'C |
| $P_c6_3cm$ | $P_{2c}6'm'm$ | P_c6_3cm | P_2c6'm'm | $P6_c\bar{2}1'_c$ | P 6c -2 1'c |
| $F_Sd\bar{3}$ | $P_Fn\bar{3}$ | F_Sd-3 | P_Fn-3 | $\bar{F}2_{uv}2_{vw}31'_n$ | -F 2uv 2vw 3 1n' |

The full set of operators for the first group ($F_Sdd2$) can be obtained by the generators: -x, -y, z, 1 (2); -x+1/4, y+1/4, z+1/4, 1 ($\bar{2}_d$); x+1/2, y+1/2, z+1/2, -1 ($1'_n$) and the F-centring translations x, y+1/2, z+1/2, 1; x+1/2, y, z+1/2,1 and x+1/2, y+1/2, z, 1.

The full set of operators for the second group ($P_C bcn$) can be obtained by the generators: $-x+1/2, -y+1/2, z+1/2, 1$ ($2_n$); $x+1/2,-y+1/2,-z,1$ ($2_{ab}$); $-x, -y, -z, 1$ ($\bar{1}$); $x+1/2, y+1/2, z, -1$ ($1'_C$).

The full set of operators for the third group ($P_c 6_3 cm$) can be obtained by the generators: $-x-y, x, z+1/2, 1$ ($6_c$); $y, x, z, 1$ ($\bar{2}$); $x, y, z+1/2, -1$ ($1'_c$).

The full set of operators for the fourth group ($F_S d\bar{3}$) can be obtained by the generators: $-x+1/4,-y+1/4,z,1$ ($2_{zuv}$), $x,-y+1/4,-z+1/4,1$ ($2_{xuv}$); $z,x,y,1$ (3 along **a**+**b**+**c**, the star symbol in 3* is not needed according to the rule (*iii*) and Table 5), the centre of symmetry $-x,-y,-z,1$ and the F-centring translations.

We provide in the Appendix-I all the standard MSG that are listed together with the Hall-symbols. The newly proposed International symbols (to be discussed in the Commission for Magnetic Structures of the IUCr) for the standard settings are also in the list.

**A console program for calculating symmetry operators from Hall-symbols**

We have developed a console program (**MHALL**) that is able to interpret Hall-symbols as described in this document. The program is based in the new version of the CRYSFML library (Gonzalez-Platas *et al*, 2020) handling general space groups (crystallographic, magnetic and superspace groups). The source code of the library as well as the program is written in the Fortran-2008 standard. The program asks for a Hall-symbol (or a set of generators in Jones faithful notation) as input and produces the full list of operators as well as the corresponding geometric symmetry element symbols (international notation) and the transformation to the standard setting as defined in (Stokes *et al*. 2013). The geometric symmetry element symbols are only provided for not too much involved settings. Two examples of running the program from a Windows console are provided in the Appendix-II. The installation of the program and examples are provided in the supplementary document (Supp-2).

The program is currently available at: https://code.ill.fr/scientific-software/crysfml/-/blob/master/BinariesDistributions/Windows/MHall.zip

**Conclusions**

The extension of the Hall symbols to MSG has been straightforward to implement from the already existing definitions with a minimal change in one axis symbol. The use of the list of standard Hall symbols for the 1651 MSG allows generating, in a faster and compact way, the full list of symmetry operators of these groups. A further extension of Hall symbols is projected for general superspace groups (crystallographic and magnetic) just adding the modulation vectors followed by the phase symbols, commonly used in the current symbols, indicating the translation component for each generator in the additional dimensions.

One of us, N.A.K., acknowledges the financial support by the FILL2030 initiative, a European Union project within the European Commission's Horizon 2020 Research and Innovation program under grant agreement N°731096.

# Appendix-I

List of the standard setting Hall symbols for the 1651 MSG and their correspondence with the conventional BNS and OG symbols. The provided symbols are ordered as in (Litvin, D.B., 2013, *Magnetic Space Group Types*, IUCr e-book). The new ordering, proposed by some member of the Commission for Magnetic Structures of the IUCr, is also provided.

| Litvin Number | BNS Number | BNS Symbol | OG Number | OG Symbol | INT Number | INT Symbol | MHall Symbol |
|---|---|---|---|---|---|---|---|
| 1 | 1.1 | P1 | 1.1.1 | P1 | 1 | P1 | P 1 |
| 2 | 1.2 | P11' | 1.2.2 | P11' | 2 | P11' | P 1' |
| 3 | 1.3 | P_S1 | 1.3.3 | P_2s1 | 3 | P11'_c[P1] | P 1 1'c |
| 4 | 2.4 | P-1 | 2.1.4 | P-1 | 4 | P-1 | -P 1 |
| 5 | 2.5 | P-11' | 2.2.5 | P-11' | 5 | P-11' | -P 1' |
| 6 | 2.6 | P-1' | 2.3.6 | P-1' | 6 | P-1' | P -1' |
| 7 | 2.7 | P_S-1 | 2.4.7 | P_2s-1 | 7 | P-11'_c[P-1] | -P 1 1'c |
| 8 | 3.1 | P2 | 3.1.8 | P2 | 8 | P2 | P 2y |
| 9 | 3.2 | P21' | 3.2.9 | P21' | 9 | P21' | P 2y 1' |
| 10 | 3.3 | P2' | 3.3.10 | P2' | 10 | P2' | P 2y' |
| 11 | 3.4 | P_a2 | 3.4.11 | P_2a2 | 11 | P21'_a[P2] | P 2y 1'a |
| 12 | 3.5 | P_b2 | 3.5.12 | P_2b2 | 12 | P21'_b[P2] | P 2y 1'b |
| 23 | 3.6 | P_C2 | 5.5.23 | C_P2 | 13 | P21'_C[C2] | P 2y 1'C |
| 15 | 4.7 | P2_1 | 4.1.15 | P2_1 | 14 | P2_1 | P 2yb |
| 16 | 4.8 | P2_11' | 4.2.16 | P2_11' | 15 | P2_11' | P 2yb 1' |
| 17 | 4.9 | P2_1' | 4.3.17 | P2_1' | 16 | P2_1' | P 2yb' |
| 18 | 4.10 | P_a2_1 | 4.4.18 | P_2a2_1 | 17 | P2_11'_a[P2_1] | P 2yb 1'a |
| 14 | 4.11 | P_b2_1 | 3.7.14 | P_2b2' | 18 | P2_11'_b[P2] | P 2yb 1'b |
| 24 | 4.12 | P_C2_1 | 5.6.24 | C_P2' | 19 | P2_11'_C[C2] | P 2yb 1'C |
| 19 | 5.13 | C2 | 5.1.19 | C2 | 20 | C2 | C 2y |
| 20 | 5.14 | C21' | 5.2.20 | C21' | 21 | C21' | C 2y 1' |
| 21 | 5.15 | C2' | 5.3.21 | C2' | 22 | C2' | C 2y' |
| 22 | 5.16 | C_c2 | 5.4.22 | C_2c2 | 23 | C21'_c[C2] | C 2y 1'c |
| 13 | 5.17 | C_a2 | 3.6.13 | P_C2 | 24 | C21'_a[P2] | C 2y 1'a |
| 25 | 6.18 | Pm | 6.1.25 | Pm | 25 | Pm | P -2y |
| 26 | 6.19 | Pm1' | 6.2.26 | Pm1' | 26 | Pm1' | P -2y 1' |
| 27 | 6.20 | Pm' | 6.3.27 | Pm' | 27 | Pm' | P -2y' |
| 28 | 6.21 | P_am | 6.4.28 | P_2am | 28 | Pm1'_a[Pm] | P -2y 1'a |
| 29 | 6.22 | P_bm | 6.5.29 | P_2bm | 29 | Pm1'_b[Pm] | P -2y 1'b |
| 42 | 6.23 | P_Cm | 8.5.42 | C_Pm | 30 | Pm1'_C[Cm] | P -2y 1'C |
| 32 | 7.24 | Pc | 7.1.32 | Pc | 31 | Pc | P -2yc |
| 33 | 7.25 | Pc1' | 7.2.33 | Pc1' | 32 | Pc1' | P -2yc 1' |
| 34 | 7.26 | Pc' | 7.3.34 | Pc' | 33 | Pc' | P -2yc' |
| 35 | 7.27 | P_ac | 7.4.35 | P_2ac | 34 | Pc1'_a[Pc] | P -2yc 1'a |
| 31 | 7.28 | P_cc | 6.7.31 | P_2cm' | 35 | Pc1'_c[Pm] | P -2yc 1'c |
| 36 | 7.29 | P_bc | 7.5.36 | P_2bc | 36 | Pc1'_b[Pc] | P -2yc 1'b |
| 48 | 7.30 | P_Cc | 9.4.48 | C_Pc | 37 | Pc1'_C[Cc] | P -2yc 1'C |
| 44 | 7.31 | P_Ac | 8.7.44 | C_Pm' | 38 | Pc1'_A[Am] | P -2yc 1'A |
| 38 | 8.32 | Cm | 8.1.38 | Cm | 39 | Cm | C -2y |
| 39 | 8.33 | Cm1' | 8.2.39 | Cm1' | 40 | Cm1' | C -2y 1' |
| 40 | 8.34 | Cm' | 8.3.40 | Cm' | 41 | Cm' | C -2y' |
| 41 | 8.35 | C_cm | 8.4.41 | C_2cm | 42 | Cm1'_c[Cm] | C -2y 1'c |
| 30 | 8.36 | C_am | 6.6.30 | P_Cm | 43 | Cm1'_a[Pm] | C -2y 1'a |
| 45 | 9.37 | Cc | 9.1.45 | Cc | 44 | Cc | C -2yc |
| 46 | 9.38 | Cc1' | 9.2.46 | Cc1' | 45 | Cc1' | C -2yc 1' |
| 47 | 9.39 | Cc' | 9.3.47 | Cc' | 46 | Cc' | C -2yc' |
| 43 | 9.40 | C_cc | 8.6.43 | C_2cm' | 47 | Cc1'_c[Cm] | C -2yc 1'c |
| 37 | 9.41 | C_ac | 7.6.37 | P_Cc | 48 | Cc1'_a[Pc] | C -2yc 1'a |
| 49 | 10.42 | P2/m | 10.1.49 | P2/m | 49 | P2/m | -P 2y |
| 50 | 10.43 | P2/m1' | 10.2.50 | P2/m1' | 50 | P2/m1' | -P 2y 1' |
| 51 | 10.44 | P2'/m | 10.3.51 | P2'/m | 51 | P2'/m | P 2y' -1' |
| 52 | 10.45 | P2/m' | 10.4.52 | P2/m' | 52 | P2/m' | P 2y -1' |
| 53 | 10.46 | P2'/m' | 10.5.53 | P2'/m' | 53 | P2'/m' | -P 2y' |
| 54 | 10.47 | P_a2/m | 10.6.54 | P_2a2/m | 54 | P2/m1'_a[P2/m] | -P 2y 1'a |
| 55 | 10.48 | P_b2/m | 10.7.55 | P_2b2/m | 55 | P2/m1'_b[P2/m] | -P 2y 1'b |
| 72 | 10.49 | P_C2/m | 12.7.72 | C_P2/m | 56 | P2/m1'_C[C2/m] | -P 2y 1'C |
| 59 | 11.50 | P2_1/m | 11.1.59 | P2_1/m | 57 | P2_1/m | -P 2yb |
| 60 | 11.51 | P2_1/m1' | 11.2.60 | P2_1/m1' | 58 | P2_1/m1' | -P 2yb 1' |
| 61 | 11.52 | P2_1'/m | 11.3.61 | P2_1'/m | 59 | P2_1'/m | P 2yb' -1' |
| 62 | 11.53 | P2_1/m' | 11.4.62 | P2_1/m' | 60 | P2_1/m' | P 2yb -1' |
| 63 | 11.54 | P2_1'/m' | 11.5.63 | P2_1'/m' | 61 | P2_1'/m' | -P 2yb' |
| 64 | 11.55 | P_a2_1/m | 11.6.64 | P_2a2_1/m | 62 | P2_1/m1'_a[P2_1/m] | -P 2yb 1'a |
| 57 | 11.56 | P_b2_1/m | 10.9.57 | P_2b2'/m | 63 | P2_1/m1'_b[P2/m] | -P 2yb 1'b |
| 74 | 11.57 | P_C2_1/m | 12.9.74 | C_P2'/m | 64 | P2_1/m1'_C[C2/m] | -P 2yb 1'C |
| 66 | 12.58 | C2/m | 12.1.66 | C2/m | 65 | C2/m | -C 2y |
| 67 | 12.59 | C2/m1' | 12.2.67 | C2/m1' | 66 | C2/m1' | -C 2y 1' |
| 68 | 12.60 | C2'/m | 12.3.68 | C2'/m | 67 | C2'/m | C 2y' -1' |
| 69 | 12.61 | C2/m' | 12.4.69 | C2/m' | 68 | C2/m' | C 2y -1' |
| 70 | 12.62 | C2'/m' | 12.5.70 | C2'/m' | 69 | C2'/m' | -C 2y' |
| 71 | 12.63 | C_c2/m | 12.6.71 | C_2c2/m | 70 | C2/m1'_c[C2/m] | -C 2y 1'c |
| 56 | 12.64 | C_a2/m | 10.8.56 | P_C2/m | 71 | C2/m1'_a[P2/m] | -C 2y 1'a |
| 77 | 13.65 | P2/c | 13.1.77 | P2/c | 72 | P2/c | -P 2yc |
| 78 | 13.66 | P2/c1' | 13.2.78 | P2/c1' | 73 | P2/c1' | -P 2yc 1' |
| 79 | 13.67 | P2'/c | 13.3.79 | P2'/c | 74 | P2'/c | P 2yc' -1' |
| 80 | 13.68 | P2/c' | 13.4.80 | P2/c' | 75 | P2/c' | P 2yc -1' |
| 81 | 13.69 | P2'/c' | 13.5.81 | P2'/c' | 76 | P2'/c' | -P 2yc' |
| 82 | 13.70 | P_a2/c | 13.6.82 | P_2a2/c | 77 | P2/c1'_a[P2/c] | -P 2yc 1'a |
| 83 | 13.71 | P_b2/c | 13.7.83 | P_2b2/c | 78 | P2/c1'_b[P2/c] | -P 2yc 1'b |
| 58 | 13.72 | P_c2/c | 10.10.58 | P_2c2/m' | 79 | P2/c1'_c[P2/m] | -P 2yc 1'c |
| 75 | 13.73 | P_A2/c | 12.10.75 | C_P2/m' | 80 | P2/c1'_A[A2/m] | -P 2yc 1'A |
| 97 | 13.74 | P_C2/c | 15.6.97 | C_P2/c | 81 | P2/c1'_C[C2/c] | -P 2yc 1'C |
| 86 | 14.75 | P2_1/c | 14.1.86 | P2_1/c | 82 | P2_1/c | -P 2ybc |
| 87 | 14.76 | P2_1/c1' | 14.2.87 | P2_1/c1' | 83 | P2_1/c1' | -P 2ybc 1' |
| 88 | 14.77 | P2_1'/c | 14.3.88 | P2_1'/c | 84 | P2_1'/c | P 2ybc' -1' |
| 89 | 14.78 | P2_1/c' | 14.4.89 | P2_1/c' | 85 | P2_1/c' | P 2ybc -1' |
| 90 | 14.79 | P2_1'/c' | 14.5.90 | P2_1'/c' | 86 | P2_1'/c' | -P 2ybc' |
| 91 | 14.80 | P_a2_1/c | 14.6.91 | P_2a2_1/c | 87 | P2_1/c1'_a[P2_1/c] | -P 2ybc 1'a |
| 85 | 14.81 | P_b2_1/c | 13.9.85 | P_2b2'/c | 88 | P2_1/c1'_b[P2/c] | -P 2ybc 1'b |
| 65 | 14.82 | P_c2_1/c | 11.7.65 | P_2c2_1/m' | 89 | P2_1/c1'_c[P2_1/m] | -P 2ybc 1'c |
| 76 | 14.83 | P_A2_1/c | 12.11.76 | C_P2'/m' | 90 | P2_1/c1'_A[A2/m] | -P 2ybc 1'A |
| 98 | 14.84 | P_C2_1/c | 15.7.98 | C_P2'/c | 91 | P2_1/c1'_C[C2/c] | -P 2ybc 1'C |
| 92 | 15.85 | C2/c | 15.1.92 | C2/c | 92 | C2/c | -C 2yc |
| 93 | 15.86 | C2/c1' | 15.2.93 | C2/c1' | 93 | C2/c1' | -C 2yc 1' |
| 94 | 15.87 | C2'/c | 15.3.94 | C2'/c | 94 | C2'/c | C 2yc' -1' |
| 95 | 15.88 | C2/c' | 15.4.95 | C2/c' | 95 | C2/c' | C 2yc -1' |
| 96 | 15.89 | C2'/c' | 15.5.96 | C2'/c' | 96 | C2'/c' | -C 2yc' |
| 73 | 15.90 | C_c2/c | 12.8.73 | C_2c2/m' | 97 | C2/c1'_c[C2/m] | -C 2yc 1'c |
| 84 | 15.91 | C_a2/c | 13.8.84 | P_C2/c | 98 | C2/c1'_a[P2/c] | -C 2yc 1'a |
| 99 | 16.1 | P222 | 16.1.99 | P222 | 99 | P222 | P 2 2 |
| 100 | 16.2 | P2221' | 16.2.100 | P2221' | 100 | P2221' | P 2 2 1' |
| 101 | 16.3 | P2'2'2 | 16.3.101 | P2'2'2 | 101 | P2'2'2 | P 2' 2' |
| 102 | 16.4 | P_a222 | 16.4.102 | P_2a222 | 102 | P2221'_a[P222] | P 2 2 1'a |
| 134 | 16.5 | P_C222 | 21.6.134 | C_P222 | 103 | P2221'_C[C222] | P 2 2 1'C |
| 148 | 16.6 | P_I222 | 23.4.148 | I_P222 | 104 | P2221'_I[I222] | P 2 2 1'I |
| 106 | 17.7 | P222_1 | 17.1.106 | P222_1 | 105 | P222_1 | P 2c 2 |
| 107 | 17.8 | P222_11' | 17.2.107 | P222_11' | 106 | P222_11' | P 2c 2 1' |
| 108 | 17.9 | P2'2'2_1 | 17.3.108 | P2'2'2_1 | 107 | P2'2'2_1 | P 2c 2' |
| 109 | 17.10 | P22'2_1' | 17.4.109 | P22'2_1' | 108 | P22'2_1' | P 2c' 2 |
| 110 | 17.11 | P_a222_1 | 17.5.110 | P_2a222_1 | 109 | P222_11'_a[P222_1] | P 2c 2 1'a |
| 105 | 17.12 | P_c222_1 | 16.7.105 | P_2c22'2' | 110 | P222_11'_c[P222] | P 2c 2 1'c |
| 138 | 17.13 | P_B222_1 | 21.10.138 | C_P22'2' | 111 | P222_11'_B[B222] | P 2c 2 1'B |
| 126 | 17.14 | P_C222_1 | 20.5.126 | C_P222_1 | 112 | P222_11'_C[C222_1] | P 2c 2 1'C |
| 154 | 17.15 | P_I222_1 | 24.5.154 | I_P2_1'2_1'2_1 | 113 | P222_11'_I[I2_12_12_1] | P 2c 2 1'I |
| 113 | 18.16 | P2_12_12 | 18.1.113 | P2_12_12 | 114 | P2_12_12 | P 2 2ab |
| 114 | 18.17 | P2_12_121' | 18.2.114 | P2_12_121' | 115 | P2_12_121' | P 2 2ab 1' |
| 115 | 18.18 | P2_1'2_1'2 | 18.3.115 | P2_1'2_1'2 | 116 | P2_1'2_1'2 | P 2 2ab' |
| 116 | 18.19 | P2_12_1'2' | 18.4.116 | P2_12_1'2' | 117 | P2_12_1'2' | P 2' 2ab |
| 112 | 18.20 | P_b2_12_12 | 17.7.112 | P_2a2'2'2_1 | 118 | P2_12_121'_b[P2_122] | P 2 2ab 1'b |
| 117 | 18.21 | P_c2_12_12 | 18.5.117 | P_2c2_12_12 | 119 | P2_12_121'_c[P2_12_12] | P 2 2ab 1'c |
| 128 | 18.22 | P_B2_12_12 | 20.7.128 | C_P22'2_1' | 120 | P2_12_121'_B[B22_12] | P 2 2ab 1'B |
| 137 | 18.23 | P_C2_12_12 | 21.9.137 | C_P2'2'2 | 121 | P2_12_121'_C[C222] | P 2 2ab 1'C |
| 149 | 18.24 | P_I2_12_12 | 23.5.149 | I_P2'2'2 | 122 | P2_12_121'_I[I222] | P 2 2ab 1'I |
| 119 | 19.25 | P2_12_12_1 | 19.1.119 | P2_12_12_1 | 123 | P2_12_12_1 | P 2ac 2ab |
| 120 | 19.26 | P2_12_12_11' | 19.2.120 | P2_12_12_11' | 124 | P2_12_12_11' | P 2ac 2ab 1' |
| 121 | 19.27 | P2_1'2_1'2_1 | 19.3.121 | P2_1'2_1'2_1 | 125 | P2_1'2_1'2_1 | P 2ac 2ab' |
| 118 | 19.28 | P_c2_12_12_1 | 18.6.118 | P_2c2_12_1'2' | 126 | P2_12_12_11'_c[P2_12_12] | P 2ac 2ab 1'c |
| 127 | 19.29 | P_C2_12_12_1 | 20.6.127 | C_P2'2'2_1 | 127 | P2_12_12_11'_C[C222_1] | P 2ac 2ab 1'C |
| 153 | 19.30 | P_I2_12_12_1 | 24.4.153 | I_P2_12_12_1 | 128 | P2_12_12_11'_I[I2_12_12_1] | P 2ac 2ab 1'I |
| 122 | 20.31 | C222_1 | 20.1.122 | C222_1 | 129 | C222_1 | C 2c 2 |
| 123 | 20.32 | C222_11' | 20.2.123 | C222_11' | 130 | C222_11' | C 2c 2 1' |
| 124 | 20.33 | C2'2'2_1 | 20.3.124 | C2'2'2_1 | 131 | C2'2'2_1 | C 2c 2' |
| 125 | 20.34 | C22'2_1' | 20.4.125 | C22'2_1' | 132 | C22'2_1' | C 2c' 2 |
| 136 | 20.35 | C_c222_1 | 21.8.136 | C_2c22'2' | 133 | C222_11'_c[C222] | C 2c 2 1'c |
| 111 | 20.36 | C_a222_1 | 17.6.111 | P_C222_1 | 134 | C222_11'_a[P222_1] | C 2c 2 1'a |
| 144 | 20.37 | C_A222_1 | 22.5.144 | F_C22'2' | 135 | C222_11'_A[F222] | C 2c 2 1'A |
| 129 | 21.38 | C222 | 21.1.129 | C222 | 136 | C222 | C 2 2 |
| 130 | 21.39 | C2221' | 21.2.130 | C2221' | 137 | C2221' | C 2 2 1' |
| 131 | 21.40 | C2'2'2 | 21.3.131 | C2'2'2 | 138 | C2'2'2 | C 2 2' |
| 132 | 21.41 | C22'2' | 21.4.132 | C22'2' | 139 | C22'2' | C 2' 2 |
| 133 | 21.42 | C_c222 | 21.5.133 | C_2c222 | 140 | C2221'_c[C222] | C 2 2 1'c |
| 103 | 21.43 | C_a222 | 16.5.103 | P_C222 | 141 | C2221'_a[P222] | C 2 2 1'a |
| 143 | 21.44 | C_A222 | 22.4.143 | F_C222 | 142 | C2221'_A[F222] | C 2 2 1'A |
| 140 | 22.45 | F222 | 22.1.140 | F222 | 143 | F222 | F 2 2 |
| 141 | 22.46 | F2221' | 22.2.141 | F2221' | 144 | F2221' | F 2 2 1' |
| 142 | 22.47 | F2'2'2 | 22.3.142 | F2'2'2 | 145 | F2'2'2 | F 2 2' |
| 104 | 22.48 | F_S222 | 16.6.104 | P_I222 | 146 | F2221'_I[P222] | F 2 2 1'n |
| 145 | 23.49 | I222 | 23.1.145 | I222 | 147 | I222 | I 2 2 |
| 146 | 23.50 | I2221' | 23.2.146 | I2221' | 148 | I2221' | I 2 2 1' |
| 147 | 23.51 | I2'2'2 | 23.3.147 | I2'2'2 | 149 | I2'2'2 | I 2 2' |
| 135 | 23.52 | I_c222 | 21.7.135 | C_I222 | 150 | I2221'_c[C222] | I 2 2 1'c |

```
150    24.53    I2_12_12_1       24.1.150    I2_12_12_1        151    I2_12_12_1              I 2b 2c
151    24.54    I2_12_12_11'     24.2.151    I2_12_12_11'      152    I2_12_12_11'            I 2b 2c   1'
152    24.55    I2_1'2_1'2_1     24.3.152    I2_1'2_1'2_1      153    I2_1'2_1'2_1            I 2b 2c'
139    24.56    I_c2_12_12_1     21.11.139   C_I2'22'          154    I2_12_12_11'_c[C222]    I 2ac 2ab 1'c
155    25.57    Pmm2             25.1.155    Pmm2              155    Pmm2                    P 2 -2
156    25.58    Pmm21'           25.2.156    Pmm21'            156    Pmm21'                  P 2 -2 1'
157    25.59    Pm'm2'           25.3.157    Pm'm2'            157    Pm'm2'                  P 2' -2'
158    25.60    Pm'm'2           25.4.158    Pm'm'2            158    Pm'm'2                  P 2 -2'
159    25.61    P_cmm2           25.5.159    P_2cmm            159    Pmm21'_c[Pmm2]          P 2 -2 1'c
160    25.62    P_amm2           25.6.160    P_2amm            160    Pmm21'_a[Pmm2]          P 2 -2 1'a
241    25.63    P_Cmm2           35.6.241    C_Pmm2            161    Pmm21'_C[Cmm2]          P 2 -2 1'C
271    25.64    P_Amm2           38.7.271    A_Pmm2            162    Pmm21'_A[Amm2]          P 2 -2 1'A
328    25.65    P_Imm2           44.5.328    I_Pmm2            163    Pmm21'_I[Imm2]          P 2 -2 1'I
168    26.66    Pmc2_1           26.1.168    Pmc2_1            164    Pmc2_1                  P 2c  -2
169    26.67    Pmc2_11'         26.2.169    Pmc2_11'          165    Pmc2_11'                P 2c  -2  1'
170    26.68    Pm'c2_1'         26.3.170    Pm'c2_1'          166    Pm'c2_1'                P 2c' -2'
171    26.69    Pmc'2_1'         26.4.171    Pmc'2_1'          167    Pmc'2_1'                P 2c' -2
172    26.70    Pm'c'2_1         26.5.172    Pm'c'2_1          168    Pm'c'2_1                P 2c  -2'
173    26.71    P_amc2_1         26.6.173    P_2amc2_1         169    Pmc2_11'_a[Pmc2_1]      P 2c -2 1'a
174    26.72    P_bmc2_1         26.7.174    P_2bmc2_1         170    Pmc2_11'_b[Pmc2_1]      P 2c -2 1'b
164    26.73    P_cmc2_1         25.10.164   P_2cmm'2'         171    Pmc2_11'_c[Pmm2]        P 2c -2 1'c
275    26.74    P_Amc2_1         38.11.275   A_Pmm'2'          172    Pmc2_11'_A[Amm2]        P 2c -2 1'A
287    26.75    P_Bmc2_1         39.10.287   A_Pb'm2'          173    Pmc2_11'_B[Bma2]        P 2c -2 1'B
254    26.76    P_Cmc2_1         36.6.254    C_Pmc2_1          174    Pmc2_11'_C[Cmc2_1]      P 2c -2 1'C
345    26.77    P_Imc2_1         46.8.345    I_Pma'2'          175    Pmc2_11'_I[Ima2]        P 2c -2 1'I
178    27.78    Pcc2             27.1.178    Pcc2              176    Pcc2                    P 2 -2c
179    27.79    Pcc21'           27.2.179    Pcc21'            177    Pcc21'                  P 2 -2c 1'
180    27.80    Pc'c2'           27.3.180    Pc'c2'            178    Pc'c2'                  P 2' -2c
181    27.81    Pc'c'2           27.4.181    Pc'c'2            179    Pc'c'2                  P 2 -2c'
165    27.82    P_ccc2           25.11.165   P_2cm'm'2         180    Pcc21'_c[Pmm2]          P 2 -2c 1'c
182    27.83    P_acc2           27.5.182    P_2acc2           181    Pcc21'_a[Pcc2]          P 2 -2c 1'a
262    27.84    P_Ccc2           37.5.262    C_Pcc2            182    Pcc21'_C[Ccc2]          P 2 -2c 1'C
289    27.85    P_Acc2           39.12.289   A_Pb'm'2          183    Pcc21'_A[Abm2]          P 2 -2c 1'A
335    27.86    P_Icc2           45.5.335    I_Pba2            184    Pcc21'_I[Iba2]          P 2 -2c 1'I
185    28.87    Pma2             28.1.185    Pma2              185    Pma2                    P 2 -2a
186    28.88    Pma21'           28.2.186    Pma21'            186    Pma21'                  P 2 -2a 1'
187    28.89    Pm'a2'           28.3.187    Pm'a2'            187    Pm'a2'                  P 2' -2a'
188    28.90    Pma'2'           28.4.188    Pma'2'            188    Pma'2'                  P 2' -2a
189    28.91    Pm'a'2           28.5.189    Pm'a'2            189    Pm'a'2                  P 2 -2a'
166    28.92    P_ama2           25.12.166   P_2am'm'2         190    Pma21'_a[Pmm2]          P 2 -2a 1'a
190    28.93    P_bma2           28.6.190    P_2bma2           191    Pma21'_b[Pma2]          P 2 -2a 1'b
191    28.94    P_cma2           28.7.191    P_2cma2           192    Pma21'_c[Pma2]          P 2 -2a 1'c
296    28.95    P_Ama2           40.6.296    A_Pma2            193    Pma21'_A[Ama2]          P 2 -2a 1'A
284    28.96    P_Bma2           39.7.284    A_Pbm2            194    Pma21'_B[Bma2]          P 2 -2a 1'B
245    28.97    P_Cma2           35.10.245   C_Pm'm2'          195    Pma21'_C[Cmm2]          P 2 -2a 1'C
343    28.98    P_Ima2           46.6.343    I_Pma2            196    Pma21'_I[Ima2]          P 2 -2a 1'I
198    29.99    Pca2_1           29.1.198    Pca2_1            197    Pca2_1                  P 2c  -2ac
199    29.100   Pca2_11'         29.2.199    Pca2_11'          198    Pca2_11'                P 2c  -2ac  1'
200    29.101   Pc'a2_1'         29.3.200    Pc'a2_1'          199    Pc'a2_1'                P 2c' -2ac'
201    29.102   Pca'2_1'         29.4.201    Pca'2_1'          200    Pca'2_1'                P 2c' -2ac
202    29.103   Pc'a'2_1         29.5.202    Pc'a'2_1          201    Pc'a'2_1                P 2c  -2ac'
177    29.104   P_aca2_1         26.10.177   P_2bm'c'2_1       202    Pca2_11'_a[Pcm2_1]      P 2c -2ac 1'a
203    29.105   P_bca2_1         29.6.203    P_2bca2_1         203    Pca2_11'_b[Pca2_1]      P 2c -2ac 1'b
194    29.106   P_cca2_1         28.10.194   P_2cm'a2'         204    Pca2_11'_c[Pma2]        P 2c -2ac 1'c
306    29.107   P_Aca2_1         41.7.306    A_Pb'a2'          205    Pca2_11'_A[Aba2]        P 2c -2ac 1'A
288    29.108   P_Bca2_1         39.11.288   A_Pbm'2'          206    Pca2_11'_B[Bma2]        P 2c -2ac 1'B
255    29.109   P_Cca2_1         36.7.255    C_Pm'c2_1'        207    Pca2_11'_C[Ccm2_1]      P 2c -2ac 1'C
336    29.110   P_Ica2_1         45.6.336    I_Pba'2'          208    Pca2_11'_I[Iba2]        P 2c -2ac 1'I
205    30.111   Pnc2             30.1.205    Pnc2              209    Pnc2                    P 2 -2bc
206    30.112   Pnc21'           30.2.206    Pnc21'            210    Pnc21'                  P 2 -2bc 1'
207    30.113   Pn'c2'           30.3.207    Pn'c2'            211    Pn'c2'                  P 2' -2bc'
208    30.114   Pnc'2'           30.4.208    Pnc'2'            212    Pnc'2'                  P 2' -2bc
209    30.115   Pn'c'2           30.5.209    Pn'c'2            213    Pn'c'2                  P 2 -2bc'
210    30.116   P_anc2           30.6.210    P_2anc2           214    Pnc21'_a[Pnc2]          P 2 -2bc 1'a
184    30.117   P_bnc2           27.7.184    P_2bc'c2'         215    Pnc21'_b[Pcc2]          P 2 -2bc 1'b
196    30.118   P_cnc2           28.12.196   P_2cm'a'2         216    Pnc21'_c[Pbm2]          P 2 -2bc 1'c
276    30.119   P_Anc2           38.12.276   A_Pm'm'2          217    Pnc21'_A[Amm2]          P 2 -2bc 1'A
308    30.120   P_Bnc2           41.9.308    A_Pb'a'2          218    Pnc21'_B[Bba2]          P 2 -2bc 1'B
263    30.121   P_Cnc2           37.6.263    C_Pc'c2'          219    Pnc21'_C[Ccc2]          P 2 -2bc 1'C
346    30.122   P_Inc2           46.9.346    I_Pm'a'2          220    Pnc21'_I[Ima2]          P 2 -2bc 1'I
212    31.123   Pmn2_1           31.1.212    Pmn2_1            221    Pmn2_1                  P 2ac  -2
213    31.124   Pmn2_11'         31.2.213    Pmn2_11'          222    Pmn2_11'                P 2ac  -2  1'
214    31.125   Pm'n2_1'         31.3.214    Pm'n2_1'          223    Pm'n2_1'                P 2ac' -2'
215    31.126   Pmn'2_1'         31.4.215    Pmn'2_1'          224    Pmn'2_1'                P 2ac' -2
216    31.127   Pm'n'2_1         31.5.216    Pm'n'2_1          225    Pm'n'2_1                P 2ac  -2'
176    31.128   P_amn2_1         26.9.176    P_2amc'2_1'       226    Pmn2_11'_a[Pmc2_1]      P 2ac -2 1'a
217    31.129   P_bmn2_1         31.6.217    P_2bmn2_1         227    Pmn2_11'_b[Pmn2_1]      P 2ac -2 1'b
195    31.130   P_cmn2_1         28.11.195   P_2cma'2'         228    Pmn2_11'_c[Pma2]        P 2ac -2 1'c
298    31.131   P_Amn2_1         40.8.298    A_Pma'2'          229    Pmn2_11'_A[Ama2]        P 2ac -2 1'A
274    31.132   P_Bmn2_1         38.10.274   A_Pm'm2'          230    Pmn2_11'_B[Bmm2]        P 2ac -2 1'B
256    31.133   P_Cmn2_1         36.8.256    C_Pmc'2_1'        231    Pmn2_11'_C[Cmc2_1]      P 2ac -2 1'C
```

```
329   31.134   P_Imn2_1       44.6.329    I_Pmm'2'          232   Pmn2_11'_I[Imm2]      P 2ac -2 1'I
219   32.135   Pba2           32.1.219    Pba2              233   Pba2                  P 2 -2ab
220   32.136   Pba21'         32.2.220    Pba21'            234   Pba21'                P 2 -2ab 1'
221   32.137   Pb'a2'         32.3.221    Pb'a2'            235   Pb'a2'                P 2' -2ab
222   32.138   Pb'a'2         32.4.222    Pb'a'2            236   Pb'a'2                P 2 -2ab'
223   32.139   P_cba2         32.5.223    P_2cba2           237   Pba21'_c[Pba2]        P 2 -2ab 1'c
193   32.140   P_bba2         28.9.193    P_2bm'a2'         238   Pba21'_b[Pma2]        P 2 -2ab 1'b
246   32.141   P_Cba2         35.11.246   C_Pm'm'2          239   Pba21'_C[Cmm2]        P 2 -2ab 1'C
305   32.142   P_Aba2         41.6.305    A_Pba2            240   Pba21'_A[Aba2]        P 2 -2ab 1'A
337   32.143   P_Iba2         45.7.337    I_Pb'a'2          241   Pba21'_I[Iba2]        P 2 -2ab 1'I
226   33.144   Pna2_1         33.1.226    Pna2_1            242   Pna2_1                P 2c -2n
227   33.145   Pna2_11'       33.2.227    Pna2_11'          243   Pna2_11'              P 2c -2n 1'
228   33.146   Pn'a2_1        33.3.228    Pn'a2_1           244   Pn'a2_1               P 2c' -2n'
229   33.147   Pna'2_1        33.4.229    Pna'2_1           245   Pna'2_1               P 2c' -2n
230   33.148   Pn'a'2_1       33.5.230    Pn'a'2_1          246   Pn'a'2_1              P 2c -2n'
218   33.149   P_ana2_1       31.7.218    P_2bm'n2_1'       247   Pna2_11'_a[Pnm2_1]    P 2c -2n 1'a
204   33.150   P_bna2_1       29.7.204    P_2bc'a'2_1       248   Pna2_11'_b[Pca2_1]    P 2c -2n 1'b
224   33.151   P_cna2_1       32.6.224    P_2cb'a2'         249   Pna2_11'_c[Pba2]      P 2c -2n 1'c
297   33.152   P_Ana2_1       40.7.297    A_Pm'a2'          250   Pna2_11'_A[Ama2]      P 2c -2n 1'A
307   33.153   P_Bna2_1       41.8.307    A_Pba'2'          251   Pna2_11'_B[Bba2]      P 2c -2n 1'B
257   33.154   P_Cna2_1       36.9.257    C_Pm'c'2_1        252   Pna2_11'_C[Ccm2_1]    P 2c -2n 1'C
344   33.155   P_Ina2_1       46.7.344    I_Pm'a2'          253   Pna2_11'_I[Ima2]      P 2c -2n 1'I
231   34.156   Pnn2           34.1.231    Pnn2              254   Pnn2                  P 2 -2n
232   34.157   Pnn21'         34.2.232    Pnn21'            255   Pnn21'                P 2 -2n 1'
233   34.158   Pn'n2'         34.3.233    Pn'n2'            256   Pn'n2'                P 2' -2n
234   34.159   Pn'n'2         34.4.234    Pn'n'2            257   Pn'n'2                P 2 -2n'
211   34.160   P_ann2         30.7.211    P_2anc'2'         258   Pnn21'_a[Pnc2]        P 2 -2n 1'a
225   34.161   P_cnn2         32.7.225    P_2cb'a'2         259   Pnn21'_c[Pba2]        P 2 -2n 1'c
299   34.162   P_Ann2         40.9.299    A_Pm'a2           260   Pnn21'_A[Ama2]        P 2 -2n 1'A
264   34.163   P_Cnn2         37.7.264    C_Pc'c'2          261   Pnn21'_C[Ccc2]        P 2 -2n 1'C
330   34.164   P_Inn2         44.7.330    I_Pm'm'2          262   Pnn21'_I[Imm2]        P 2 -2n 1'I
236   35.165   Cmm2           35.1.236    Cmm2              263   Cmm2                  C 2 -2
237   35.166   Cmm21'         35.2.237    Cmm21'            264   Cmm21'                C 2 -2 1'
238   35.167   Cm'm2'         35.3.238    Cm'm2'            265   Cm'm2'                C 2' -2
239   35.168   Cm'm'2         35.4.239    Cm'm'2            266   Cm'm'2                C 2 -2'
240   35.169   C_cmm2         35.5.240    C_2cmm2           267   Cmm21'_c[Cmm2]        C 2 -2 1'c
161   35.170   C_amm2         25.7.161    P_Cmm2            268   Cmm21'_a[Pmm2]        C 2 -2 1'a
313   35.171   C_Amm2         42.5.313    F_Cmm2            269   Cmm21'_A[Fmm2]        C 2 -2 1'A
249   36.172   Cmc2_1         36.1.249    Cmc2_1            270   Cmc2_1                C 2c -2
250   36.173   Cmc2_11'       36.2.250    Cmc2_11'          271   Cmc2_11'              C 2c -2 1'
251   36.174   Cm'c2_1'       36.3.251    Cm'c2_1'          272   Cm'c2_1'              C 2c' -2'
252   36.175   Cmc'2_1'       36.4.252    Cmc'2_1'          273   Cmc'2_1'              C 2c' -2
253   36.176   Cm'c'2_1       36.5.253    Cm'c'2_1          274   Cm'c'2_1              C 2c -2'
243   36.177   C_cmc2_1       35.8.243    C_2cm'm2'         275   Cmc2_11'_c[Cmm2]      C 2c -2 1'c
175   36.178   C_amc2_1       26.8.175    P_Cmc2_1          276   Cmc2_11'_a[Pmc2_1]    C 2c -2 1'a
315   36.179   C_Amc2_1       42.7.315    F_Cmm'2'          277   Cmc2_11'_A[Fmm2]      C 2c -2 1'A
258   37.180   Ccc2           37.1.258    Ccc2              278   Ccc2                  C 2 -2c
259   37.181   Ccc21'         37.2.259    Ccc21'            279   Ccc21'                C 2 -2c 1'
260   37.182   Cc'c2'         37.3.260    Cc'c2'            280   Cc'c2'                C 2' -2c
261   37.183   Cc'c'2         37.4.261    Cc'c'2            281   Cc'c'2                C 2 -2c'
244   37.184   C_ccc2         35.9.244    C_2cm'm'2         282   Ccc21'_c[Cmm2]        C 2 -2c 1'c
183   37.185   C_acc2         27.6.183    P_Ccc2            283   Ccc21'_a[Pcc2]        C 2 -2c 1'a
316   37.186   C_Acc2         42.8.316    F_Cm'm'2          284   Ccc21'_A[Fmm2]        C 2 -2c 1'A
265   38.187   Amm2           38.1.265    Amm2              285   Amm2                  A 2 -2
266   38.188   Amm21'         38.2.266    Amm21'            286   Amm21'                A 2 -2 1'
267   38.189   Am'm2'         38.3.267    Am'm2'            287   Am'm2'                A 2' -2'
268   38.190   Amm'2'         38.4.268    Amm'2'            288   Amm'2'                A 2' -2
269   38.191   Am'm'2         38.5.269    Am'm'2            289   Am'm'2                A 2 -2'
270   38.192   A_amm2         38.6.270    A_2amm2           290   Amm21'_a[Amm2]        A 2 -2 1'a
162   38.193   A_bmm2         25.8.162    P_Amm2            291   Amm21'_b[Pmm2]        A 2 -2 1'b
314   38.194   A_Bmm2         42.6.314    F_Amm2            292   Amm21'_B[Fmm2]        A 2 -2 1'B
278   39.195   Abm2           39.1.278    Abm2              293   Abm2                  A 2 -2c
279   39.196   Abm21'         39.2.279    Abm21'            294   Abm21'                A 2 -2c 1'
280   39.197   Ab'm2'         39.3.280    Ab'm2'            295   Ab'm2'                A 2' -2c'
281   39.198   Abm'2'         39.4.281    Abm'2'            296   Abm'2'                A 2' -2c
282   39.199   Ab'm'2         39.5.282    Ab'm'2            297   Ab'm'2                A 2 -2c'
283   39.200   A_abm2         39.6.283    A_2abm2           298   Abm21'_a[Abm2]        A 2 -2c 1'a
167   39.201   A_bbm2         25.13.167   P_Am'm2           299   Abm21'_b[Pmm2]        A 2 -2c 1'b
317   39.202   A_Bbm2         42.9.317    F_Am'm2           300   Abm21'_B[Fmm2]        A 2 -2c 1'B
291   40.203   Ama2           40.1.291    Ama2              301   Ama2                  A 2 -2a
292   40.204   Ama21'         40.2.292    Ama21'            302   Ama21'                A 2 -2a 1'
293   40.205   Am'a2'         40.3.293    Am'a2'            303   Am'a2'                A 2 -2a'
294   40.206   Ama'2'         40.4.294    Ama'2'            304   Ama'2'                A 2 -2a
295   40.207   Am'a'2         40.5.295    Am'a'2            305   Am'a'2                A 2 -2a'
273   40.208   A_ama2         38.9.273    A_2amm'2'         306   Ama21'_a[Amm2]        A 2 -2a 1'a
192   40.209   A_bma2         28.8.192    P_Ama2            307   Ama21'_b[Pma2]        A 2 -2a 1'b
318   40.210   A_Bma2         42.10.318   F_Amm'2'          308   Ama21'_B[Fmm2]        A 2 -2a 1'B
300   41.211   Aba2           41.1.300    Aba2              309   Aba2                  A 2 -2ac
301   41.212   Aba21'         41.2.301    Aba21'            310   Aba21'                A 2 -2ac 1'
302   41.213   Ab'a2'         41.3.302    Ab'a2'            311   Ab'a2'                A 2 -2ac'
303   41.214   Aba'2'         41.4.303    Aba'2'            312   Aba'2'                A 2' -2ac
```

```
304   41.215   Ab'a'2    41.5.304   Ab'a'2        313   Ab'a'2           A 2 -2ac'
286   41.216   A_aba2    39.9.286   A_2ab'm'2     314   Aba21'_a[Abm2]   A 2 -2ac 1'a
197   41.217   A_bba2    28.13.197  P_Am'a'2      315   Aba21'_b[Pma2]   A 2 -2ac 1'b
319   41.218   A_Bba2    42.11.319  F_Am'm'2      316   Aba21'_B[Fmm2]   A 2 -2ac 1'B
309   42.219   Fmm2      42.1.309   Fmm2          317   Fmm2             F 2 -2
310   42.220   Fmm21'    42.2.310   Fmm21'        318   Fmm21'           F 2 -2 1'
311   42.221   Fm'm2'    42.3.311   Fm'm2'        319   Fm'm2'           F 2' -2
312   42.222   Fm'm'2    42.4.312   Fm'm'2        320   Fm'm'2           F 2 -2'
163   42.223   F_Smm2    25.9.163   P_Imm2        321   Fmm21'_I[Pmm2]   F 2 -2 1'n
320   43.224   Fdd2      43.1.320   Fdd2          322   Fdd2             F 2 -2d
321   43.225   Fdd21'    43.2.321   Fdd21'        323   Fdd21'           F 2 -2d 1'
322   43.226   Fd'd2'    43.3.322   Fd'd2'        324   Fd'd2'           F 2' -2d
323   43.227   Fd'd'2    43.4.323   Fd'd'2        325   Fd'd'2           F 2 -2d'
235   43.228   F_Sdd2    34.5.235   P_Inn2        326   Fdd21'_I[Pnn2]   F 2 -2d 1'n
324   44.229   Imm2      44.1.324   Imm2          327   Imm2             I 2 -2
325   44.230   Imm21'    44.2.325   Imm21'        328   Imm21'           I 2 -2 1'
326   44.231   Im'm2'    44.3.326   Im'm2'        329   Im'm2'           I 2' -2
327   44.232   Im'm'2    44.4.327   Im'm'2        330   Im'm'2           I 2 -2'
242   44.233   I_cmm2    35.7.242   C_Imm2        331   Imm21'_c[Cmm2]   I 2 -2 1'c
272   44.234   I_amm2    38.8.272   A_Imm2        332   Imm21'_a[Amm2]   I 2 -2 1'a
331   45.235   Iba2      45.1.331   Iba2          333   Iba2             I 2 -2c
332   45.236   Iba21'    45.2.332   Iba21'        334   Iba21'           I 2 -2c 1'
333   45.237   Ib'a2'    45.3.333   Ib'a2'        335   Ib'a2'           I 2' -2c'
334   45.238   Ib'a'2    45.4.334   Ib'a'2        336   Ib'a'2           I 2 -2c'
248   45.239   I_cba2    35.13.248  C_Im'm'2      337   Iba21'_c[Cmm2]   I 2 -2c 1'c
290   45.240   I_aba2    39.13.290  A_Ib'm'2      338   Iba21'_a[Abm2]   I 2 -2c 1'a
338   46.241   Ima2      46.1.338   Ima2          339   Ima2             I 2 -2a
339   46.242   Ima21'    46.2.339   Ima21'        340   Ima21'           I 2 -2a 1'
340   46.243   Im'a2'    46.3.340   Im'a2'        341   Im'a2'           I 2' -2a'
341   46.244   Ima'2'    46.4.341   Ima'2'        342   Ima'2'           I 2' -2a
342   46.245   Im'a'2    46.5.342   Im'a'2        343   Im'a'2           I 2 -2a'
247   46.246   I_cma2    35.12.247  C_Im'm2'      344   Ima21'_c[Cmm2]   I 2 -2a 1'c
277   46.247   I_ama2    38.13.277  A_Im'm'2      345   Ima21'_a[Amm2]   I 2 -2a 1'a
285   46.248   I_bma2    39.8.285   A_Ibm2        346   Ima21'_b[Bma2]   I 2 -2a 1'b
347   47.249   Pmmm      47.1.347   Pmmm          347   Pmmm             -P 2 2
348   47.250   Pmmm1'    47.2.348   Pmmm1'        348   Pmmm1'           -P 2 2 1'
349   47.251   Pm'mm     47.3.349   Pm'mm         349   Pm'mm            P 2' 2 -1'
350   47.252   Pm'm'm    47.4.350   Pm'm'm        350   Pm'm'm           -P 2 2'
351   47.253   Pm'm'm'   47.5.351   Pm'm'm'       351   Pm'm'm'          P 2 2 -1'
352   47.254   P_ammm    47.6.352   P_2ammm       352   Pmmm1'_a[Pmmm]   -P 2 2 1'a
553   47.255   P_Cmmm    65.9.553   C_Pmmm        353   Pmmm1'_C[Cmmm]   -P 2 2 1'C
626   47.256   P_Immm    71.6.626   I_Pmmm        354   Pmmm1'_I[Immm]   -P 2 2 1'I
358   48.257   Pnnn      48.1.358   Pnnn          355   Pnnn             -P 2ab 2bc
359   48.258   Pnnn1'    48.2.359   Pnnn1'        356   Pnnn1'           -P 2ab 2bc 1'
360   48.259   Pn'nn     48.3.360   Pn'nn         357   Pn'nn            P 2ab' 2bc -1'
361   48.260   Pn'n'n    48.4.361   Pn'n'n        358   Pn'n'n           -P 2ab 2bc'
362   48.261   Pn'n'n'   48.5.362   Pn'n'n'       359   Pn'n'n'          P 2ab 2bc -1'
386   48.262   P_cnnn    50.10.386  P_2cb'a'n     360   Pnnn1'_c[Pban]   -P 2ab 2bc 1'c
576   48.263   P_Cnnn    66.13.576  C_Pc'c'm'     361   Pnnn1'_C[Cccm]   -P 2ab 2bc 1'C
629   48.264   P_Innn    71.9.629   I_Pm'm'm'     362   Pnnn1'_I[Immm]   -P 2ab 2bc 1'I
364   49.265   Pccm      49.1.364   Pccm          363   Pccm             -P 2 2c
365   49.266   Pccm1'    49.2.365   Pccm1'        364   Pccm1'           -P 2 2c 1'
366   49.267   Pc'cm     49.3.366   Pc'cm         365   Pc'cm            P 2' 2c -1'
367   49.268   Pccm'     49.4.367   Pccm'         366   Pccm'            P 2 2c' -1'
368   49.269   Pc'c'm    49.5.368   Pc'c'm        367   Pc'c'm           -P 2 2c'
369   49.270   Pc'cm'    49.6.369   Pc'cm'        368   Pc'cm'           -P 2' 2c'
370   49.271   Pc'c'm'   49.7.370   Pc'c'm'       369   Pc'c'm'          P 2 2c -1'
371   49.272   P_accm    49.8.371   P_2accm       370   Pccm1'_a[Pccm]   -P 2 2c 1'a
356   49.273   P_cccm    47.10.356  P_2cm'm'm     371   Pccm1'_c[Pmmm]   -P 2 2c 1'c
585   49.274   P_Bccm    67.9.585   C_Pmma        372   Pccm1'_B[Bmcm]   -P 2 2c 1'B
571   49.275   P_Cccm    66.8.571   C_Pccm        373   Pccm1'_C[Cccm]   -P 2 2c 1'C
637   49.276   P_Iccm    72.8.637   I_Pbam        374   Pccm1'_I[Ibam]   -P 2 2c 1'I
377   50.277   Pban      50.1.377   Pban          375   Pban             -P 2ab 2b
378   50.278   Pban1'    50.2.378   Pban1'        376   Pban1'           -P 2ab 2b 1'
379   50.279   Pb'an     50.3.379   Pb'an         377   Pb'an            P 2ab' 2b -1'
380   50.280   Pban'     50.4.380   Pban'         378   Pban'            P 2ab 2b' -1'
381   50.281   Pb'a'n    50.5.381   Pb'a'n        379   Pb'a'n           -P 2ab 2b'
382   50.282   Pb'an'    50.6.382   Pb'an'        380   Pb'an'           -P 2ab' 2b'
383   50.283   Pb'a'n'   50.7.383   Pb'a'n'       381   Pb'a'n'          P 2ab 2b -1'
375   50.284   P_aban    49.12.375  P_2ac'c'm'    382   Pban1'_a[Pbmb]   -P 2ab 2b 1'a
384   50.285   P_cban    50.8.384   P_2cban       383   Pban1'_c[Pban]   -P 2ab 2b 1'c
601   50.286   P_Aban    68.8.601   C_Pcca        384   Pban1'_A[Acaa]   -P 2ab 2b 1'A
561   50.287   P_Cban    65.17.561  C_Pm'm'm'     385   Pban1'_C[Cmmm]   -P 2ab 2b 1'C
642   50.288   P_Iban    72.13.642  I_Pb'a'm'     386   Pban1'_I[Ibam]   -P 2ab 2b 1'I
387   51.289   Pmma      51.1.387   Pmma          387   Pmma             -P 2a 2a
388   51.290   Pmma1'    51.2.388   Pmma1'        388   Pmma1'           -P 2a 2a 1'
389   51.291   Pm'ma     51.3.389   Pm'ma         389   Pm'ma            P 2a' 2a -1'
390   51.292   Pmm'a     51.4.390   Pmm'a         390   Pmm'a            P 2a' 2a' -1'
391   51.293   Pmma'     51.5.391   Pmma'         391   Pmma'            P 2a 2a' -1'
392   51.294   Pm'm'a    51.6.392   Pm'm'a        392   Pm'm'a           -P 2a 2a'
393   51.295   Pmm'a'    51.7.393   Pmm'a'        393   Pmm'a'           -P 2a' 2a
```

```
394    51.296    Pm'ma'       51.8.394     Pm'ma'         394    Pm'ma'            -P 2a'  2a'
395    51.297    Pm'm'a'      51.9.395     Pm'm'a'        395    Pm'm'a'            P 2a   2a   -1'
355    51.298    P_amma       47.9.355     P_2ammm'       396    Pmma1'_a[Pmmm]    -P 2a  2a  1'a
396    51.299    P_bmma       51.10.396    P_2bmma        397    Pmma1'_b[Pmma]    -P 2a  2a  1'b
397    51.300    P_cmma       51.11.397    P_2cmma        398    Pmma1'_c[Pmma]    -P 2a  2a  1'c
520    51.301    P_Amma       63.10.520    C_Pmcm         399    Pmma1'_A[Amma]    -P 2a  2a  1'A
557    51.302    P_Bmma       65.13.557    C_Pm'mm        400    Pmma1'_B[Bmmm]    -P 2a  2a  1'B
590    51.303    P_Cmma       67.14.590    C_Pmm'a        401    Pmma1'_C[Cmma]    -P 2a  2a  1'C
657    51.304    P_Imma       74.8.657     I_Pmma         402    Pmma1'_I[Immb]    -P 2a  2a  1'I
406    52.305    Pnna         52.1.406     Pnna           403    Pnna              -P 2a   2bc
407    52.306    Pnna1'       52.2.407     Pnna1'         404    Pnna1'            -P 2a   2bc    1'
408    52.307    Pn'na        52.3.408     Pn'na          405    Pn'na              P 2a'  2bc  -1'
409    52.308    Pnn'a        52.4.409     Pnn'a          406    Pnn'a              P 2a'  2bc' -1'
410    52.309    Pnna'        52.5.410     Pnna'          407    Pnna'              P 2a   2bc' -1'
411    52.310    Pn'n'a       52.6.411     Pn'n'a         408    Pn'n'a            -P 2a   2bc'
412    52.311    Pnn'a'       52.7.412     Pnn'a'         409    Pnn'a'            -P 2a'  2bc
413    52.312    Pn'na'       52.8.413     Pn'na'         410    Pn'na'            -P 2a'  2bc'
414    52.313    Pn'n'a'      52.9.414     Pn'n'a'        411    Pn'n'a'            P 2a   2bc  -1'
427    52.314    P_anna       53.13.427    P_2bm'na'      412    Pnna1'_a[Pncm]    -P 2a  2bc 1'a
385    52.315    P_bnna       50.9.385     P_2cb'an       413    Pnna1'_b[Pcna]    -P 2a  2bc 1'b
440    52.316    P_cnna       54.13.440    P_2bc'ca'      414    Pnna1'_c[Pbaa]    -P 2a  2bc 1'c
575    52.317    P_Anna       66.12.575    C_Pcc'm'       415    Pnna1'_A[Amaa]    -P 2a  2bc 1'A
527    52.318    P_Bnna       63.17.527    C_Pm'c'm'      416    Pnna1'_B[Bbmm]    -P 2a  2bc 1'B
604    52.319    P_Cnna       68.11.604    C_Pcc'a'       417    Pnna1'_C[Ccca]    -P 2a  2bc 1'C
658    52.320    P_Inna       74.9.658     I_Pm'm'a       418    Pnna1'_I[Immb]    -P 2a  2bc 1'I
415    53.321    Pmna         53.1.415     Pmna           419    Pmna              -P 2ac  2
416    53.322    Pmna1'       53.2.416     Pmna1'         420    Pmna1'            -P 2ac  2    1'
417    53.323    Pm'na        53.3.417     Pm'na          421    Pm'na              P 2ac' 2   -1'
418    53.324    Pmn'a        53.4.418     Pmn'a          422    Pmn'a              P 2ac' 2'  -1'
419    53.325    Pmna'        53.5.419     Pmna'          423    Pmna'              P 2ac  2'  -1'
420    53.326    Pm'n'a       53.6.420     Pm'n'a         424    Pm'n'a            -P 2ac  2'
421    53.327    Pmn'a'       53.7.421     Pmn'a'         425    Pmn'a'            -P 2ac' 2
422    53.328    Pm'na'       53.8.422     Pm'na'         426    Pm'na'            -P 2ac' 2'
423    53.329    Pm'n'a'      53.9.423     Pm'n'a'        427    Pm'n'a'            P 2ac  2   -1'
401    53.330    P_amna       51.15.401    P_2bm'ma'      428    Pmna1'_a[Pmcm]    -P 2ac 2 1'a
424    53.331    P_bmna       53.10.424    P_2bmna        429    Pmna1'_b[Pmna]    -P 2ac 2 1'b
374    53.332    P_cmna       49.11.374    P_2ac'c'm      430    Pmna1'_c[Pmaa]    -P 2ac 2 1'c
572    53.333    P_Amna       66.9.572     C_Pc'cm        431    Pmna1'_A[Amaa]    -P 2ac 2 1'A
560    53.334    P_Bmna       65.16.560    C_Pmm'm'       432    Pmna1'_B[Bmmm]    -P 2ac 2 1'B
542    53.335    P_Cmna       64.15.542    C_Pmc'a'       433    Pmna1'_C[Cmca]    -P 2ac 2 1'C
659    53.336    P_Imna       74.10.659    I_Pmm'a'       434    Pmna1'_I[Imma]    -P 2ac 2 1'I
428    54.337    Pcca         54.1.428     Pcca           435    Pcca              -P 2a   2ac
429    54.338    Pcca1'       54.2.429     Pcca1'         436    Pcca1'            -P 2a   2ac    1'
430    54.339    Pc'ca        54.3.430     Pc'ca          437    Pc'ca              P 2a'  2ac  -1'
431    54.340    Pcc'a        54.4.431     Pcc'a          438    Pcc'a              P 2a'  2ac' -1'
432    54.341    Pcca'        54.5.432     Pcca'          439    Pcca'              P 2a   2ac' -1'
433    54.342    Pc'c'a       54.6.433     Pc'c'a         440    Pc'c'a            -P 2a   2ac'
434    54.343    Pcc'a'       54.7.434     Pcc'a'         441    Pcc'a'            -P 2a'  2ac
435    54.344    Pc'ca'       54.8.435     Pc'ca'         442    Pc'ca'            -P 2a'  2ac'
436    54.345    Pc'c'a'      54.9.436     Pc'c'a'        443    Pc'c'a'            P 2a   2ac  -1'
373    54.346    P_acca       49.10.373    P_2accm'       444    Pcca1'_a[Pccm]    -P 2a  2ac 1'a
437    54.347    P_bcca       54.10.437    P_2bcca        445    Pcca1'_b[Pcca]    -P 2a  2ac 1'b
404    54.348    P_ccca       51.18.404    P_2cm'm'a      446    Pcca1'_c[Pmma]    -P 2a  2ac 1'c
538    54.349    P_Acca       64.11.538    C_Pm'ca        447    Pcca1'_A[Abma]    -P 2a  2ac 1'A
589    54.350    P_Bcca       67.13.589    C_Pm'ma        448    Pcca1'_B[Bmcm]    -P 2a  2ac 1'B
602    54.351    P_Ccca       68.9.602     C_Pc'ca        449    Pcca1'_C[Cccb]    -P 2a  2ac 1'C
649    54.352    P_Icca       73.7.649     I_Pb'ca        450    Pcca1'_I[Icab]    -P 2a  2ac 1'I
441    55.353    Pbam         55.1.441     Pbam           451    Pbam              -P 2   2ab
442    55.354    Pbam1'       55.2.442     Pbam1'         452    Pbam1'            -P 2   2ab  1'
443    55.355    Pb'am        55.3.443     Pb'am          453    Pb'am              P 2'  2ab  -1'
444    55.356    Pbam'        55.4.444     Pbam'          454    Pbam'              P 2   2ab' -1'
445    55.357    Pb'a'm       55.5.445     Pb'a'm         455    Pb'a'm            -P 2   2ab'
446    55.358    Pb'am'       55.6.446     Pb'am'         456    Pb'am'            -P 2'  2ab'
447    55.359    Pb'a'm'      55.7.447     Pb'a'm'        457    Pb'a'm'            P 2   2ab  -1'
402    55.360    P_abam       51.16.402    P_2cm'ma       458    Pbam1'_a[Pbmm]    -P 2 2ab 1'a
448    55.361    P_cbam       55.8.448     P_2cbam        459    Pbam1'_c[Pbam]    -P 2 2ab 1'c
537    55.362    P_Abam       64.10.537    C_Pmca         460    Pbam1'_A[Acam]    -P 2 2ab 1'A
559    55.363    P_Cbam       65.15.559    C_Pm'm'm       461    Pbam1'_C[Cmmm]    -P 2 2ab 1'C
640    55.364    P_Ibam       72.11.640    I_Pb'a'm       462    Pbam1'_I[Ibam]    -P 2 2ab 1'I
451    56.365    Pccn         56.1.451     Pccn           463    Pccn              -P 2ab  2ac
452    56.366    Pccn1'       56.2.452     Pccn1'         464    Pccn1'            -P 2ab  2ac    1'
453    56.367    Pc'cn        56.3.453     Pc'cn          465    Pc'cn              P 2ab' 2ac  -1'
454    56.368    Pccn'        56.4.454     Pccn'          466    Pccn'              P 2ab  2ac' -1'
455    56.369    Pc'c'n       56.5.455     Pc'c'n         467    Pc'c'n            -P 2ab  2ac'
456    56.370    Pc'cn'       56.6.456     Pc'cn'         468    Pc'cn'            -P 2ab' 2ac'
457    56.371    Pc'c'n'      56.7.457     Pc'c'n'        469    Pc'c'n'            P 2ab  2ac  -1'
439    56.372    P_bccn       54.12.439    P_2bcca'       470    Pccn1'_b[Pcca]    -P 2ab 2ac 1'b
487    56.373    P_cccn       59.10.487    P_2cm'm'n      471    Pccn1'_c[Pmmn]    -P 2ab 2ac 1'c
541    56.374    P_Accn       64.14.541    C_Pm'c'a       472    Pccn1'_A[Abma]    -P 2ab 2ac 1'A
573    56.375    P_Cccn       66.10.573    C_Pccm         473    Pccn1'_C[Cccm]    -P 2ab 2ac 1'C
639    56.376    P_Iccn       72.10.639    I_Pbam'        474    Pccn1'_I[Ibam]    -P 2ab 2ac 1'I
```

```
458   57.377    Pbcm          57.1.458    Pbcm          475   Pbcm              -P 2c   2b
459   57.378    Pbcm1'        57.2.459    Pbcm1'        476   Pbcm1'            -P 2c   2b   1'
460   57.379    Pb'cm         57.3.460    Pb'cm         477   Pb'cm             P 2c'  2b  -1'
461   57.380    Pbc'm         57.4.461    Pbc'm         478   Pbc'm             P 2c'  2b' -1'
462   57.381    Pbcm'         57.5.462    Pbcm'         479   Pbcm'             P 2c   2b' -1'
463   57.382    Pb'c'm        57.6.463    Pb'c'm        480   Pb'c'm            -P 2c   2b'
464   57.383    Pbc'm'        57.7.464    Pbc'm'        481   Pbc'm'            -P 2c'  2b
465   57.384    Pb'cm'        57.8.465    Pb'cm'        482   Pb'cm'            -P 2c'  2b'
466   57.385    Pb'c'm'       57.9.466    Pb'c'm'       483   Pb'c'm'           P 2c   2b  -1'
467   57.386    P_abcm        57.10.467   P_2abcm       484   Pbcm1'_a[Pbcm]    -P 2c 2b 1'a
403   57.387    P_bbcm        51.17.403   P_2cmm'a      485   Pbcm1'_b[Pmcm]    -P 2c 2b 1'b
399   57.388    P_cbcm        51.13.399   P_2bm'ma      486   Pbcm1'_c[Pbmm]    -P 2c 2b 1'c
591   57.389    P_Abcm        67.15.591   C_Pmma'       487   Pbcm1'_A[Acmm]    -P 2c 2b 1'A
540   57.390    P_Bbcm        64.13.540   C_Pmca'       488   Pbcm1'_B[Bbcm]    -P 2c 2b 1'B
521   57.391    P_Cbcm        63.11.521   C_Pm'cm       489   Pbcm1'_C[Cmcm]    -P 2c 2b 1'C
638   57.392    P_Ibcm        72.9.638    I_Pb'am       490   Pbcm1'_I[Ibam]    -P 2c 2b 1'I
471   58.393    Pnnm          58.1.471    Pnnm          491   Pnnm              -P 2 2n
472   58.394    Pnnm1'        58.2.472    Pnnm1'        492   Pnnm1'            -P 2 2n 1'
473   58.395    Pn'nm         58.3.473    Pn'nm         493   Pn'nm             P 2'  2n  -1'
474   58.396    Pnnm'         58.4.474    Pnnm'         494   Pnnm'             P 2 2n' -1'
475   58.397    Pn'n'm        58.5.475    Pn'n'm        495   Pn'n'm            -P 2 2n'
476   58.398    Pnn'm'        58.6.476    Pnn'm'        496   Pnn'm'            -P 2' 2n
477   58.399    Pn'n'm'       58.7.477    Pn'n'm'       497   Pn'n'm'           P 2 2n -1'
426   58.400    P_annm        53.12.426   P_2bmna'      498   Pnnm1'_a[Pncm]    -P 2 2n 1'a
450   58.401    P_cnnm        55.10.450   P_2cb'a'm     499   Pnnm1'_c[Pbam]    -P 2 2n 1'c
525   58.402    P_Bnnm        63.15.525   C_Pmc'm       500   Pnnm1'_B[Bbmm]    -P 2 2n 1'B
574   58.403    P_Cnnm        66.11.574   C_Pc'c'm      501   Pnnm1'_C[Cccm]    -P 2 2n 1'C
628   58.404    P_Innm        71.8.628    I_Pm'm'm      502   Pnnm1'_I[Immm]    -P 2 2n 1'I
478   59.405    Pmmn          59.1.478    Pmmn          503   Pmmn              -P 2ab  2a
479   59.406    Pmmn1'        59.2.479    Pmmn1'        504   Pmmn1'            -P 2ab  2a   1'
480   59.407    Pm'mn         59.3.480    Pm'mn         505   Pm'mn             P 2ab' 2a  -1'
481   59.408    Pmmn'         59.4.481    Pmmn'         506   Pmmn'             P 2ab  2a' -1'
482   59.409    Pm'm'n        59.5.482    Pm'm'n        507   Pm'm'n            -P 2ab  2a'
483   59.410    Pmm'n'        59.6.483    Pmm'n'        508   Pmm'n'            -P 2ab' 2a
484   59.411    Pm'm'n'       59.7.484    Pm'm'n'       509   Pm'm'n'           P 2ab  2a  -1'
400   59.412    P_bmmn        51.14.400   P_2bmma'      510   Pmmn1'_b[Pmma]    -P 2ab 2a 1'b
485   59.413    P_cmmn        59.8.485    P_2cmmn       511   Pmmn1'_c[Pmmn]    -P 2ab 2a 1'c
522   59.414    P_Bmmn        63.12.522   C_Pmc'm       512   Pmmn1'_B[Bbcb]    -P 2ab 2a 1'B
558   59.415    P_Cmmn        65.14.558   C_Pmmm'       513   Pmmn1'_C[Cmmm]    -P 2ab 2a 1'C
627   59.416    P_Immn        71.7.627    I_Pm'mm       514   Pmmn1'_I[Immm]    -P 2ab 2a 1'I
488   60.417    Pbcn          60.1.488    Pbcn          515   Pbcn              -P 2n   2ab
489   60.418    Pbcn1'        60.2.489    Pbcn1'        516   Pbcn1'            -P 2n   2ab   1'
490   60.419    Pb'cn         60.3.490    Pb'cn         517   Pb'cn             P 2n'  2ab  -1'
491   60.420    Pbc'n         60.4.491    Pbc'n         518   Pbc'n             P 2n'  2ab' -1'
492   60.421    Pbcn'         60.5.492    Pbcn'         519   Pbcn'             P 2n   2ab' -1'
493   60.422    Pb'c'n        60.6.493    Pb'c'n        520   Pb'c'n            -P 2n   2ab'
494   60.423    Pbc'n'        60.7.494    Pbc'n'        521   Pbc'n'            -P 2n'  2ab
495   60.424    Pb'cn'        60.8.495    Pb'cn'        522   Pb'cn'            -P 2n'  2ab'
496   60.425    Pb'c'n'       60.9.496    Pb'c'n'       523   Pb'c'n'           P 2n   2ab  -1'
438   60.426    P_abcn        54.11.438   P_2bc'ca      524   Pbcn1'_a[Pbcb]    -P 2n 2ab 1'a
470   60.427    P_bbcn        57.13.470   P_2abc'm'     525   Pbcn1'_b[Pmca]    -P 2n 2ab 1'b
425   60.428    P_cbcn        53.11.425   P_2bm'na      526   Pbcn1'_c[Pbmn]    -P 2n 2ab 1'c
544   60.429    P_Abcn        64.17.544   C_Pm'c'a'     527   Pbcn1'_A[Abma]    -P 2n 2ab 1'A
603   60.430    P_Bbcn        68.10.603   C_Pcca        528   Pbcn1'_B[Bbcb]    -P 2n 2ab 1'B
526   60.431    P_Cbcn        63.16.526   C_Pm'cm'      529   Pbcn1'_C[Cmcm]    -P 2n 2ab 1'C
641   60.432    P_Ibcn        72.12.641   I_Pb'am'      530   Pbcn1'_I[Ibam]    -P 2n 2ab 1'I
497   61.433    Pbca          61.1.497    Pbca          531   Pbca              -P 2ac  2ab
498   61.434    Pbca1'        61.2.498    Pbca1'        532   Pbca1'            -P 2ac  2ab  1'
499   61.435    Pb'ca         61.3.499    Pb'ca         533   Pb'ca             P 2ac' 2ab -1'
500   61.436    Pb'c'a        61.4.500    Pb'c'a        534   Pb'c'a            -P 2ac  2ab'
501   61.437    Pb'c'a'       61.5.501    Pb'c'a'       535   Pb'c'a'           P 2ac  2ab -1'
469   61.438    P_abca        57.12.469   P_2abcm       536   Pbca1'_a[Pbcm]    -P 2ac 2ab 1'a
543   61.439    P_Cbca        64.16.543   C_Pm'ca'      537   Pbca1'_C[Cmca]    -P 2ac 2ab 1'C
648   61.440    P_Ibca        73.6.648    I_Pbca        538   Pbca1'_I[Icab]    -P 2ac 2ab 1'I
502   62.441    Pnma          62.1.502    Pnma          539   Pnma              -P 2ac  2n
503   62.442    Pnma1'        62.2.503    Pnma1'        540   Pnma1'            -P 2ac  2n   1'
504   62.443    Pn'ma         62.3.504    Pn'ma         541   Pn'ma             P 2ac' 2n  -1'
505   62.444    Pnm'a         62.4.505    Pnm'a         542   Pnm'a             P 2ac  2n' -1'
506   62.445    Pnma'         62.5.506    Pnma'         543   Pnma'             P 2ac  2n' -1'
507   62.446    Pn'm'a        62.6.507    Pn'm'a        544   Pn'm'a            -P 2ac  2n'
508   62.447    Pnm'a'        62.7.508    Pnm'a'        545   Pnm'a'            -P 2ac' 2n
509   62.448    Pn'ma'        62.8.509    Pn'ma'        546   Pn'ma'            -P 2ac' 2n'
510   62.449    Pn'm'a'       62.9.510    Pn'm'a'       547   Pn'm'a'           P 2ac  2n  -1'
486   62.450    P_anma        59.9.486    P_2cm'mn      548   Pnma1'_a[Pnmm]    -P 2ac 2n 1'a
449   62.451    P_bnma        55.9.449    P_2cb'am      549   Pnma1'_b[Pcma]    -P 2ac 2n 1'b
468   62.452    P_cnma        57.11.468   P_2abc'm      550   Pnma1'_c[Pbma]    -P 2ac 2n 1'c
523   62.453    P_Anma        63.13.523   C_Pmcm'       551   Pnma1'_A[Amma]    -P 2ac 2n 1'A
524   62.454    P_Bnma        63.14.524   C_Pm'c'm      552   Pnma1'_B[Bbmm]    -P 2ac 2n 1'B
539   62.455    P_Cnma        64.12.539   C_Pmc'a       553   Pnma1'_C[Ccmb]    -P 2ac 2n 1'C
660   62.456    P_Inma        74.11.660   I_Pm'ma       554   Pnma1'_I[Imma]    -P 2ac 2n 1'I
511   63.457    Cmcm          63.1.511    Cmcm          555   Cmcm              -C 2c   2
```

| | | | | | | | |
|---|---|---|---|---|---|---|---|
| 512 | 63.458 | Cmcm1' | 63.2.512 | Cmcm1' | 556 | Cmcm1' | -C 2c 2 1' |
| 513 | 63.459 | Cm'cm | 63.3.513 | Cm'cm | 557 | Cm'cm | C 2c' 2 -1' |
| 514 | 63.460 | Cmc'm | 63.4.514 | Cmc'm | 558 | Cmc'm | C 2c' 2' -1' |
| 515 | 63.461 | Cmcm' | 63.5.515 | Cmcm' | 559 | Cmcm' | C 2c 2' -1' |
| 516 | 63.462 | Cm'c'm | 63.6.516 | Cm'c'm | 560 | Cm'c'm | -C 2c 2' |
| 517 | 63.463 | Cmc'm' | 63.7.517 | Cmc'm' | 561 | Cmc'm' | -C 2c' 2 |
| 518 | 63.464 | Cm'cm' | 63.8.518 | Cm'cm' | 562 | Cm'cm' | -C 2c' 2' |
| 519 | 63.465 | Cm'c'm' | 63.9.519 | Cm'c'm' | 563 | Cm'c'm' | C 2c 2 -1' |
| 556 | 63.466 | C_cmcm | 65.12.556 | C_2cmm'm | 564 | Cmcm1'_c[Cmmm] | -C 2c 2 1'c |
| 398 | 63.467 | C_amcm | 51.12.398 | P_Amma | 565 | Cmcm1'_a[Pmcm] | -C 2c 2 1'a |
| 611 | 63.468 | C_Amcm | 69.7.611 | F_Cm'mm | 566 | Cmcm1'_A[Fmmm] | -C 2c 2 1'A |
| 528 | 64.469 | Cmca | 64.1.528 | Cmca | 567 | Cmca | -C 2bc 2 |
| 529 | 64.470 | Cmca1' | 64.2.529 | Cmca1' | 568 | Cmca1' | -C 2bc 2 1' |
| 530 | 64.471 | Cm'ca | 64.3.530 | Cm'ca | 569 | Cm'ca | C 2bc' 2 -1' |
| 531 | 64.472 | Cmc'a | 64.4.531 | Cmc'a | 570 | Cmc'a | C 2bc' 2' -1' |
| 532 | 64.473 | Cmca' | 64.5.532 | Cmca' | 571 | Cmca' | C 2bc 2' -1' |
| 533 | 64.474 | Cm'c'a | 64.6.533 | Cm'c'a | 572 | Cm'c'a | -C 2bc 2' |
| 534 | 64.475 | Cmc'a' | 64.7.534 | Cmc'a' | 573 | Cmc'a' | -C 2bc' 2 |
| 535 | 64.476 | Cm'ca' | 64.8.535 | Cm'ca' | 574 | Cm'ca' | -C 2bc' 2' |
| 536 | 64.477 | Cm'c'a' | 64.9.536 | Cm'c'a' | 575 | Cm'c'a' | C 2bc 2 -1' |
| 587 | 64.478 | C_cmca | 67.11.587 | C_2cm'ma | 576 | Cmca1'_c[Cmmb] | -C 2bc 2 1'c |
| 405 | 64.479 | C_amca | 51.19.405 | P_Am'ma | 577 | Cmca1'_a[Pmcm] | -C 2bc 2 1'a |
| 614 | 64.480 | C_Amca | 69.10.614 | F_Cmm'm' | 578 | Cmca1'_A[Fmmm] | -C 2bc 2 1'A |
| 545 | 65.481 | Cmmm | 65.1.545 | Cmmm | 579 | Cmmm | -C 2 2 |
| 546 | 65.482 | Cmmm1' | 65.2.546 | Cmmm1' | 580 | Cmmm1' | -C 2 2 1' |
| 547 | 65.483 | Cm'mm | 65.3.547 | Cm'mm | 581 | Cm'mm | C 2' 2 -1' |
| 548 | 65.484 | Cmmm' | 65.4.548 | Cmmm' | 582 | Cmmm' | C 2 2' -1' |
| 549 | 65.485 | Cm'm'm | 65.5.549 | Cm'm'm | 583 | Cm'm'm | -C 2 2' |
| 550 | 65.486 | Cmm'm' | 65.6.550 | Cmm'm' | 584 | Cmm'm' | -C 2' 2 |
| 551 | 65.487 | Cm'm'm' | 65.7.551 | Cm'm'm' | 585 | Cm'm'm' | C 2 2 -1' |
| 552 | 65.488 | C_cmmm | 65.8.552 | C_2cmmm | 586 | Cmmm1'_c[Cmmm] | -C 2 2 1'c |
| 353 | 65.489 | C_ammm | 47.7.353 | P_Cmmm | 587 | Cmmm1'_a[Pmmm] | -C 2 2 1'a |
| 610 | 65.490 | C_Ammm | 69.6.610 | F_Cmmm | 588 | Cmmm1'_A[Fmmm] | -C 2 2 1'A |
| 564 | 66.491 | Cccm | 66.1.564 | Cccm | 589 | Cccm | -C 2 2c |
| 565 | 66.492 | Cccm1' | 66.2.565 | Cccm1' | 590 | Cccm1' | -C 2 2c 1' |
| 566 | 66.493 | Cc'cm | 66.3.566 | Cc'cm | 591 | Cc'cm | C 2' 2c -1' |
| 567 | 66.494 | Cccm' | 66.4.567 | Cccm' | 592 | Cccm' | C 2 2c' -1' |
| 568 | 66.495 | Cc'c'm | 66.5.568 | Cc'c'm | 593 | Cc'c'm | -C 2 2c' |
| 569 | 66.496 | Ccc'm' | 66.6.569 | Ccc'm' | 594 | Ccc'm' | -C 2' 2c |
| 570 | 66.497 | Cc'c'm' | 66.7.570 | Cc'c'm' | 595 | Cc'c'm' | C 2 2c -1' |
| 555 | 66.498 | C_cccm | 65.11.555 | C_2cm'm'm | 596 | Cccm1'_c[Cmmm] | -C 2 2c 1'c |
| 372 | 66.499 | C_accm | 49.9.372 | P_Cccm | 597 | Cccm1'_a[Pccm] | -C 2 2c 1'a |
| 613 | 66.500 | C_Accm | 69.9.613 | F_Cm'm'm | 598 | Cccm1'_A[Fmmm] | -C 2 2c 1'A |
| 577 | 67.501 | Cmma | 67.1.577 | Cmma | 599 | Cmma | -C 2b 2 |
| 578 | 67.502 | Cmma1' | 67.2.578 | Cmma1' | 600 | Cmma1' | -C 2b 2 1' |
| 579 | 67.503 | Cm'ma | 67.3.579 | Cm'ma | 601 | Cm'ma | C 2b' 2 -1' |
| 580 | 67.504 | Cmma' | 67.4.580 | Cmma' | 602 | Cmma' | C 2b 2' -1' |
| 581 | 67.505 | Cm'm'a | 67.5.581 | Cm'm'a | 603 | Cm'm'a | -C 2b 2' |
| 582 | 67.506 | Cmm'a' | 67.6.582 | Cmm'a' | 604 | Cmm'a' | -C 2b' 2 |
| 583 | 67.507 | Cm'm'a' | 67.7.583 | Cm'm'a' | 605 | Cm'm'a' | C 2b 2 -1' |
| 584 | 67.508 | C_cmma | 67.8.584 | C_2cmma | 606 | Cmma1'_c[Cmma] | -C 2b 2 1'c |
| 357 | 67.509 | C_amma | 47.11.357 | P_Cmmm' | 607 | Cmma1'_a[Pmmm] | -C 2b 2 1'a |
| 612 | 67.510 | C_Amma | 69.8.612 | F_Cmmm' | 608 | Cmma1'_A[Fmmm] | -C 2b 2 1'A |
| 594 | 68.511 | Ccca | 68.1.594 | Ccca | 609 | Ccca | -C 2b 2bc |
| 595 | 68.512 | Ccca1' | 68.2.595 | Ccca1' | 610 | Ccca1' | -C 2b 2bc 1' |
| 596 | 68.513 | Cc'ca | 68.3.596 | Cc'ca | 611 | Cc'ca | C 2b' 2bc -1' |
| 597 | 68.514 | Ccca' | 68.4.597 | Ccca' | 612 | Ccca' | C 2b 2bc' -1' |
| 598 | 68.515 | Cc'c'a | 68.5.598 | Cc'c'a | 613 | Cc'c'a | -C 2b 2bc' |
| 599 | 68.516 | Ccc'a' | 68.6.599 | Ccc'a' | 614 | Ccc'a' | -C 2b' 2bc |
| 600 | 68.517 | Cc'c'a' | 68.7.600 | Cc'c'a' | 615 | Cc'c'a' | C 2b 2bc -1' |
| 588 | 68.518 | C_ccca | 67.12.588 | C_2cm'm'a | 616 | Ccca1'_c[Cmmb] | -C 2b 2bc 1'c |
| 376 | 68.519 | C_acca | 49.13.376 | P_Cccm | 617 | Ccca1'_a[Pccm] | -C 2b 2bc 1'a |
| 615 | 68.520 | C_Acca | 69.11.615 | F_Cm'm'm' | 618 | Ccca1'_A[Fmmm] | -C 2b 2bc 1'A |
| 605 | 69.521 | Fmmm | 69.1.605 | Fmmm | 619 | Fmmm | -F 2 2 |
| 606 | 69.522 | Fmmm1' | 69.2.606 | Fmmm1' | 620 | Fmmm1' | -F 2 2 1' |
| 607 | 69.523 | Fm'mm | 69.3.607 | Fm'mm | 621 | Fm'mm | F 2' 2 -1' |
| 608 | 69.524 | Fm'm'm | 69.4.608 | Fm'm'm | 622 | Fm'm'm | -F 2 2' |
| 609 | 69.525 | Fm'm'm' | 69.5.609 | Fm'm'm' | 623 | Fm'm'm' | F 2 2 -1' |
| 354 | 69.526 | F_Smmm | 47.8.354 | P_Immm | 624 | Fmmm1'_I[Pmmm] | -F 2 2 1'n |
| 616 | 70.527 | Fddd | 70.1.616 | Fddd | 625 | Fddd | -F 2uv 2vw |
| 617 | 70.528 | Fddd1' | 70.2.617 | Fddd1' | 626 | Fddd1' | -F 2uv 2vw 1' |
| 618 | 70.529 | Fd'dd | 70.3.618 | Fd'dd | 627 | Fd'dd | F 2uv' 2vw -1' |
| 619 | 70.530 | Fd'd'd | 70.4.619 | Fd'd'd | 628 | Fd'd'd | -F 2uv 2vw' |
| 620 | 70.531 | Fd'd'd' | 70.5.620 | Fd'd'd' | 629 | Fd'd'd' | F 2uv 2vw -1' |
| 363 | 70.532 | F_Sddd | 48.6.363 | P_Innn | 630 | Fddd1'_I[Pnnn] | -F 2uv 2vw 1'n |
| 621 | 71.533 | Immm | 71.1.621 | Immm | 631 | Immm | -I 2 2 |
| 622 | 71.534 | Immm1' | 71.2.622 | Immm1' | 632 | Immm1' | -I 2 2 1' |
| 623 | 71.535 | Im'mm | 71.3.623 | Im'mm | 633 | Im'mm | I 2' 2 -1' |
| 624 | 71.536 | Im'm'm | 71.4.624 | Im'm'm | 634 | Im'm'm | -I 2 2' |
| 625 | 71.537 | Im'm'm' | 71.5.625 | Im'm'm' | 635 | Im'm'm' | I 2 2 -1' |
| 554 | 71.538 | I_cmmm | 65.10.554 | C_Immm | 636 | Immm1'_c[Cmmm] | -I 2 2 1'c |

```
630   72.539    Ibam         72.1.630    Ibam         637   Ibam               -I 2 2c
631   72.540    Ibam1'       72.2.631    Ibam1'       638   Ibam1'             -I 2 2c 1'
632   72.541    Ib'am        72.3.632    Ib'am        639   Ib'am              I 2' 2c -1'
633   72.542    Ibam'        72.4.633    Ibam'        640   Ibam'              I 2 2c' -1'
634   72.543    Ib'a'm       72.5.634    Ib'a'm       641   Ib'a'm             -I 2 2c'
635   72.544    Iba'm'       72.6.635    Iba'm'       642   Iba'm'             -I 2' 2c
636   72.545    Ib'a'm'      72.7.636    Ib'a'm'      643   Ib'a'm'            I 2 2c -1'
563   72.546    I_cbam       65.19.563   C_Im'm'm     644   Ibam1'_c[Cmmm]     -I 2 2c 1'c
586   72.547    I_bbam       67.10.586   C_Imma       645   Ibam1'_b[Bmcm]     -I 2 2c 1'b
643   73.548    Ibca         73.1.643    Ibca         646   Ibca               -I 2b 2c
644   73.549    Ibca1'       73.2.644    Ibca1'       647   Ibca1'             -I 2b 2c 1'
645   73.550    Ib'ca        73.3.645    Ib'ca        648   Ib'ca              I 2b' 2c -1'
646   73.551    Ib'c'a       73.4.646    Ib'c'a       649   Ib'c'a             -I 2b 2c'
647   73.552    Ib'c'a'      73.5.647    Ib'c'a'      650   Ib'c'a'            I 2b 2c -1'
593   73.553    I_cbca       67.17.593   C_Im'ma'     651   Ibca1'_c[Cmma]     -I 2b 2c 1'c
650   74.554    Imma         74.1.650    Imma         652   Imma               -I 2b 2
651   74.555    Imma1'       74.2.651    Imma1'       653   Imma1'             -I 2b 2  1'
652   74.556    Im'ma        74.3.652    Im'ma        654   Im'ma              I 2b' 2 -1'
653   74.557    Imma'        74.4.653    Imma'        655   Imma'              I 2b 2' -1'
654   74.558    Im'm'a       74.5.654    Im'm'a       656   Im'm'a             -I 2b 2'
655   74.559    Imm'a'       74.6.655    Imm'a'       657   Imm'a'             -I 2b' 2
656   74.560    Im'm'a'      74.7.656    Im'm'a'      658   Im'm'a'            I 2b 2  -1'
592   74.561    I_cmma       67.16.592   C_Imm'a      659   Imma1'_c[Cmma]     -I 2b 2 1'c
562   74.562    I_bmma       65.18.562   C_Im'mm      660   Imma1'_b[Bmmm]     -I 2b 2 1'b
661   75.1      P4           75.1.661    P4           661   P4                 P 4
662   75.2      P41'         75.2.662    P41'         662   P41'               P 4 1'
663   75.3      P4'          75.3.663    P4'          663   P4'                P 4'
664   75.4      P_c4         75.4.664    P_2c4        664   P41'_c[P4]         P 4 1'c
665   75.5      P_C4         75.5.665    P_P4         665   P41'_C[rP4]        P 4 1'C
686   75.6      P_I4         79.4.686    I_P4         666   P41'_I[I4]         P 4 1'I
668   76.7      P4_1         76.1.668    P4_1         667   P4_1               P 4w
669   76.8      P4_11'       76.2.669    P4_11'       668   P4_11'             P 4w 1'
670   76.9      P4_1'        76.3.670    P4_1'        669   P4_1'              P 4w'
675   76.10     P_c4_1       77.4.675    P_2c4_2      670   P4_11'_c[P4_2]     P 4w 1'c
671   76.11     P_C4_1       76.4.671    P_P4_1       671   P4_11'_C[rP4_1]    P 4w 1'C
691   76.12     P_I4_1       80.4.691    I_P4_1       672   P4_11'_I[I4_1]     P 4w 1'I
672   77.13     P4_2         77.1.672    P4_2         673   P4_2               P 4c
673   77.14     P4_21'       77.2.673    P4_21'       674   P4_21'             P 4c 1'
674   77.15     P4_2'        77.3.674    P4_2'        675   P4_2'              P 4c'
667   77.16     P_c4_2       75.7.667    P_2c4'       676   P4_21'_c[P4]       P 4c 1'c
676   77.17     P_C4_2       77.5.676    P_P4_2       677   P4_21'_C[rP4_2]    P 4c 1'C
687   77.18     P_I4_2       79.5.687    I_P4'        678   P4_21'_I[I4]       P 4c 1'I
679   78.19     P4_3         78.1.679    P4_3         679   P4_3               P 4cw
680   78.20     P4_31'       78.2.680    P4_31'       680   P4_31'             P 4cw  1'
681   78.21     P4_3'        78.3.681    P4_3'        681   P4_3'              P 4cw'
678   78.22     P_c4_3       77.7.678    P_2c4_2'     682   P4_31'_c[P4_2]     P 4cw 1'c
682   78.23     P_C4_3       78.4.682    P_P4_3       683   P4_31'_C[rP4_3]    P 4cw 1'C
692   78.24     P_I4_3       80.5.692    I_P4_1'      684   P4_31'_I[I4_1]     P 4cw 1'I
683   79.25     I4           79.1.683    I4           685   I4                 I 4
684   79.26     I41'         79.2.684    I41'         686   I41'               I 4 1'
685   79.27     I4'          79.3.685    I4'          687   I4'                I 4'
666   79.28     I_c4         75.6.666    P_I4         688   I41'_c[rP4]        I 4 1'c
688   80.29     I4_1         80.1.688    I4_1         689   I4_1               I 4bw
689   80.30     I4_11'       80.2.689    I4_11'       690   I4_11'             I 4bw  1'
690   80.31     I4_1'        80.3.690    I4_1'        691   I4_1'              I 4bw'
677   80.32     I_c4_1       77.6.677    P_I4_2       692   I4_11'_c[rP4_2]    I 4bw 1'c
693   81.33     P-4          81.1.693    P-4          693   P-4                P -4
694   81.34     P-41'        81.2.694    P-41'        694   P-41'              P -4  1'
695   81.35     P-4'         81.3.695    P-4'         695   P-4'               P -4'
696   81.36     P_c-4        81.4.696    P_2c-4       696   P-41'_c[P-4]       P -4 1'c
697   81.37     P_C-4        81.5.697    P_P-4        697   P-41'_C[rP-4]      P -4 1'C
702   81.38     P_I-4        82.4.702    I_P-4        698   P-41'_I[I-4]       P -4 1'I
699   82.39     I-4          82.1.699    I-4          699   I-4                I -4
700   82.40     I-41'        82.2.700    I-41'        700   I-41'              I -4  1'
701   82.41     I-4'         82.3.701    I-4'         701   I-4'               I -4'
698   82.42     I_c-4        81.6.698    P_I-4        702   I-41'_c[rP-4]      I -4 1'c
703   83.43     P4/m         83.1.703    P4/m         703   P4/m               -P 4
704   83.44     P4/m1'       83.2.704    P4/m1'       704   P4/m1'             -P 4 1'
705   83.45     P4'/m        83.3.705    P4'/m        705   P4'/m              -P 4'
706   83.46     P4/m'        83.4.706    P4/m'        706   P4/m'              P 4 -1'
707   83.47     P4'/m'       83.5.707    P4'/m'       707   P4'/m'             P 4' -1'
708   83.48     P_c4/m       83.6.708    P_2c4/m      708   P4/m1'_c[P4/m]     -P 4 1'c
709   83.49     P_C4/m       83.7.709    P_P4/m       709   P4/m1'_C[rP4/m]    -P 4 1'C
738   83.50     P_I4/m       87.6.738    I_P4/m       710   P4/m1'_I[I4/m]     -P 4 1'I
713   84.51     P4_2/m       84.1.713    P4_2/m       711   P4_2/m             -P 4c
714   84.52     P4_2/m1'     84.2.714    P4_2/m1'     712   P4_2/m1'           -P 4c   1'
715   84.53     P4_2'/m      84.3.715    P4_2'/m      713   P4_2'/m            -P 4c'
716   84.54     P4_2/m'      84.4.716    P4_2/m'      714   P4_2/m'            P 4c -1'
717   84.55     P4_2'/m'     84.5.717    P4_2'/m'     715   P4_2'/m'           P 4c' -1'
711   84.56     P_c4_2/m     83.9.711    P_2c4'/m     716   P4_2/m1'_c[P4/m]   -P 4c 1'c
718   84.57     P_C4_2/m     84.6.718    P_P4_2/m     717   P4_2/m1'_C[rP4_2/m] -P 4c 1'C
```

```
739   84.58    P_I4_2/m      87.7.739   I_P4'/m       718  P4_2/m1'_I[I4/m]      -P 4c 1'I
720   85.59    P4/n          85.1.720   P4/n          719  P4/n                  -P 4a
721   85.60    P4/n1'        85.2.721   P4/n1'        720  P4/n1'                -P 4a    1'
722   85.61    P4'/n         85.3.722   P4'/n         721  P4'/n                 -P 4a'
723   85.62    P4/n'         85.4.723   P4/n'         722  P4/n'                 P 4a -1'
724   85.63    P4'/n'        85.5.724   P4'/n'        723  P4'/n'                P 4a' -1'
725   85.64    P_c4/n        85.6.725   P_2c4/n       724  P4/n1'_c[P4/n]        -P 4a 1'c
712   85.65    P_C4/n        83.10.712  P_P4/m'       725  P4/n1'_C[rP4/m]       -P 4a 1'C
740   85.66    P_I4/n        87.8.740   I_P4/m'       726  P4/n1'_I[I4/m]        -P 4a 1'I
727   86.67    P4_2/n        86.1.727   P4_2/n        727  P4_2/n                -P 4bc
728   86.68    P4_2/n1'      86.2.728   P4_2/n1'      728  P4_2/n1'              -P 4bc   1'
729   86.69    P4_2'/n       86.3.729   P4_2'/n       729  P4_2'/n               -P 4bc'
730   86.70    P4_2/n'       86.4.730   P4_2/n'       730  P4_2/n'               P 4bc -1'
731   86.71    P4_2'/n'      86.5.731   P4_2'/n'      731  P4_2'/n'              P 4bc' -1'
726   86.72    P_c4_2/n      85.7.726   P_2c4'/n      732  P4_2/n1'_c[P4/n]      -P 4bc 1'c
719   86.73    P_C4_2/n      84.7.719   P_P4_2/m'     733  P4_2/n1'_C[rP4_2/m]   -P 4bc 1'C
741   86.74    P_I4_2/n      87.9.741   I_P4'/m'      734  P4_2/n1'_I[I4/m]      -P 4bc 1'I
733   87.75    I4/m          87.1.733   I4/m          735  I4/m                  -I 4
734   87.76    I4/m1'        87.2.734   I4/m1'        736  I4/m1'                -I 4 1'
735   87.77    I4'/m         87.3.735   I4'/m         737  I4'/m                 -I 4'
736   87.78    I4/m'         87.4.736   I4/m'         738  I4/m'                 I 4 -1'
737   87.79    I4'/m'        87.5.737   I4'/m'        739  I4'/m'                I 4' -1'
710   87.80    I_c4/m        83.8.710   P_I4/m        740  I4/m1'_c[rP4/m]       -I 4 1'c
742   88.81    I4_1/a        88.1.742   I4_1/a        741  I4_1/a                -I 4ad
743   88.82    I4_1/a1'      88.2.743   I4_1/a1'      742  I4_1/a1'              -I 4ad   1'
744   88.83    I4_1'/a       88.3.744   I4_1'/a       743  I4_1'/a               -I 4ad'
745   88.84    I4_1/a'       88.4.745   I4_1/a'       744  I4_1/a'               I 4ad -1'
746   88.85    I4_1'/a'      88.5.746   I4_1'/a'      745  I4_1'/a'              I 4ad' -1'
732   88.86    I_c4_1/a      86.6.732   P_I4_2/n      746  I4_1/a1'_c[rP4_2/n]   -I 4ad 1'c
747   89.87    P422          89.1.747   P422          747  P422                  P 4 2
748   89.88    P4221'        89.2.748   P4221'        748  P4221'                P 4 2 1'
749   89.89    P4'22'        89.3.749   P4'22'        749  P4'22'                P 4' 2
750   89.90    P42'2'        89.4.750   P42'2'        750  P42'2'                P 4 2'
751   89.91    P4'2'2        89.5.751   P4'2'2        751  P4'2'2                P 4' 2'
752   89.92    P_c422        89.6.752   P_2c422       752  P4221'_c[P422]        P 4 2 1'c
753   89.93    P_C422        89.7.753   P_P422        753  P4221'_C[rP422]       P 4 2 1'C
810   89.94    P_I422        97.6.810   I_P422        754  P4221'_I[I422]        P 4 2 1'I
757   90.95    P42_12        90.1.757   P42_12        755  P42_12                P 4ab  2ab
758   90.96    P42_121'      90.2.758   P42_121'      756  P42_121'              P 4ab  2ab  1'
759   90.97    P4'2_12'      90.3.759   P4'2_12'      757  P4'2_12'              P 4ab' 2ab
760   90.98    P42_1'2'      90.4.760   P42_1'2'      758  P42_1'2'              P 4ab  2ab'
761   90.99    P4'2_1'2      90.5.761   P4'2_1'2      759  P4'2_1'2              P 4ab' 2ab'
762   90.100   P_c42_12      90.6.762   P_2c42_12     760  P42_121'_c[P42_12]    P 4ab 2ab 1'c
756   90.101   P_C42_12      89.10.756  P_P4'22'      761  P42_121'_C[rP422]     P 4ab 2ab 1'C
812   90.102   P_I42_12      97.8.812   I_P42'2'      762  P42_121'_I[I422]      P 4ab 2ab 1'I
764   91.103   P4_122        91.1.764   P4_122        763  P4_122                P 4w   2c
765   91.104   P4_1221'      91.2.765   P4_1221'      764  P4_1221'              P 4w   2c  1'
766   91.105   P4_1'22'      91.3.766   P4_1'22'      765  P4_1'22'              P 4w'  2c
767   91.106   P4_12'2'      91.4.767   P4_12'2'      766  P4_12'2'              P 4w   2c'
768   91.107   P4_1'2'2      91.5.768   P4_1'2'2      767  P4_1'2'2              P 4w'  2c'
781   91.108   P_c4_122      93.6.781   P_2c4_222     768  P4_1221'_c[P4_222]    P 4w 2c 1'c
769   91.109   P_C4_122      91.6.769   P_P4_122      769  P4_1221'_C[rP4_122]   P 4w 2c 1'C
819   91.110   P_I4_122      98.6.819   I_P4_122      770  P4_1221'_I[I4_122]    P 4w 2c 1'I
771   92.111   P4_12_12      92.1.771   P4_12_12      771  P4_12_12              P 4abw   2nw
772   92.112   P4_12_121'    92.2.772   P4_12_121'    772  P4_12_121'            P 4abw   2nw   1'
773   92.113   P4_1'2_12'    92.3.773   P4_1'2_12'    773  P4_1'2_12'            P 4abw'  2nw
774   92.114   P4_12_1'2'    92.4.774   P4_12_1'2'    774  P4_12_1'2'            P 4abw   2nw'
775   92.115   P4_1'2_1'2    92.5.775   P4_1'2_1'2    775  P4_1'2_1'2            P 4abw'  2nw'
791   92.116   P_c4_12_12    94.6.791   P_2c4_22_12   776  P4_12_121'_c[P4_22_12] P 4abw 2nw 1'c
770   92.117   P_C4_12_12    91.7.770   P_P4_1'22'    777  P4_12_121'_C[rP4_122] P 4abw 2nw 1'C
821   92.118   P_I4_12_12    98.8.821   I_P4_12'2'    778  P4_12_121'_I[I4_122]  P 4abw 2nw 1'I
776   93.119   P4_222        93.1.776   P4_222        779  P4_222                P 4c   2
777   93.120   P4_2221'      93.2.777   P4_2221'      780  P4_2221'              P 4c   2  1'
778   93.121   P4_2'22'      93.3.778   P4_2'22'      781  P4_2'22'              P 4c'  2
779   93.122   P4_22'2'      93.4.779   P4_22'2'      782  P4_22'2'              P 4c   2'
780   93.123   P4_2'2'2      93.5.780   P4_2'2'2      783  P4_2'2'2              P 4c'  2'
755   93.124   P_c4_222      89.9.755   P_2c4'22'     784  P4_2221'_c[P422]      P 4c 2 1'c
782   93.125   P_C4_222      93.7.782   P_P4_222      785  P4_2221'_C[rP4_222]   P 4c 2 1'C
811   93.126   P_I4_222      97.7.811   I_P4'22'      786  P4_2221'_I[I422]      P 4c 2 1'I
786   94.127   P4_22_12      94.1.786   P4_22_12      787  P4_22_12              P 4n   2n
787   94.128   P4_22_121'    94.2.787   P4_22_121'    788  P4_22_121'            P 4n   2n  1'
788   94.129   P4_2'2_12'    94.3.788   P4_2'2_12'    789  P4_2'2_12'            P 4n'  2n
789   94.130   P4_22_1'2'    94.4.789   P4_22_1'2'    790  P4_22_1'2'            P 4n   2n'
790   94.131   P4_2'2_1'2    94.5.790   P4_2'2_1'2    791  P4_2'2_1'2            P 4n'  2n'
763   94.132   P_c4_22_12    90.7.763   P_2c4'2_1'2'  792  P4_22_121'_c[P42_12]  P 4n 2n 1'c
785   94.133   P_C4_22_12    93.10.785  P_P4_2'22'    793  P4_22_121'_C[rP4_222] P 4n 2n 1'C
813   94.134   P_I4_22_12    97.9.813   I_P4'2'2      794  P4_22_121'_I[I422]    P 4n 2n 1'I
793   95.135   P4_322        95.1.793   P4_322        795  P4_322                P 4cw   2c
794   95.136   P4_3221'      95.2.794   P4_3221'      796  P4_3221'              P 4cw   2c   1'
795   95.137   P4_3'22'      95.3.795   P4_3'22'      797  P4_3'22'              P 4cw'  2c
796   95.138   P4_32'2'      95.4.796   P4_32'2'      798  P4_32'2'              P 4cw   2c'
```

```
797   95.139    P4_3'2'2         95.5.797   P4_3'2'2          799  P4_3'2'2              P 4cw' 2c'
784   95.140    P_c4_322         93.9.784   P_2c4_2'22'       800  P4_3221'_c[P4_222]    P 4cw  2c 1'c
798   95.141    P_C4_322         95.6.798   P_P4_322          801  P4_3221'_C[rP4_322]   P 4cw  2c 1'C
820   95.142    P_I4_322         98.7.820   I_P4_1'22'        802  P4_3221'_I[I4_122]    P 4cw  2c 1'I
800   96.143    P4_32_12         96.1.800   P4_32_12          803  P4_32_12              P 4nw  2abw
801   96.144    P4_32_121'       96.2.801   P4_32_121'        804  P4_32_121'            P 4nw  2abw  1'
802   96.145    P4_3'2_12'       96.3.802   P4_3'2_12'        805  P4_3'2_12'            P 4nw' 2abw
803   96.146    P4_32_1'2'       96.4.803   P4_32_1'2'        806  P4_32_1'2'            P 4nw  2abw'
804   96.147    P4_3'2_1'2       96.5.804   P4_3'2_1'2        807  P4_3'2_1'2            P 4nw' 2abw'
792   96.148    P_c4_32_12       94.7.792   P_2c4_2'2_1'2     808  P4_32_121'_c[P4_22_12] P 4nw 2abw 1'c
799   96.149    P_C4_32_12       95.7.799   P_P4_3'22'        809  P4_32_121'_C[rP4_322] P 4nw 2abw 1'C
822   96.150    P_I4_32_12       98.9.822   I_P4_1'2'2        810  P4_32_121'_I[I4_122]  P 4nw 2abw 1'I
805   97.151    I422             97.1.805   I422              811  I422                  I 4 2
806   97.152    I4221'           97.2.806   I4221'            812  I4221'                I 4 2 1'
807   97.153    I4'22'           97.3.807   I4'22'            813  I4'22'                I 4' 2
808   97.154    I42'2'           97.4.808   I42'2'            814  I42'2'                I 4 2'
809   97.155    I4'2'2           97.5.809   I4'2'2            815  I4'2'2                I 4' 2'
754   97.156    I_c422           89.8.754   P_I422            816  I4221'_c[rP422]       I 4 2 1'c
814   98.157    I4_122           98.1.814   I4_122            817  I4_122                I 4bw  2bw
815   98.158    I4_1221'         98.2.815   I4_1221'          818  I4_1221'              I 4bw  2bw  1'
816   98.159    I4_1'22'         98.3.816   I4_1'22'          819  I4_1'22'              I 4bw' 2bw
817   98.160    I4_12'2'         98.4.817   I4_12'2'          820  I4_12'2'              I 4bw  2bw'
818   98.161    I4_1'2'2         98.5.818   I4_1'2'2          821  I4_1'2'2              I 4bw' 2bw'
783   98.162    I_c4_122         93.8.783   P_I4_222          822  I4_1221'_c[rP4_222]   I 4bw 2bw 1'c
823   99.163    P4mm             99.1.823   P4mm              823  P4mm                  P 4 -2
824   99.164    P4mm1'           99.2.824   P4mm1'            824  P4mm1'                P 4 -2 1'
825   99.165    P4'm'm           99.3.825   P4'm'm            825  P4'm'm                P 4' -2'
826   99.166    P4'mm'           99.4.826   P4'mm'            826  P4'mm'                P 4' -2
827   99.167    P4m'm'           99.5.827   P4m'm'            827  P4m'm'                P 4 -2'
828   99.168    P_c4mm           99.6.828   P_2c4mm           828  P4mm1'_c[P4mm]        P 4 -2 1'c
829   99.169    P_C4mm           99.7.829   P_P4mm            829  P4mm1'_C[rP4mm]       P 4 -2 1'C
888   99.170    P_I4mm          107.6.888   I_P4mm            830  P4mm1'_I[I4mm]        P 4 -2 1'I
836  100.171    P4bm            100.1.836   P4bm              831  P4bm                  P 4 -2ab
837  100.172    P4bm1'          100.2.837   P4bm1'            832  P4bm1'                P 4 -2ab 1'
838  100.173    P4'b'm          100.3.838   P4'b'm            833  P4'b'm                P 4' -2ab'
839  100.174    P4'bm'          100.4.839   P4'bm'            834  P4'bm'                P 4' -2ab
840  100.175    P4b'm'          100.5.840   P4b'm'            835  P4b'm'                P 4 -2ab'
841  100.176    P_c4bm          100.6.841   P_2c4bm           836  P4bm1'_c[P4bm]        P 4 -2ab 1'c
834  100.177    P_C4bm           99.12.834  P_P4'mm'          837  P4bm1'_C[rP4mm]       P 4 -2ab 1'C
897  100.178    P_I4bm          108.6.897   I_P4cm            838  P4bm1'_I[I4cm]        P 4 -2ab 1'I
845  101.179    P4_2cm          101.1.845   P4_2cm            839  P4_2cm                P 4c  -2c
846  101.180    P4_2cm1'        101.2.846   P4_2cm1'          840  P4_2cm1'              P 4c  -2c  1'
847  101.181    P4_2'c'm        101.3.847   P4_2'c'm          841  P4_2'c'm              P 4c' -2c'
848  101.182    P4_2'cm'        101.4.848   P4_2'cm'          842  P4_2'cm'              P 4c' -2c
849  101.183    P4_2c'm'        101.5.849   P4_2c'm'          843  P4_2c'm'              P 4c  -2c'
831  101.184    P_c4_2cm         99.9.831   P_2c4'm'm         844  P4_2cm1'_c[P4mm]      P 4c -2c 1'c
876  101.185    P_C4_2cm        105.6.876   P_P4_2mc          845  P4_2cm1'_C[rP4_2mc]   P 4c -2c 1'C
898  101.186    P_I4_2cm        108.7.898   I_P4'c'm          846  P4_2cm1'_I[I4cm]      P 4c -2c 1'I
852  102.187    P4_2nm          102.1.852   P4_2nm            847  P4_2nm                P 4n  -2n
853  102.188    P4_2nm1'        102.2.853   P4_2nm1'          848  P4_2nm1'              P 4n  -2n  1'
854  102.189    P4_2'n'm        102.3.854   P4_2'n'm          849  P4_2'n'm              P 4n' -2n'
855  102.190    P4_2'nm'        102.4.855   P4_2'nm'          850  P4_2'nm'              P 4n' -2n
856  102.191    P4_2n'm'        102.5.856   P4_2n'm'          851  P4_2n'm'              P 4n  -2n'
842  102.192    P_c4_2nm        100.7.842   P_2c4'b'm         852  P4_2nm1'_c[P4bm]      P 4n -2n 1'c
877  102.193    P_C4_2nm        105.7.877   P_P4_2'mc'        853  P4_2nm1'_C[rP4_2mc]   P 4n -2n 1'C
889  102.194    P_I4_2nm        107.7.889   I_P4'm'm          854  P4_2nm1'_I[I4mm]      P 4n -2n 1'I
859  103.195    P4cc            103.1.859   P4cc              855  P4cc                  P 4 -2c
860  103.196    P4cc1'          103.2.860   P4cc1'            856  P4cc1'                P 4 -2c 1'
861  103.197    P4'c'c          103.3.861   P4'c'c            857  P4'c'c                P 4' -2c'
862  103.198    P4'cc'          103.4.862   P4'cc'            858  P4'cc'                P 4' -2c
863  103.199    P4c'c'          103.5.863   P4c'c'            859  P4c'c'                P 4 -2c'
833  103.200    P_c4cc           99.11.833  P_2c4m'm'         860  P4cc1'_c[P4mm]        P 4 -2c 1'c
864  103.201    P_C4cc          103.6.864   P_P4cc            861  P4cc1'_C[rP4cc]       P 4 -2c 1'C
900  103.202    P_I4cc          108.9.900   I_P4c'm'          862  P4cc1'_I[I4cm]        P 4 -2c 1'I
866  104.203    P4nc            104.1.866   P4nc              863  P4nc                  P 4 -2n
867  104.204    P4nc1'          104.2.867   P4nc1'            864  P4nc1'                P 4 -2n 1'
868  104.205    P4'n'c          104.3.868   P4'n'c            865  P4'n'c                P 4' -2n'
869  104.206    P4'nc'          104.4.869   P4'nc'            866  P4'nc'                P 4' -2n
870  104.207    P4n'c'          104.5.870   P4n'c'            867  P4n'c'                P 4 -2n'
844  104.208    P_c4nc          100.9.844   P_2c4b'm'         868  P4nc1'_c[P4bm]        P 4 -2n 1'c
865  104.209    P_C4nc          103.7.865   P_P4'cc'          869  P4nc1'_C[rP4cc]       P 4 -2n 1'C
891  104.210    P_I4nc          107.9.891   I_P4m'm'          870  P4nc1'_I[I4mm]        P 4 -2n 1'I
871  105.211    P4_2mc          105.1.871   P4_2mc            871  P4_2mc                P 4c  -2
872  105.212    P4_2mc1'        105.2.872   P4_2mc1'          872  P4_2mc1'              P 4c  -2  1'
873  105.213    P4_2'm'c        105.3.873   P4_2'm'c          873  P4_2'm'c              P 4c' -2'
874  105.214    P4_2'mc'        105.4.874   P4_2'mc'          874  P4_2'mc'              P 4c' -2
875  105.215    P4_2m'c'        105.5.875   P4_2m'c'          875  P4_2m'c'              P 4c  -2'
832  105.216    P_c4_2mc         99.10.832  P_2c4'mm'         876  P4_2mc1'_c[P4mm]      P 4c -2 1'c
850  105.217    P_C4_2mc        101.6.850   P_P4_2cm          877  P4_2mc1'_C[rP4_2cm]   P 4c -2 1'C
890  105.218    P_I4_2mc        107.8.890   I_P4'mm'          878  P4_2mc1'_I[I4mm]      P 4c -2 1'I
878  106.219    P4_2bc          106.1.878   P4_2bc            879  P4_2bc                P 4c  -2ab
```

```
879   106.220   P4_2bc1'       106.2.879    P4_2bc1'        880   P4_2bc1'                    P 4c  -2ab  1'
880   106.221   P4_2'b'c       106.3.880    P4_2'b'c        881   P4_2'b'c                    P 4c' -2ab'
881   106.222   P4_2'bc'       106.4.881    P4_2'bc'        882   P4_2'bc'                    P 4c' -2ab
882   106.223   P4_2b'c'       106.5.882    P4_2b'c'        883   P4_2b'c'                    P 4c  -2ab'
843   106.224   P_c4_2bc       100.8.843    P_2c4'bm'       884   P4_2bc1'_c[P4bm]            P 4c  -2ab 1'c
851   106.225   P_C4_2bc       101.7.851    P_P4_2'cm'      885   P4_2bc1'_C[rP4_2cm]         P 4c  -2ab 1'C
899   106.226   P_I4_2bc       108.8.899    I_P4'cm'        886   P4_2bc1'_I[I4cm]            P 4c  -2ab 1'I
883   107.227   I4mm           107.1.883    I4mm            887   I4mm                        I 4 -2
884   107.228   I4mm1'         107.2.884    I4mm1'          888   I4mm1'                      I 4 -2 1'
885   107.229   I4'm'm         107.3.885    I4'm'm          889   I4'm'm                      I 4' -2'
886   107.230   I4'mm'         107.4.886    I4'mm'          890   I4'mm'                      I 4' -2
887   107.231   I4m'm'         107.5.887    I4m'm'          891   I4m'm'                      I 4 -2'
830   107.232   I_c4mm         99.8.830     P_I4mm          892   I4mm1'_c[rP4mm]             I 4 -2 1'c
892   108.233   I4cm           108.1.892    I4cm            893   I4cm                        I 4 -2c
893   108.234   I4cm1'         108.2.893    I4cm1'          894   I4cm1'                      I 4 -2c 1'
894   108.235   I4'c'm         108.3.894    I4'c'm          895   I4'c'm                      I 4' -2c'
895   108.236   I4'cm'         108.4.895    I4'cm'          896   I4'cm'                      I 4' -2c
896   108.237   I4c'm'         108.5.896    I4c'm'          897   I4c'm'                      I 4 -2c'
835   108.238   I_c4cm         99.13.835    P_I4m'm'        898   I4cm1'_c[rP4mm]             I 4 -2c 1'c
901   109.239   I4_1md         109.1.901    I4_1md          899   I4_1md                      I 4bw  -2
902   109.240   I4_1md1'       109.2.902    I4_1md1'        900   I4_1md1'                    I 4bw  -2  1'
903   109.241   I4_1'm'd       109.3.903    I4_1'm'd        901   I4_1'm'd                    I 4bw' -2'
904   109.242   I4_1'md'       109.4.904    I4_1'md'        902   I4_1'md'                    I 4bw' -2
905   109.243   I4_1m'd'       109.5.905    I4_1m'd'        903   I4_1m'd'                    I 4bw  -2'
857   109.244   I_c4_1md       102.6.857    P_I4_2nm        904   I4_1md1'_c[rP4_2nm]         I 4bw -2 1'c
906   110.245   I4_1cd         110.1.906    I4_1cd          905   I4_1cd                      I 4bw  -2c
907   110.246   I4_1cd1'       110.2.907    I4_1cd1'        906   I4_1cd1'                    I 4bw  -2c  1'
908   110.247   I4_1'c'd       110.3.908    I4_1'c'd        907   I4_1'c'd                    I 4bw' -2c'
909   110.248   I4_1'cd'       110.4.909    I4_1'cd'        908   I4_1'cd'                    I 4bw' -2c
910   110.249   I4_1c'd'       110.5.910    I4_1c'd'        909   I4_1c'd'                    I 4bw  -2c'
858   110.250   I_c4_1cd       102.7.858    P_I4_2n'm'      910   I4_1cd1'_c[rP4_2nm]         I 4bw -2c 1'c
911   111.251   P-42m          111.1.911    P-42m           911   P-42m                       P -4   2
912   111.252   P-42m1'        111.2.912    P-42m1'         912   P-42m1'                     P -4   2  1'
913   111.253   P-4'2'm        111.3.913    P-4'2'm         913   P-4'2'm                     P -4'  2'
914   111.254   P-4'2m'        111.4.914    P-4'2m'         914   P-4'2m'                     P -4'  2
915   111.255   P-42'm'        111.5.915    P-42'm'         915   P-42'm'                     P -4   2'
916   111.256   P_c-42m        111.6.916    P_2c-42m        916   P-42m1'_c[P-42m]            P -4 2 1'c
947   111.257   P_C-42m        115.7.947    P_P-4m2         917   P-42m1'_C[rP-4m2]           P -4 2 1'C
990   111.258   P_I-42m        121.6.990    I_P-42m         918   P-42m1'_I[I-42m]            P -4 2 1'I
922   112.259   P-42c          112.1.922    P-42c           919   P-42c                       P -4   2c
923   112.260   P-42c1'        112.2.923    P-42c1'         920   P-42c1'                     P -4   2c  1'
924   112.261   P-4'2'c        112.3.924    P-4'2'c         921   P-4'2'c                     P -4'  2c'
925   112.262   P-4'2c'        112.4.925    P-4'2c'         922   P-4'2c'                     P -4'  2c
926   112.263   P-42'c'        112.5.926    P-42'c'         923   P-42'c'                     P -4   2c'
919   112.264   P_c-42c        111.9.919    P_2c-42'm'      924   P-42c1'_c[P-42m]            P -4 2c 1'c
956   112.265   P_C-42c        116.6.956    P_P-4c2         925   P-42c1'_C[rP-4c2]           P -4 2c 1'C
992   112.266   P_I-42c        121.8.992    I_P-4'2m'       926   P-42c1'_I[I-42m]            P -4 2c 1'I
929   113.267   P-42_1m        113.1.929    P-42_1m         927   P-42_1m                     P -4   2ab
930   113.268   P-42_1m1'      113.2.930    P-42_1m1'       928   P-42_1m1'                   P -4   2ab  1'
931   113.269   P-4'2_1'm      113.3.931    P-4'2_1'm       929   P-4'2_1'm                   P -4'  2ab'
932   113.270   P-4'2_1m'      113.4.932    P-4'2_1m'       930   P-4'2_1m'                   P -4'  2ab
933   113.271   P-42_1'm'      113.5.933    P-42_1'm'       931   P-42_1'm'                   P -4   2ab'
934   113.272   P_c-42_1m      113.6.934    P_2c-42_1m      932   P-42_1m1'_c[P-42_1m]        P -4 2ab 1'c
950   113.273   P_C-42_1m      115.10.950   P_P-4'm2'       933   P-42_1m1'_C[rP-4m2]         P -4 2ab 1'C
991   113.274   P_I-42_1m      121.7.991    I_P-4'2'm       934   P-42_1m1'_I[I-42m]          P -4 2ab 1'I
936   114.275   P-42_1c        114.1.936    P-42_1c         935   P-42_1c                     P -4   2n
937   114.276   P-42_1c1'      114.2.937    P-42_1c1'       936   P-42_1c1'                   P -4   2n  1'
938   114.277   P-4'2_1'c      114.3.938    P-4'2_1'c       937   P-4'2_1'c                   P -4'  2n'
939   114.278   P-4'2_1c'      114.4.939    P-4'2_1c'       938   P-4'2_1c'                   P -4'  2n
940   114.279   P-42_1'c'      114.5.940    P-42_1'c'       939   P-42_1'c'                   P -4   2n'
935   114.280   P_c-42_1c      113.7.935    P_2c-4'2_1m'    940   P-42_1c1'_c[P-42_1m]        P -4 2n 1'c
957   114.281   P_C-42_1c      116.7.957    P_P-4'c2'       941   P-42_1c1'_C[rP-4c2]         P -4 2n 1'C
993   114.282   P_I-42_1c      121.9.993    I_P-42'm'       942   P-42_1c1'_I[I-42m]          P -4 2n 1'I
941   115.283   P-4m2          115.1.941    P-4m2           943   P-4m2                       P -4   -2
942   115.284   P-4m21'        115.2.942    P-4m21'         944   P-4m21'                     P -4   -2  1'
943   115.285   P-4'm'2        115.3.943    P-4'm'2         945   P-4'm'2                     P -4'  -2'
944   115.286   P-4'm2'        115.4.944    P-4'm2'         946   P-4'm2'                     P -4'  -2
945   115.287   P-4m'2'        115.5.945    P-4m'2'         947   P-4m'2'                     P -4   -2'
946   115.288   P_c-4m2        115.6.946    P_2c-4m2        948   P-4m21'_c[P-4m2]            P -4 -2 1'c
917   115.289   P_C-4m2        111.7.917    P_P-42m         949   P-4m21'_C[rP-42m]           P -4 -2 1'C
976   115.290   P_I-4m2        119.6.976    I_P-4m2         950   P-4m21'_I[I-4m2]            P -4 -2 1'I
951   116.291   P-4c2          116.1.951    P-4c2           951   P-4c2                       P -4   -2c
952   116.292   P-4c21'        116.2.952    P-4c21'         952   P-4c21'                     P -4   -2c  1'
953   116.293   P-4'c'2        116.3.953    P-4'c'2         953   P-4'c'2                     P -4'  -2c'
954   116.294   P-4'c2'        116.4.954    P-4'c2'         954   P-4'c2'                     P -4'  -2c
955   116.295   P-4c'2'        116.5.955    P-4c'2'         955   P-4c'2'                     P -4   -2c'
949   116.296   P_c-4c2        115.9.949    P_2c-4'm'2      956   P-4c21'_c[P-4m2]            P -4 -2c 1'c
927   116.297   P_C-4c2        112.6.927    P_P-42c         957   P-4c21'_C[rP-42c]           P -4 -2c 1'C
983   116.298   P_I-4c2        120.6.983    I_P-4c2         958   P-4c21'_I[I-4c2]            P -4 -2c 1'I
958   117.299   P-4b2          117.1.958    P-4b2           959   P-4b2                       P -4   -2ab
959   117.300   P-4b21'        117.2.959    P-4b21'         960   P-4b21'                     P -4   -2ab  1'
```

```
960   117.301  P-4'b'2       117.3.960   P-4'b'2       961   P-4'b'2              P -4' -2ab'
961   117.302  P-4'b2'       117.4.961   P-4'b2'       962   P-4'b2'              P -4' -2ab
962   117.303  P-4b'2'       117.5.962   P-4b'2'       963   P-4b'2'              P -4  -2ab'
963   117.304  P_c-4b2       117.6.963   P_2c-4b2      964   P-4b21'_c[P-4b2]     P -4  -2ab 1'c
920   117.305  P_C-4b2       111.10.920  P_P-4'2m'     965   P-4b21'_C[rP-42m]    P -4  -2ab 1'C
984   117.306  P_I-4b2       120.7.984   I_P-4c'2'     966   P-4b21'_I[I-4c2]     P -4  -2ab 1'I
965   118.307  P-4n2         118.1.965   P-4n2         967   P-4n2                P -4  -2n
966   118.308  P-4n21'       118.2.966   P-4n21'       968   P-4n21'              P -4  -2n  1'
967   118.309  P-4'n'2       118.3.967   P-4'n'2       969   P-4'n'2              P -4' -2n
968   118.310  P-4'n2'       118.4.968   P-4'n2'       970   P-4'n2'              P -4' -2n
969   118.311  P-4n'2'       118.5.969   P-4n'2'       971   P-4n'2'              P -4  -2n'
964   118.312  P_c-4n2       117.7.964   P_2c-4'b'2    972   P-4n21'_c[P-4b2]     P -4  -2n 1'c
928   118.313  P_C-4n2       112.7.928   P_P-4'2c'     973   P-4n21'_C[rP-42c]    P -4  -2n 1'C
977   118.314  P_I-4n2       119.7.977   I_P-4'm'2     974   P-4n21'_I[I-4m2]     P -4  -2n 1'I
971   119.315  I-4m2         119.1.971   I-4m2         975   I-4m2                I -4  -2
972   119.316  I-4m21'       119.2.972   I-4m21'       976   I-4m21'              I -4  -2   1'
973   119.317  I-4'm'2       119.3.973   I-4'm'2       977   I-4'm'2              I -4' -2'
974   119.318  I-4'm2'       119.4.974   I-4'm2'       978   I-4'm2'              I -4' -2
975   119.319  I-4m'2'       119.5.975   I-4m'2'       979   I-4m'2'              I -4  -2'
918   119.320  I_c-4m2       111.8.918   P_I-42m       980   I-4m21'_c[rP-42m]    I -4  -2 1'c
978   120.321  I-4c2         120.1.978   I-4c2         981   I-4c2                I -4  -2c
979   120.322  I-4c21'       120.2.979   I-4c21'       982   I-4c21'              I -4  -2c  1'
980   120.323  I-4'c'2       120.3.980   I-4'c'2       983   I-4'c'2              I -4' -2c'
981   120.324  I-4'c2'       120.4.981   I-4'c2'       984   I-4'c2'              I -4' -2c
982   120.325  I-4c'2'       120.5.982   I-4c'2'       985   I-4c'2'              I -4  -2c'
921   120.326  I_c-4c2       111.11.921  P_I-4'2m'     986   I-4c21'_c[rP-42m]    I -4  -2c 1'c
985   121.327  I-42m         121.1.985   I-42m         987   I-42m                I -4   2
986   121.328  I-42m1'       121.2.986   I-42m1'       988   I-42m1'              I -4   2   1'
987   121.329  I-4'2'm       121.3.987   I-4'2'm       989   I-4'2'm              I -4' 2'
988   121.330  I-4'2m'       121.4.988   I-4'2m'       990   I-4'2m'              I -4' 2
989   121.331  I-42'm'       121.5.989   I-42'm'       991   I-42'm'              I -4   2'
948   121.332  I_c-42m       115.8.948   P_I-4m2       992   I-42m1'_c[rP-4m2]    I -4   2 1'c
994   122.333  I-42d         122.1.994   I-42d         993   I-42d                I -4   2bw
995   122.334  I-42d1'       122.2.995   I-42d1'       994   I-42d1'              I -4   2bw  1'
996   122.335  I-4'2'd       122.3.996   I-4'2'd       995   I-4'2'd              I -4' 2bw'
997   122.336  I-4'2d'       122.4.997   I-4'2d'       996   I-4'2d'              I -4' 2bw
998   122.337  I-42'd'       122.5.998   I-42'd'       997   I-42'd'              I -4   2bw'
970   122.338  I_c-42d       118.6.970   P_I-4n2       998   I-42d1'_c[rP-4n2]    I -4   2bw 1'c
999   123.339  P4/mmm        123.1.999   P4/mmm        999   P4/mmm               -P 4 2
1000  123.340  P4/mmm1'      123.2.1000  P4/mmm1'      1000  P4/mmm1'             -P 4 2 1'
1001  123.341  P4/m'mm       123.3.1001  P4/m'mm       1001  P4/m'mm              P 4 2' -1'
1002  123.342  P4'/mm'm      123.4.1002  P4'/mm'm      1002  P4'/mm'm             -P 4' 2'
1003  123.343  P4'/mmm'      123.5.1003  P4'/mmm'      1003  P4'/mmm'             -P 4' 2
1004  123.344  P4'/m'm'm     123.6.1004  P4'/m'm'm     1004  P4'/m'm'm            P 4' 2  -1'
1005  123.345  P4/mm'm'      123.7.1005  P4/mm'm'      1005  P4/mm'm'             -P 4 2'
1006  123.346  P4'/m'mm'     123.8.1006  P4'/m'mm'     1006  P4'/m'mm'            P 4' 2' -1'
1007  123.347  P4/m'm'm'     123.9.1007  P4/m'm'm'     1007  P4/m'm'm'            P 4 2 -1'
1008  123.348  P_c4/mmm      123.10.1008 P_2c4/mmm     1008  P4/mmm1'_c[P4/mmm]   -P 4 2 1'c
1009  123.349  P_C4/mmm      123.11.1009 P_P4/mmm      1009  P4/mmm1'_C[rP4/mmm]  -P 4 2 1'C
1188  123.350  P_I4/mmm      139.10.1188 I_P4/mmm      1010  P4/mmm1'_I[I4/mmm]   -P 4 2 1'I
1018  124.351  P4/mcc        124.1.1018  P4/mcc        1011  P4/mcc               -P 4 2c
1019  124.352  P4/mcc1'      124.2.1019  P4/mcc1'      1012  P4/mcc1'             -P 4 2c 1'
1020  124.353  P4/m'cc       124.3.1020  P4/m'cc       1013  P4/m'cc              P 4 2c' -1'
1021  124.354  P4'/mc'c      124.4.1021  P4'/mc'c      1014  P4'/mc'c             -P 4' 2c'
1022  124.355  P4'/mcc'      124.5.1022  P4'/mcc'      1015  P4'/mcc'             -P 4' 2c
1023  124.356  P4'/m'c'c     124.6.1023  P4'/m'c'c     1016  P4'/m'c'c            P 4' 2c  -1'
1024  124.357  P4/mc'c'      124.7.1024  P4/mc'c'      1017  P4/mc'c'             -P 4 2c'
1025  124.358  P4'/m'cc'     124.8.1025  P4'/m'cc'     1018  P4'/m'cc'            P 4' 2c' -1'
1026  124.359  P4/m'c'c'     124.9.1026  P4/m'c'c'     1019  P4/m'c'c'            P 4 2c -1'
1013  124.360  P_c4/mcc      123.15.1013 P_2c4/mm'm'   1020  P4/mcc1'_c[P4/mmm]   -P 4 2c 1'c
1027  124.361  P_C4/mcc      124.10.1027 P_P4/mcc      1021  P4/mcc1'_C[rP4/mcc]  -P 4 2c 1'C
1205  124.362  P_I4/mcc      140.10.1205 I_P4/mcm      1022  P4/mcc1'_I[I4/mcm]   -P 4 2c 1'I
1031  125.363  P4/nbm        125.1.1031  P4/nbm        1023  P4/nbm               -P 4a  2b
1032  125.364  P4/nbm1'      125.2.1032  P4/nbm1'      1024  P4/nbm1'             -P 4a  2b   1'
1033  125.365  P4/n'bm       125.3.1033  P4/n'bm       1025  P4/n'bm              P 4a  2b' -1'
1034  125.366  P4'/nb'm      125.4.1034  P4'/nb'm      1026  P4'/nb'm             -P 4a' 2b'
1035  125.367  P4'/nbm'      125.5.1035  P4'/nbm'      1027  P4'/nbm'             -P 4a' 2b
1036  125.368  P4'/n'b'm     125.6.1036  P4'/n'b'm     1028  P4'/n'b'm            P 4a' 2b  -1'
1037  125.369  P4/nb'm'      125.7.1037  P4/nb'm'      1029  P4/nb'm'             -P 4a  2b'
1038  125.370  P4'/n'bm'     125.8.1038  P4'/n'bm'     1030  P4'/n'bm'            P 4a' 2b' -1'
1039  125.371  P4/n'b'm'     125.9.1039  P4/n'b'm'     1031  P4/n'b'm'            P 4a  2b  -1'
1040  125.372  P_c4/nbm      125.10.1040 P_2c4/nbm     1032  P4/nbm1'_c[P4/nbm]   -P 4a  2b 1'c
1016  125.373  P_C4/nbm      123.18.1016 P_P4'/m'mm    1033  P4/nbm1'_C[rP4/mmm]  -P 4a  2b 1'C
1212  125.374  P_I4/nbm      140.17.1212 I_P4/m'c'm'   1034  P4/nbm1'_I[I4/mcm]   -P 4a  2b 1'I
1044  126.375  P4/nnc        126.1.1044  P4/nnc        1035  P4/nnc               -P 4a  2bc
1045  126.376  P4/nnc1'      126.2.1045  P4/nnc1'      1036  P4/nnc1'             -P 4a  2bc  1'
1046  126.377  P4/n'nc       126.3.1046  P4/n'nc       1037  P4/n'nc              P 4a  2bc' -1'
1047  126.378  P4'/nn'c      126.4.1047  P4'/nn'c      1038  P4'/nn'c             -P 4a' 2bc'
1048  126.379  P4'/nnc'      126.5.1048  P4'/nnc'      1039  P4'/nnc'             -P 4a' 2bc
1049  126.380  P4'/n'n'c     126.6.1049  P4'/n'n'c     1040  P4'/n'n'c            P 4a' 2bc  -1'
1050  126.381  P4/nn'c'      126.7.1050  P4/nn'c'      1041  P4/nn'c'             -P 4a  2bc'
```

```
1051   126.382   P4'/n'nc'      126.8.1051   P4'/n'nc'       1042   P4'/n'nc'                  P 4a' 2bc'  -1'
1052   126.383   P4'/n'n'c      126.9.1052   P4'/n'n'c       1043   P4'/n'n'c                  P 4a  2bc   -1'
1043   126.384   P_c4/nnc       125.13.1043  P_2c4/nb'm'     1044   P4/nnc1'_c[P4/nbm]         -P 4a 2bc 1'c
1030   126.385   P_C4/nnc       124.13.1030  P_P4'/m'cc      1045   P4/nnc1'_C[rP4/mcc]        -P 4a 2bc 1'C
1195   126.386   P_I4/nnc       139.17.1195  I_P4'/m'm'm'    1046   P4/nnc1'_I[I4/mmm]         -P 4a 2bc 1'I
1053   127.387   P4/mbm         127.1.1053   P4/mbm          1047   P4/mbm                     -P 4 2ab
1054   127.388   P4/mbm1'       127.2.1054   P4/mbm1'        1048   P4/mbm1'                   -P 4 2ab 1'
1055   127.389   P4'/m'bm       127.3.1055   P4'/m'bm        1049   P4'/m'bm                   P 4 2ab' -1'
1056   127.390   P4'/mb'm       127.4.1056   P4'/mb'm        1050   P4'/mb'm                   -P 4' 2ab'
1057   127.391   P4'/mbm'       127.5.1057   P4'/mbm'        1051   P4'/mbm'                   -P 4' 2ab
1058   127.392   P4'/m'b'm      127.6.1058   P4'/m'b'm       1052   P4'/m'b'm                  P 4' 2ab -1'
1059   127.393   P4/mb'm'       127.7.1059   P4/mb'm'        1053   P4/mb'm'                   -P 4 2ab'
1060   127.394   P4'/m'bm'      127.8.1060   P4'/m'bm'       1054   P4'/m'bm'                  P 4' 2ab' -1'
1061   127.395   P4'/m'b'm'     127.9.1061   P4'/m'b'm'      1055   P4'/m'b'm'                 P 4 2ab -1'
1062   127.396   P_c4/mbm       127.10.1062  P_2c4/mbm       1056   P4/mbm1'_c[P4/mbm]         -P 4 2ab 1'c
1015   127.397   P_C4/mbm       123.17.1015  P_P4'/mmm'      1057   P4/mbm1'_C[rP4/mmm]        -P 4 2ab 1'C
1210   127.398   P_I4/mbm       140.15.1210  I_P4'/mc'm'     1058   P4/mbm1'_I[I4/mcm]         -P 4 2ab 1'I
1066   128.399   P4/mnc         128.1.1066   P4/mnc          1059   P4/mnc                     -P 4 2n
1067   128.400   P4/mnc1'       128.2.1067   P4/mnc1'        1060   P4/mnc1'                   -P 4 2n 1'
1068   128.401   P4'/m'nc       128.3.1068   P4'/m'nc        1061   P4'/m'nc                   P 4 2n' -1'
1069   128.402   P4'/mn'c       128.4.1069   P4'/mn'c        1062   P4'/mn'c                   -P 4' 2n'
1070   128.403   P4'/mnc'       128.5.1070   P4'/mnc'        1063   P4'/mnc'                   -P 4' 2n
1071   128.404   P4'/m'n'c      128.6.1071   P4'/m'n'c       1064   P4'/m'n'c                  P 4' 2n  -1'
1072   128.405   P4/mn'c'       128.7.1072   P4/mn'c'        1065   P4/mn'c'                   -P 4 2n'
1073   128.406   P4'/m'nc'      128.8.1073   P4'/m'nc'       1066   P4'/m'nc'                  P 4' 2n' -1'
1074   128.407   P4'/m'n'c'     128.9.1074   P4'/m'n'c'      1067   P4'/m'n'c'                 P 4 2n -1'
1065   128.408   P_c4/mnc       127.13.1065  P_2c4/mb'm'     1068   P4/mnc1'_c[P4/mbm]         -P 4 2n 1'c
1029   128.409   P_C4/mnc       124.12.1029  P_P4'/mcc'      1069   P4/mnc1'_C[rP4/mcc]        -P 4 2n 1'C
1193   128.410   P_I4/mnc       139.15.1193  I_P4'/mm'm'     1070   P4/mnc1'_I[I4/mmm]         -P 4 2n 1'I
1075   129.411   P4/nmm         129.1.1075   P4/nmm          1071   P4/nmm                     -P 4a  2a
1076   129.412   P4/nmm1'       129.2.1076   P4/nmm1'        1072   P4/nmm1'                   -P 4a  2a   1'
1077   129.413   P4'/n'mm       129.3.1077   P4'/n'mm        1073   P4'/n'mm                   P 4a  2a' -1'
1078   129.414   P4'/nm'm       129.4.1078   P4'/nm'm        1074   P4'/nm'm                   -P 4a' 2a'
1079   129.415   P4'/nmm'       129.5.1079   P4'/nmm'        1075   P4'/nmm'                   -P 4a' 2a
1080   129.416   P4'/n'm'm      129.6.1080   P4'/n'm'm       1076   P4'/n'm'm                  P 4a' 2a  -1'
1081   129.417   P4/nm'm'       129.7.1081   P4/nm'm'        1077   P4/nm'm'                   -P 4a  2a'
1082   129.418   P4'/n'mm'      129.8.1082   P4'/n'mm'       1078   P4'/n'mm'                  P 4a' 2a' -1'
1083   129.419   P4'/n'm'm'     129.9.1083   P4'/n'm'm'      1079   P4'/n'm'm'                 P 4a  2a  -1'
1084   129.420   P_c4/nmm       129.10.1084  P_2c4/nmm       1080   P4/nmm1'_c[P4/nmm]         -P 4a 2a 1'c
1014   129.421   P_C4/nmm       123.16.1014  P_P4/m'mm       1081   P4/nmm1'_C[rP4/mmm]        -P 4a 2a 1'C
1189   129.422   P_I4/nmm       139.11.1189  I_P4/m'mm       1082   P4/nmm1'_I[I4/mmm]         -P 4a 2a 1'I
1088   130.423   P4/ncc         130.1.1088   P4/ncc          1083   P4/ncc                     -P 4a  2ac
1089   130.424   P4/ncc1'       130.2.1089   P4/ncc1'        1084   P4/ncc1'                   -P 4a  2ac   1'
1090   130.425   P4'/n'cc       130.3.1090   P4'/n'cc        1085   P4'/n'cc                   P 4a  2ac' -1'
1091   130.426   P4'/nc'c       130.4.1091   P4'/nc'c        1086   P4'/nc'c                   -P 4a' 2ac'
1092   130.427   P4'/ncc'       130.5.1092   P4'/ncc'        1087   P4'/ncc'                   -P 4a' 2ac
1093   130.428   P4'/n'c'c      130.6.1093   P4'/n'c'c       1088   P4'/n'c'c                  P 4a' 2ac  -1'
1094   130.429   P4/nc'c'       130.7.1094   P4/nc'c'        1089   P4/nc'c'                   -P 4a  2ac'
1095   130.430   P4'/n'cc'      130.8.1095   P4'/n'cc'       1090   P4'/n'cc'                  P 4a' 2ac' -1'
1096   130.431   P4'/n'c'c'     130.9.1096   P4'/n'c'c'      1091   P4'/n'c'c'                 P 4a  2ac  -1'
1087   130.432   P_c4/ncc       129.13.1087  P_2c4/nm'm'     1092   P4/ncc1'_c[P4/nmm]         -P 4a 2ac 1'c
1028   130.433   P_C4/ncc       124.11.1028  P_P4/m'cc      1093   P4/ncc1'_C[rP4/mcc]        -P 4a 2ac 1'C
1206   130.434   P_I4/ncc       140.11.1206  I_P4/m'cm      1094   P4/ncc1'_I[I4/mcm]         -P 4a 2ac 1'I
1097   131.435   P4_2/mmc       131.1.1097   P4_2/mmc        1095   P4_2/mmc                   -P 4c  2
1098   131.436   P4_2/mmc1'     131.2.1098   P4_2/mmc1'      1096   P4_2/mmc1'                 -P 4c  2   1'
1099   131.437   P4_2/m'mc      131.3.1099   P4_2/m'mc       1097   P4_2/m'mc                  P 4c  2' -1'
1100   131.438   P4_2'/mm'c     131.4.1100   P4_2'/mm'c      1098   P4_2'/mm'c                 -P 4c' 2'
1101   131.439   P4_2'/mmc'     131.5.1101   P4_2'/mmc'      1099   P4_2'/mmc'                 -P 4c' 2
1102   131.440   P4_2'/m'm'c    131.6.1102   P4_2'/m'm'c     1100   P4_2'/m'm'c                P 4c' 2  -1'
1103   131.441   P4_2/mm'c'     131.7.1103   P4_2/mm'c'      1101   P4_2/mm'c'                 -P 4c  2'
1104   131.442   P4_2'/m'mc'    131.8.1104   P4_2'/m'mc'     1102   P4_2'/m'mc'                P 4c' 2' -1'
1105   131.443   P4_2'/m'm'c'   131.9.1105   P4_2'/m'm'c'    1103   P4_2'/m'm'c'               P 4c  2  -1'
1012   131.444   P_c4_2/mmc     123.14.1012  P_2c4'/mmm      1104   P4_2/mmc1'_c[P4/mmm]       -P 4c 2 1'c
1119   131.445   P_C4_2/mmc     132.10.1119  P_P4_2/mcm      1105   P4_2/mmc1'_C[rP4_2/mcm]    -P 4c 2 1'C
1191   131.446   P_I4_2/mmc     139.13.1191  I_P4'/mmm'      1106   P4_2/mmc1'_I[I4/mmm]       -P 4c 2 1'I
1110   132.447   P4_2/mcm       132.1.1110   P4_2/mcm        1107   P4_2/mcm                   -P 4c  2c
1111   132.448   P4_2/mcm1'     132.2.1111   P4_2/mcm1'      1108   P4_2/mcm1'                 -P 4c  2c   1'
1112   132.449   P4_2/m'cm      132.3.1112   P4_2/m'cm       1109   P4_2/m'cm                  P 4c  2c' -1'
1113   132.450   P4_2'/mc'm     132.4.1113   P4_2'/mc'm      1110   P4_2'/mc'm                 -P 4c' 2c'
1114   132.451   P4_2'/mcm'     132.5.1114   P4_2'/mcm'      1111   P4_2'/mcm'                 -P 4c' 2c
1115   132.452   P4_2'/m'c'm    132.6.1115   P4_2'/m'c'm     1112   P4_2'/m'c'm                P 4c  2c  -1'
1116   132.453   P4_2/mc'm'     132.7.1116   P4_2/mc'm'      1113   P4_2/mc'm'                 -P 4c  2c'
1117   132.454   P4_2'/m'cm'    132.8.1117   P4_2'/m'cm'     1114   P4_2'/m'cm'                P 4c' 2c' -1'
1118   132.455   P4_2'/m'c'm'   132.9.1118   P4_2'/m'c'm'    1115   P4_2'/m'c'm'               P 4c  2c  -1'
1011   132.456   P_c4_2/mcm     123.13.1011  P_2c4'/mm'm     1116   P4_2/mcm1'_c[P4/mmm]       -P 4c 2c 1'c
1106   132.457   P_C4_2/mcm     131.10.1106  P_P4_2/mmc      1117   P4_2/mcm1'_C[rP4_2/mmc]    -P 4c 2c 1'C
1208   132.458   P_I4_2/mcm     140.13.1208  I_P4'/mcm'      1118   P4_2/mcm1'_I[I4/mcm]       -P 4c 2c 1'I
1123   133.459   P4_2/nbc       133.1.1123   P4_2/nbc        1119   P4_2/nbc                   -P 4ac 2b
1124   133.460   P4_2/nbc1'     133.2.1124   P4_2/nbc1'      1120   P4_2/nbc1'                 -P 4ac 2b  1'
1125   133.461   P4_2/n'bc      133.3.1125   P4_2/n'bc       1121   P4_2/n'bc                  P 4ac  2b' -1'
1126   133.462   P4_2'/nb'c     133.4.1126   P4_2'/nb'c      1122   P4_2'/nb'c                 -P 4ac' 2b'
```

```
1127   133.463  P4_2'/nbc'      133.5.1127   P4_2'/nbc'      1123   P4_2'/nbc'                    -P 4ac' 2b
1128   133.464  P4_2'/n'b'c     133.6.1128   P4_2'/n'b'c     1124   P4_2'/n'b'c                    P 4ac' 2b  -1'
1129   133.465  P4_2'/nb'c'     133.7.1129   P4_2'/nb'c'     1125   P4_2'/nb'c'                   -P 4ac  2b'
1130   133.466  P4_2'/n'bc      133.8.1130   P4_2'/n'bc      1126   P4_2'/n'bc                     P 4ac  2b' -1'
1131   133.467  P4_2/n'b'c'     133.9.1131   P4_2/n'b'c'     1127   P4_2/n'b'c'                    P 4ac  2b  -1'
1042   133.468  P_c4_2/nbc      125.12.1042  P_2c4'/nbm      1128   P4_2/nbc1'_c[P4/nbm]          -P 4ac 2b 1'c
1122   133.469  P_C4_2/nbc      132.13.1122  P_P4_2'/m'cm    1129   P4_2/nbc1'_C[rP4_2/mcm]       -P 4ac 2b 1'C
1209   133.470  P_I4_2/nbc      140.14.1209  I_P4'/m'c'm     1130   P4_2/nbc1'_I[I4/mcm]          -P 4ac 2b 1'I
1132   134.471  P4_2/nnm        134.1.1132   P4_2/nnm        1131   P4_2/nnm                      -P 4ac  2bc
1133   134.472  P4_2/nnm1'      134.2.1133   P4_2/nnm1'      1132   P4_2/nnm1'                    -P 4ac  2bc   1'
1134   134.473  P4_2/n'nm       134.3.1134   P4_2/n'nm       1133   P4_2/n'nm                      P 4ac  2bc' -1'
1135   134.474  P4_2'/nn'm      134.4.1135   P4_2'/nn'm      1134   P4_2'/nn'm                    -P 4ac' 2bc'
1136   134.475  P4_2'/nnm'      134.5.1136   P4_2'/nnm'      1135   P4_2'/nnm'                    -P 4ac' 2bc
1137   134.476  P4_2'/n'n'm     134.6.1137   P4_2'/n'n'm     1136   P4_2'/n'n'm                    P 4ac' 2bc  -1'
1138   134.477  P4_2/nn'm'      134.7.1138   P4_2/nn'm'      1137   P4_2/nn'm'                    -P 4ac  2bc'
1139   134.478  P4_2'/n'nm'     134.8.1139   P4_2'/n'nm'     1138   P4_2'/n'nm'                    P 4ac' 2bc' -1'
1140   134.479  P4_2/n'n'm'     134.9.1140   P4_2/n'n'm'     1139   P4_2/n'n'm'                    P 4ac  2bc  -1'
1041   134.480  P_c4_2/nnm      125.11.1041  P_2c4'/nb'm     1140   P4_2/nnm1'_c[P4/nbm]          -P 4ac 2bc 1'c
1109   134.481  P_C4_2/nnm      131.13.1109  P_P4_2'/m'mc    1141   P4_2/nnm1'_C[rP4_2/mmc]       -P 4ac 2bc 1'C
1192   134.482  P_I4_2/nnm      139.14.1192  I_P4'/m'm'm     1142   P4_2/nnm1'_I[I4/mmm]          -P 4ac 2bc 1'I
1143   135.483  P4_2/mbc        135.1.1143   P4_2/mbc        1143   P4_2/mbc                      -P 4c  2ab
1144   135.484  P4_2/mbc1'      135.2.1144   P4_2/mbc1'      1144   P4_2/mbc1'                    -P 4c  2ab   1'
1145   135.485  P4_2/m'bc       135.3.1145   P4_2/m'bc       1145   P4_2/m'bc                      P 4c  2ab' -1'
1146   135.486  P4_2'/mb'c      135.4.1146   P4_2'/mb'c      1146   P4_2'/mb'c                    -P 4c' 2ab'
1147   135.487  P4_2'/mbc'      135.5.1147   P4_2'/mbc'      1147   P4_2'/mbc'                    -P 4c' 2ab
1148   135.488  P4_2'/m'b'c     135.6.1148   P4_2'/m'b'c     1148   P4_2'/m'b'c                    P 4c' 2ab  -1'
1149   135.489  P4_2/mb'c'      135.7.1149   P4_2/mb'c'      1149   P4_2/mb'c'                    -P 4c  2ab'
1150   135.490  P4_2'/m'bc'     135.8.1150   P4_2'/m'bc'     1150   P4_2'/m'bc'                    P 4c' 2ab' -1'
1151   135.491  P4_2/m'b'c'     135.9.1151   P4_2/m'b'c'     1151   P4_2/m'b'c'                    P 4c  2ab  -1'
1064   135.492  P_c4_2/mbc      127.12.1064  P_2c4'/mbm      1152   P4_2/mbc1'_c[P4/mbm]          -P 4c 2ab 1'c
1121   135.493  P_C4_2/mbc      132.12.1121  P_P4_2'/mcm'    1153   P4_2/mbc1'_C[rP4_2/mcm]       -P 4c 2ab 1'C
1207   135.494  P_I4_2/mbc      140.12.1207  I_P4'/mc'm      1154   P4_2/mbc1'_I[I4/mcm]          -P 4c 2ab 1'I
1152   136.495  P4_2/mnm        136.1.1152   P4_2/mnm        1155   P4_2/mnm                      -P 4n   2n
1153   136.496  P4_2/mnm1'      136.2.1153   P4_2/mnm1'      1156   P4_2/mnm1'                    -P 4n   2n    1'
1154   136.497  P4_2/m'nm       136.3.1154   P4_2/m'nm       1157   P4_2/m'nm                      P 4n   2n' -1'
1155   136.498  P4_2'/mn'm      136.4.1155   P4_2'/mn'm      1158   P4_2'/mn'm                    -P 4n'  2n'
1156   136.499  P4_2'/mnm'      136.5.1156   P4_2'/mnm'      1159   P4_2'/mnm'                    -P 4n'  2n
1157   136.500  P4_2'/m'n'm     136.6.1157   P4_2'/m'n'm     1160   P4_2'/m'n'm                    P 4n'  2n  -1'
1158   136.501  P4_2/mn'm'      136.7.1158   P4_2/mn'm'      1161   P4_2/mn'm'                    -P 4n   2n'
1159   136.502  P4_2'/m'nm'     136.8.1159   P4_2'/m'nm'     1162   P4_2'/m'nm'                    P 4n'  2n' -1'
1160   136.503  P4_2/m'n'm'     136.9.1160   P4_2/m'n'm'     1163   P4_2/m'n'm'                    P 4n   2n  -1'
1063   136.504  P_c4_2/mnm      127.11.1063  P_2c4'/mb'm     1164   P4_2/mnm1'_c[P4/mbm]          -P 4n 2n 1'c
1108   136.505  P_C4_2/mnm      131.12.1108  P_P4_2'/mm'c    1165   P4_2/mnm1'_C[rP4_2/mmc]       -P 4n 2n 1'C
1190   136.506  P_I4_2/mnm      139.12.1190  I_P4'/mm'm      1166   P4_2/mnm1'_I[I4/mmm]          -P 4n 2n 1'I
1161   137.507  P4_2/nmc        137.1.1161   P4_2/nmc        1167   P4_2/nmc                      -P 4ac  2a
1162   137.508  P4_2/nmc1'      137.2.1162   P4_2/nmc1'      1168   P4_2/nmc1'                    -P 4ac  2a    1'
1163   137.509  P4_2/n'mc       137.3.1163   P4_2/n'mc       1169   P4_2/n'mc                      P 4ac  2a' -1'
1164   137.510  P4_2'/nm'c      137.4.1164   P4_2'/nm'c      1170   P4_2'/nm'c                    -P 4ac' 2a'
1165   137.511  P4_2'/nmc'      137.5.1165   P4_2'/nmc'      1171   P4_2'/nmc'                    -P 4ac' 2a
1166   137.512  P4_2'/n'm'c     137.6.1166   P4_2'/n'm'c     1172   P4_2'/n'm'c                    P 4ac' 2a  -1'
1167   137.513  P4_2/nm'c'      137.7.1167   P4_2/nm'c'      1173   P4_2/nm'c'                    -P 4ac  2a'
1168   137.514  P4_2'/n'mc'     137.8.1168   P4_2'/n'mc'     1174   P4_2'/n'mc'                    P 4ac' 2a' -1'
1169   137.515  P4_2/n'm'c'     137.9.1169   P4_2/n'm'c'     1175   P4_2/n'm'c'                    P 4ac  2a  -1'
1086   137.516  P_c4_2/nmc      129.12.1086  P_2c4'/nmm      1176   P4_2/nmc1'_c[P4/nmm]          -P 4ac 2a 1'c
1120   137.517  P_C4_2/nmc      132.11.1120  P_P4_2'/m'cm    1177   P4_2/nmc1'_C[rP4_2/mcm]       -P 4ac 2a 1'C
1194   137.518  P_I4_2/nmc      139.16.1194  I_P4'/m'mm      1178   P4_2/nmc1'_I[I4/mmm]          -P 4ac 2a 1'I
1170   138.519  P4_2/ncm        138.1.1170   P4_2/ncm        1179   P4_2/ncm                      -P 4ac  2ac
1171   138.520  P4_2/ncm1'      138.2.1171   P4_2/ncm1'      1180   P4_2/ncm1'                    -P 4ac  2ac   1'
1172   138.521  P4_2/n'cm       138.3.1172   P4_2/n'cm       1181   P4_2/n'cm                      P 4ac  2ac' -1'
1173   138.522  P4_2'/nc'm      138.4.1173   P4_2'/nc'm      1182   P4_2'/nc'm                    -P 4ac' 2ac'
1174   138.523  P4_2'/ncm'      138.5.1174   P4_2'/ncm'      1183   P4_2'/ncm'                    -P 4ac' 2ac
1175   138.524  P4_2'/n'c'm     138.6.1175   P4_2'/n'c'm     1184   P4_2'/n'c'm                    P 4ac' 2ac  -1'
1176   138.525  P4_2/nc'm'      138.7.1176   P4_2/nc'm'      1185   P4_2/nc'm'                    -P 4ac  2ac'
1177   138.526  P4_2'/n'cm'     138.8.1177   P4_2'/n'cm'     1186   P4_2'/n'cm'                    P 4ac' 2ac' -1'
1178   138.527  P4_2/n'c'm'     138.9.1178   P4_2/n'c'm'     1187   P4_2/n'c'm'                    P 4ac  2ac  -1'
1085   138.528  P_c4_2/ncm      129.11.1085  P_2c4'/nm'm     1188   P4_2/ncm1'_c[P4/nmm]          -P 4ac 2ac 1'c
1107   138.529  P_C4_2/ncm      131.11.1107  P_P4_2'/m'mc    1189   P4_2/ncm1'_C[rP4_2/mmc]       -P 4ac 2ac 1'C
1211   138.530  P_I4_2/ncm      140.16.1211  I_P4'/m'c'm     1190   P4_2/ncm1'_I[I4/mcm]          -P 4ac 2ac 1'I
1179   139.531  I4/mmm          139.1.1179   I4/mmm          1191   I4/mmm                        -I 4 2
1180   139.532  I4/mmm1'        139.2.1180   I4/mmm1'        1192   I4/mmm1'                      -I 4 2 1'
1181   139.533  I4/m'mm         139.3.1181   I4/m'mm         1193   I4/m'mm                        I 4 2' -1'
1182   139.534  I4'/mm'm        139.4.1182   I4'/mm'm        1194   I4'/mm'm                      -I 4' 2'
1183   139.535  I4'/mmm'        139.5.1183   I4'/mmm'        1195   I4'/mmm'                      -I 4' 2
1184   139.536  I4'/m'm'm       139.6.1184   I4'/m'm'm       1196   I4'/m'm'm                      I 4 2  -1'
1185   139.537  I4/mm'm'        139.7.1185   I4/mm'm'        1197   I4/mm'm'                      -I 4 2'
1186   139.538  I4'/m'mm'       139.8.1186   I4'/m'mm'       1198   I4'/m'mm'                      I 4 2' -1'
1187   139.539  I4/m'm'm'       139.9.1187   I4/m'm'm'       1199   I4/m'm'm'                      I 4 2  -1'
1010   139.540  I_c4/mmm        123.12.1010  P_I4/mmm        1200   I4/mmm1'_c[rP4/mmm]           -I 4 2 1'c
1196   140.541  I4/mcm          140.1.1196   I4/mcm          1201   I4/mcm                        -I 4 2c
1197   140.542  I4/mcm1'        140.2.1197   I4/mcm1'        1202   I4/mcm1'                      -I 4 2c 1'
1198   140.543  I4/m'cm         140.3.1198   I4/m'cm         1203   I4/m'cm                        I 4 2c' -1'
```

```
1199   140.544   I4'/mc'm       140.4.1199   I4'/mc'm       1204   I4'/mc'm                  -I 4' 2c'
1200   140.545   I4'/mcm'       140.5.1200   I4'/mcm'       1205   I4'/mcm'                  -I 4' 2c
1201   140.546   I4'/m'c'm      140.6.1201   I4'/m'c'm      1206   I4'/m'c'm                 I 4' 2c -1'
1202   140.547   I4/mc'm'       140.7.1202   I4/mc'm'       1207   I4/mc'm'                  -I 4 2c'
1203   140.548   I4'/m'cm'      140.8.1203   I4'/m'cm'      1208   I4'/m'cm'                 I 4' 2c' -1'
1204   140.549   I4/m'c'm'      140.9.1204   I4/m'c'm'      1209   I4/m'c'm'                 I 4 2c -1'
1017   140.550   I_c4/mcm       123.19.1017  P_I4/mm'm'     1210   I4/mcm1'_c[rP4/mmm]       -I 4 2c 1'c
1213   141.551   I4_1/amd       141.1.1213   I4_1/amd       1211   I4_1/amd                  -I 4bd 2
1214   141.552   I4_1/amd1'     141.2.1214   I4_1/amd1'     1212   I4_1/amd1'                -I 4bd 2   1'
1215   141.553   I4_1/a'md      141.3.1215   I4_1/a'md      1213   I4_1/a'md                 I 4bd  2' -1'
1216   141.554   I4_1'/am'd     141.4.1216   I4_1'/am'd     1214   I4_1'/am'd                -I 4bd' 2'
1217   141.555   I4_1'/amd'     141.5.1217   I4_1'/amd'     1215   I4_1'/amd'                -I 4bd 2
1218   141.556   I4_1'/a'm'd    141.6.1218   I4_1'/a'm'd    1216   I4_1'/a'm'd               I 4bd' 2 -1'
1219   141.557   I4_1/am'd'     141.7.1219   I4_1/am'd'     1217   I4_1/am'd'                -I 4bd  2'
1220   141.558   I4_1'/a'md'    141.8.1220   I4_1'/a'md'    1218   I4_1'/a'md'               I 4bd' 2' -1'
1221   141.559   I4_1/a'm'd'    141.9.1221   I4_1/a'm'd'    1219   I4_1/a'm'd'               I 4bd  2 -1'
1141   141.560   I_c4_1/amd     134.10.1141  P_I4/nnm       1220   I4_1/amd1'_c[rP4_2/nnm]   -I 4bd 2 1'c
1222   142.561   I4_1/acd       142.1.1222   I4_1/acd       1221   I4_1/acd                  -I 4bd 2c
1223   142.562   I4_1/acd1'     142.2.1223   I4_1/acd1'     1222   I4_1/acd1'                -I 4bd 2c   1'
1224   142.563   I4_1/a'cd      142.3.1224   I4_1/a'cd      1223   I4_1/a'cd                 I 4bd 2c' -1'
1225   142.564   I4_1'/ac'd     142.4.1225   I4_1'/ac'd     1224   I4_1'/ac'd                -I 4bd' 2c'
1226   142.565   I4_1'/acd'     142.5.1226   I4_1'/acd'     1225   I4_1'/acd'                -I 4bd' 2c
1227   142.566   I4_1'/a'c'd    142.6.1227   I4_1'/a'c'd    1226   I4_1'/a'c'd               I 4bd' 2c -1'
1228   142.567   I4_1/ac'd'     142.7.1228   I4_1/ac'd'     1227   I4_1/ac'd'                -I 4bd 2c'
1229   142.568   I4_1'/a'cd'    142.8.1229   I4_1'/a'cd'    1228   I4_1'/a'cd'               I 4bd' 2c' -1'
1230   142.569   I4_1/a'c'd'    142.9.1230   I4_1/a'c'd'    1229   I4_1/a'c'd'               I 4bd 2c -1'
1142   142.570   I_c4_1/acd     134.11.1142  P_I4_2/nn'm'   1230   I4_1/acd1'_c[rP4_2/nnm]   -I 4bd 2c 1'c
1231   143.1     P3             143.1.1231   P3             1231   P3                        P 3
1232   143.2     P31'           143.2.1232   P31'           1232   P31'                      P 3 1'
1233   143.3     P_c3           143.3.1233   P_2c3          1233   P31'_c[P3]                P 3 1'c
1234   144.4     P3_1           144.1.1234   P3_1           1234   P3_1                      P 31
1235   144.5     P3_11'         144.2.1235   P3_11'         1235   P3_11'                    P 31 1'
1239   144.6     P_c3_1         145.3.1239   P_2c3_1        1236   P3_11'_c[P3_2]            P 31 1'c
1237   145.7     P3_2           145.1.1237   P3_2           1237   P3_2                      P 32
1238   145.8     P3_21'         145.2.1238   P3_21'         1238   P3_21'                    P 32 1'
1236   145.9     P_c3_2         144.3.1236   P_2c3_2        1239   P3_21'_c[P3_1]            P 32 1'c
1240   146.10    R3             146.1.1240   R3             1240   R3                        R 3
1241   146.11    R31'           146.2.1241   R31'           1241   R31'                      R 3 1'
1242   146.12    R_I3           146.3.1242   R_R3           1242   R31'_c[R3]                R 3 1'I
1243   147.13    P-3            147.1.1243   P-3            1243   P-3                       -P 3
1244   147.14    P-31'          147.2.1244   P-31'          1244   P-31'                     -P 3 1'
1245   147.15    P-3'           147.3.1245   P-3'           1245   P-3'                      P -3'
1246   147.16    P_c-3          147.4.1246   P_2c-3         1246   P-31'_c[P-3]              -P 3 1'c
1247   148.17    R-3            148.1.1247   R-3            1247   R-3                       -R 3
1248   148.18    R-31'          148.2.1248   R-31'          1248   R-31'                     -R 3 1'
1249   148.19    R-3'           148.3.1249   R-3'           1249   R-3'                      R -3'
1250   148.20    R_I-3          148.4.1250   R_R-3          1250   R-31'_c[R-3]              -R 3 1'I
1251   149.21    P312           149.1.1251   P312           1251   P312                      P 3 2
1252   149.22    P3121'         149.2.1252   P3121'         1252   P3121'                    P 3 2 1'
1253   149.23    P312'          149.3.1253   P312'          1253   P312'                     P 3 2'
1254   149.24    P_c312         149.4.1254   P_2c312        1254   P3121'_c[P312]            P 3 2 1'c
1255   150.25    P321           150.1.1255   P321           1255   P321                      P 3 2"
1256   150.26    P3211'         150.2.1256   P3211'         1256   P3211'                    P 3 2" 1'
1257   150.27    P32'1          150.3.1257   P32'1          1257   P32'1                     P 3 2"'
1258   150.28    P_c321         150.4.1258   P_2c321        1258   P3211'_c[P321]            P 3 2" 1'c
1259   151.29    P3_112         151.1.1259   P3_112         1259   P3_112                    P 31 2c (0 0 1)
1260   151.30    P3_1121'       151.2.1260   P3_1121'       1260   P3_1121'                  P 31 2c 1' (0 0 1)
1261   151.31    P3_112'        151.3.1261   P3_112'        1261   P3_112'                   P 31 2c' (0 0 1)
1270   151.32    P_c3_112       153.4.1270   P_2c3_112      1262   P3_1121'_c[P3_212]        P 31 2c 1'c (0 0 1)
1263   152.33    P3_121         152.1.1263   P3_121         1263   P3_121                    P 31 2"
1264   152.34    P3_1211'       152.2.1264   P3_1211'       1264   P3_1211'                  P 31 2" 1'
1265   152.35    P3_12'1        152.3.1265   P3_12'1        1265   P3_12'1                   P 31 2"'
1274   152.36    P_c3_121       154.4.1274   P_2c3_121      1266   P3_1211'_c[P3_221]        P 31 2" 1'c
1267   153.37    P3_212         153.1.1267   P3_212         1267   P3_212                    P 32 2c (0 0 -1)
1268   153.38    P3_2121'       153.2.1268   P3_2121'       1268   P3_2121'                  P 32 2c 1' (0 0 -1)
1269   153.39    P3_212'        153.3.1269   P3_212'        1269   P3_212'                   P 32 2c' (0 0 -1)
1262   153.40    P_c3_212       151.4.1262   P_2c3_212      1270   P3_2121'_c[P3_112]        P 32 2c 1'c (0 0 -1)
1271   154.41    P3_221         154.1.1271   P3_221         1271   P3_221                    P 32 2"
1272   154.42    P3_2211'       154.2.1272   P3_2211'       1272   P3_2211'                  P 32 2" 1'
1273   154.43    P3_22'1        154.3.1273   P3_22'1        1273   P3_22'1                   P 32 2"'
1266   154.44    P_c3_221       152.4.1266   P_2c3_221      1274   P3_2211'_c[P3_121]        P 32 2" 1'c
1275   155.45    R32            155.1.1275   R32            1275   R32                       R 3 2"
1276   155.46    R321'          155.2.1276   R321'          1276   R321'                     R 3 2" 1'
1277   155.47    R32'           155.3.1277   R32'           1277   R32'                      R 3 2"'
1278   155.48    R_I32          155.4.1278   R_R32          1278   R321'_c[R32]              R 3 2" 1'I
1279   156.49    P3m1           156.1.1279   P3m1           1279   P3m1                      P 3 -2"
1280   156.50    P3m11'         156.2.1280   P3m11'         1280   P3m11'                    P 3 -2" 1'
1281   156.51    P3m'1          156.3.1281   P3m'1          1281   P3m'1                     P 3 -2"'
1282   156.52    P_c3m1         156.4.1282   P_2c3m1        1282   P3m11'_c[P3m1]            P 3 -2" 1'c
1284   157.53    P31m           157.1.1284   P31m           1283   P31m                      P 3 -2
1285   157.54    P31m1'         157.2.1285   P31m1'         1284   P31m1'                    P 3 -2 1'
```

```
1286   157.55   P31m'        157.3.1286   P31m'        1285   P31m'              P 3 -2'
1287   157.56   P_c31m       157.4.1287   P_2c31m      1286   P31m1'_c[P31m]     P 3 -2 1'c
1289   158.57   P-3c1        158.1.1289   P-3c1        1287   P-3c1              P 3 -2"c
1290   158.58   P-3c11'      158.2.1290   P-3c11'      1288   P-3c11'            P 3 -2"c 1'
1291   158.59   P-3c'1       158.3.1291   P-3c'1       1289   P-3c'1             P 3 -2"c'
1283   158.60   P_c-3c1      156.5.1283   P_2c-3m'1    1290   P-3c11'_c[P-3m1]   P 3 -2"c 1'c
1292   159.61   P-31c        159.1.1292   P-31c        1291   P-31c              P 3 -2c
1293   159.62   P-31c1'      159.2.1293   P-31c1'      1292   P-31c1'            P 3 -2c 1'
1294   159.63   P-31c'       159.3.1294   P-31c'       1293   P-31c'             P 3 -2c'
1288   159.64   P_c-31c      157.5.1288   P_2c-31m'    1294   P-31c1'_c[P-31m]   P 3 -2c 1'c
1295   160.65   R-3m         160.1.1295   R-3m         1295   R-3m               R 3 -2"
1296   160.66   R-3m1'       160.2.1296   R-3m1'       1296   R-3m1'             R 3 -2" 1'
1297   160.67   R-3m'        160.3.1297   R-3m'        1297   R-3m'              R 3 -2"'
1298   160.68   R_I-3m       160.4.1298   R_R-3m       1298   R-3m1'_c[R-3m]     R 3 -2" 1'I
1300   161.69   R-3c         161.1.1300   R-3c         1299   R-3c               R 3 -2"c
1301   161.70   R-3c1'       161.2.1301   R-3c1'       1300   R-3c1'             R 3 -2"c 1'
1302   161.71   R-3c'        161.3.1302   R-3c'        1301   R-3c'              R 3 -2"c'
1299   161.72   R_I-3c       160.5.1299   R_R-3m'      1302   R-3c1'_c[R-3m]     R 3 -2"c 1'I
1303   162.73   P-31m        162.1.1303   P-31m        1303   P-31m              -P 3 2
1304   162.74   P-31m1'      162.2.1304   P-31m1'      1304   P-31m1'            -P 3 2 1'
1305   162.75   P-3'1m       162.3.1305   P-3'1m       1305   P-3'1m             P 3 2' -1'
1306   162.76   P-3'1m'      162.4.1306   P-3'1m'      1306   P-3'1m'            P 3 2 -1'
1307   162.77   P-31m'       162.5.1307   P-31m'       1307   P-31m'             -P 3 2'
1308   162.78   P_c-31m      162.6.1308   P_2c-31m     1308   P-31m1'_c[P-31m]   -P 3 2 1'c
1310   163.79   P-31c        163.1.1310   P-31c        1309   P-31c              -P 3 2c
1311   163.80   P-31c1'      163.2.1311   P-31c1'      1310   P-31c1'            -P 3 2c 1'
1312   163.81   P-3'1c       163.3.1312   P-3'1c       1311   P-3'1c             P 3 2c' -1'
1313   163.82   P-3'1c'      163.4.1313   P-3'1c'      1312   P-3'1c'            P 3 2c -1'
1314   163.83   P-31c'       163.5.1314   P-31c'       1313   P-31c'             -P 3 2c'
1309   163.84   P_c-31c      162.7.1309   P_2c-31m'    1314   P-31c1'_c[P-31m]   -P 3 2c 1'c
1315   164.85   P-3m1        164.1.1315   P-3m1        1315   P-3m1              -P 3 2"
1316   164.86   P-3m11'      164.2.1316   P-3m11'      1316   P-3m11'            -P 3 2" 1'
1317   164.87   P-3'm1       164.3.1317   P-3'm1       1317   P-3'm1             P 3 2"' -1'
1318   164.88   P-3'm'1      164.4.1318   P-3'm'1      1318   P-3'm'1            P 3 2" -1'
1319   164.89   P-3m'1       164.5.1319   P-3m'1       1319   P-3m'1             -P 3 2"'
1320   164.90   P_c-3m1      164.6.1320   P_2c-3m1     1320   P-3m11'_c[P-3m1]   -P 3 2" 1'c
1322   165.91   P-3c1        165.1.1322   P-3c1        1321   P-3c1              -P 3 2"c
1323   165.92   P-3c11'      165.2.1323   P-3c11'      1322   P-3c11'            -P 3 2"c 1'
1324   165.93   P-3'c1       165.3.1324   P-3'c1       1323   P-3'c1             P 3 2"c' -1'
1325   165.94   P-3'c'1      165.4.1325   P-3'c'1      1324   P-3'c'1            P 3 2"c -1'
1326   165.95   P-3c'1       165.5.1326   P-3c'1       1325   P-3c'1             -P 3 2"c'
1321   165.96   P_c-3c1      164.7.1321   P_2c-3m'1    1326   P-3c11'_c[P-3m1]   -P 3 2"c 1'c
1327   166.97   R-3m         166.1.1327   R-3m         1327   R-3m               -R 3 2"
1328   166.98   R-3m1'       166.2.1328   R-3m1'       1328   R-3m1'             -R 3 2" 1'
1329   166.99   R-3'm        166.3.1329   R-3'm        1329   R-3'm              R 3 2"' -1'
1330   166.100  R-3'm'       166.4.1330   R-3'm'       1330   R-3'm'             R 3 2" -1'
1331   166.101  R-3m'        166.5.1331   R-3m'        1331   R-3m'              -R 3 2"'
1332   166.102  R_I-3m       166.6.1332   R_R-3m       1332   R-3m1'_c[R-3m]     -R 3 2" 1'I
1334   167.103  R-3c         167.1.1334   R-3c         1333   R-3c               -R 3 2"c
1335   167.104  R-3c1'       167.2.1335   R-3c1'       1334   R-3c1'             -R 3 2"c 1'
1336   167.105  R-3'c        167.3.1336   R-3'c        1335   R-3'c              R 3 2"c' -1'
1337   167.106  R-3'c'       167.4.1337   R-3'c'       1336   R-3'c'             R 3 2"c -1'
1338   167.107  R-3c'        167.5.1338   R-3c'        1337   R-3c'              -R 3 2"c'
1333   167.108  R_I-3c       166.7.1333   R_R-3m'      1338   R-3c1'_c[R-3m]     -R 3 2"c 1'I
1339   168.109  P6           168.1.1339   P6           1339   P6                 P 6
1340   168.110  P61'         168.2.1340   P61'         1340   P61'               P 6 1'
1341   168.111  P6'          168.3.1341   P6'          1341   P6'                P 6'
1342   168.112  P_c6         168.4.1342   P_2c6        1342   P61'_c[P6]         P 6 1'c
1344   169.113  P6_1         169.1.1344   P6_1         1343   P6_1               P 61
1345   169.114  P6_11'       169.2.1345   P6_11'       1344   P6_11'             P 61  1'
1346   169.115  P6_1'        169.3.1346   P6_1'        1345   P6_1'              P 61'
1353   169.116  P_c6_1       171.4.1353   P_2c6_2      1346   P6_11'_c[P6_2]     P 61 1'c
1347   170.117  P6_5         170.1.1347   P6_5         1347   P6_5               P 65
1348   170.118  P6_51'       170.2.1348   P6_51'       1348   P6_51'             P 65  1'
1349   170.119  P6_5'        170.3.1349   P6_5'        1349   P6_5'              P 65'
1359   170.120  P_c6_5       172.5.1359   P_2c6_4'     1350   P6_51'_c[P6_4]     P 65 1'c
1350   171.121  P6_2         171.1.1350   P6_2         1351   P6_2               P 62
1351   171.122  P6_21'       171.2.1351   P6_21'       1352   P6_21'             P 62  1'
1352   171.123  P6_2'        171.3.1352   P6_2'        1353   P6_2'              P 62'
1358   171.124  P_c6_2       172.4.1358   P_2c6_4      1354   P6_21'_c[P6_4]     P 62 1'c
1355   172.125  P6_4         172.1.1355   P6_4         1355   P6_4               P 64
1356   172.126  P6_41'       172.2.1356   P6_41'       1356   P6_41'             P 64  1'
1357   172.127  P6_4'        172.3.1357   P6_4'        1357   P6_4'              P 64'
1354   172.128  P_c6_4       171.5.1354   P_2c6_2'     1358   P6_41'_c[P6_2]     P 64 1'c
1360   173.129  P6_3         173.1.1360   P6_3         1359   P6_3               P 6c
1361   173.130  P6_31'       173.2.1361   P6_31'       1360   P6_31'             P 6c  1'
1362   173.131  P6_3'        173.3.1362   P6_3'        1361   P6_3'              P 6c'
1343   173.132  P_c6_3       168.5.1343   P_2c6'       1362   P6_31'_c[P6]       P 6c 1'c
1363   174.133  P-6          174.1.1363   P-6          1363   P-6                P -6
1364   174.134  P-61'        174.2.1364   P-61'        1364   P-61'              P -6  1'
1365   174.135  P-6'         174.3.1365   P-6'         1365   P-6'               P -6'
```

```
1366    174.136  P_c-6        174.4.1366  P_2c-6       1366  P-61'_c[P-6]         P -6 1'c
1367    175.137  P6/m         175.1.1367  P6/m         1367  P6/m                 -P 6
1368    175.138  P6/m1'       175.2.1368  P6/m1'       1368  P6/m1'               -P 6 1'
1369    175.139  P6'/m        175.3.1369  P6'/m        1369  P6'/m                P 6' -1'
1370    175.140  P6/m'        175.4.1370  P6/m'        1370  P6/m'                P 6 -1'
1371    175.141  P6'/m'       175.5.1371  P6'/m'       1371  P6'/m'               -P 6'
1372    175.142  P_c6/m       175.6.1372  P_2c6/m      1372  P6/m1'_c[P6/m]       -P 6 1'c
1374    176.143  P6_3/m       176.1.1374  P6_3/m       1373  P6_3/m               -P 6c
1375    176.144  P6_3/m1'     176.2.1375  P6_3/m1'     1374  P6_3/m1'             -P 6c   1'
1376    176.145  P6_3'/m      176.3.1376  P6_3'/m      1375  P6_3'/m              P 6c' -1'
1377    176.146  P6_3/m'      176.4.1377  P6_3/m'      1376  P6_3/m'              P 6c  -1'
1378    176.147  P6_3'/m'     176.5.1378  P6_3'/m'     1377  P6_3'/m'             -P 6c'
1373    176.148  P_c6_3/m     175.7.1373  P_2c6'/m     1378  P6_3/m1'_c[P6/m]     -P 6c 1'c
1379    177.149  P622         177.1.1379  P622         1379  P622                 P 6 2
1380    177.150  P6221'       177.2.1380  P6221'       1380  P6221'               P 6 2 1'
1381    177.151  P6'2'2       177.3.1381  P6'2'2       1381  P6'2'2               P 6' 2
1382    177.152  P6'22'       177.4.1382  P6'22'       1382  P6'22'               P 6' 2'
1383    177.153  P62'2'       177.5.1383  P62'2'       1383  P62'2'               P 6 2'
1384    177.154  P_c622       177.6.1384  P_2c622      1384  P6221'_c[P622]       P 6 2 1'c
1386    178.155  P6_122       178.1.1386  P6_122       1385  P6_122               P 61  2 (0 0 -1)
1387    178.156  P6_1221'     178.2.1387  P6_1221'     1386  P6_1221'             P 61  2  1' (0 0 -1)
1388    178.157  P6_1'2'2     178.3.1388  P6_1'2'2     1387  P6_1'2'2             P 61' 2 (0 0 -1)
1389    178.158  P6_1'22'     178.4.1389  P6_1'22'     1388  P6_1'22'             P 61' 2' (0 0 -1)
1390    178.159  P6_12'2'     178.5.1390  P6_12'2'     1389  P6_12'2'             P 61  2' (0 0 -1)
1401    178.160  P_c6_122     180.6.1401  P_2c6_222    1390  P6_1221'_c[P6_222]   P 61 2 1'c (0 0 -1)
1391    179.161  P6_522       179.1.1391  P6_522       1391  P6_522               P 65  2 (0 0 1)
1392    179.162  P6_5221'     179.2.1392  P6_5221'     1392  P6_5221'             P 65  2 1' (0 0 1)
1393    179.163  P6_5'2'2     179.3.1393  P6_5'2'2     1393  P6_5'2'2             P 65' 2 (0 0 1)
1394    179.164  P6_5'22'     179.4.1394  P6_5'22'     1394  P6_5'22'             P 65' 2' (0 0 1)
1395    179.165  P6_52'2'     179.5.1395  P6_52'2'     1395  P6_52'2'             P 65  2' (0 0 1)
1409    179.166  P_c6_522     181.7.1409  P_2c6_4'2'2  1396  P6_5221'_c[P6_422]   P 65 2 1'c (0 0 1)
1396    180.167  P6_222       180.1.1396  P6_222       1397  P6_222               P 62  2c (0 0 1)
1397    180.168  P6_2221'     180.2.1397  P6_2221'     1398  P6_2221'             P 62  2c 1' (0 0 1)
1398    180.169  P6_2'2'2     180.3.1398  P6_2'2'2     1399  P6_2'2'2             P 62' 2c (0 0 1)
1399    180.170  P6_2'22'     180.4.1399  P6_2'22'     1400  P6_2'22'             P 62' 2c' (0 0 1)
1400    180.171  P6_22'2'     180.5.1400  P6_22'2'     1401  P6_22'2'             P 62  2c' (0 0 1)
1408    180.172  P_c6_222     181.6.1408  P_2c6_422    1402  P6_2221'_c[P6_422]   P 62 2c 1'c (0 0 1)
1403    181.173  P6_422       181.1.1403  P6_422       1403  P6_422               P 64  2c (0 0 -1)
1404    181.174  P6_4221'     181.2.1404  P6_4221'     1404  P6_4221'             P 64  2c 1' (0 0 -1)
1405    181.175  P6_4'2'2     181.3.1405  P6_4'2'2     1405  P6_4'2'2             P 64' 2c (0 0 -1)
1406    181.176  P6_4'22'     181.4.1406  P6_4'22'     1406  P6_4'22'             P 64' 2c' (0 0 -1)
1407    181.177  P6_42'2'     181.5.1407  P6_42'2'     1407  P6_42'2'             P 64  2c' (0 0 -1)
1402    181.178  P_c6_422     180.7.1402  P_2c6_2'22'  1408  P6_4221'_c[P6_222]   P 64 2c 1'c (0 0 -1)
1410    182.179  P6_322       182.1.1410  P6_322       1409  P6_322               P 6c  2c
1411    182.180  P6_3221'     182.2.1411  P6_3221'     1410  P6_3221'             P 6c  2c  1'
1412    182.181  P6_3'2'2     182.3.1412  P6_3'2'2     1411  P6_3'2'2             P 6c' 2c
1413    182.182  P6_3'22'     182.4.1413  P6_3'22'     1412  P6_3'22'             P 6c' 2c'
1414    182.183  P6_32'2'     182.5.1414  P6_32'2'     1413  P6_32'2'             P 6c  2c'
1385    182.184  P_c6_322     177.7.1385  P_2c6'22'    1414  P6_3221'_c[P622]     P 6c 2c 1'c
1415    183.185  P6mm         183.1.1415  P6mm         1415  P6mm                 P 6 -2
1416    183.186  P6mm1'       183.2.1416  P6mm1'       1416  P6mm1'               P 6 -2 1'
1417    183.187  P6'm'm       183.3.1417  P6'm'm       1417  P6'm'm               P 6' -2
1418    183.188  P6'mm'       183.4.1418  P6'mm'       1418  P6'mm'               P 6' -2'
1419    183.189  P6m'm'       183.5.1419  P6m'm'       1419  P6m'm'               P 6 -2'
1420    183.190  P_c6mm       183.6.1420  P_2c6mm      1420  P6mm1'_c[P6mm]       P 6 -2 1'c
1424    184.191  P6cc         184.1.1424  P6cc         1421  P6cc                 P 6 -2c
1425    184.192  P6cc1'       184.2.1425  P6cc1'       1422  P6cc1'               P 6 -2c 1'
1426    184.193  P6'c'c       184.3.1426  P6'c'c       1423  P6'c'c               P 6' -2c
1427    184.194  P6'cc'       184.4.1427  P6'cc'       1424  P6'cc'               P 6' -2c'
1428    184.195  P6c'c'       184.5.1428  P6c'c'       1425  P6c'c'               P 6 -2c'
1423    184.196  P_c6cc       183.9.1423  P_2c6m'm'    1426  P6cc1'_c[P6mm]       P 6 -2c 1'c
1429    185.197  P6_3cm       185.1.1429  P6_3cm       1427  P6_3cm               P 6c  -2
1430    185.198  P6_3cm1'     185.2.1430  P6_3cm1'     1428  P6_3cm1'             P 6c  -2  1'
1431    185.199  P6_3'c'm     185.3.1431  P6_3'c'm     1429  P6_3'c'm             P 6c' -2
1432    185.200  P6_3'cm'     185.4.1432  P6_3'cm'     1430  P6_3'cm'             P 6c' -2'
1433    185.201  P6_3c'm'     185.5.1433  P6_3c'm'     1431  P6_3c'm'             P 6c  -2'
1421    185.202  P_c6_3cm     183.7.1421  P_2c6'm'm    1432  P6_3cm1'_c[P6mm]     P 6c -2 1'c
1434    186.203  P6_3mc       186.1.1434  P6_3mc       1433  P6_3mc               P 6c  -2c
1435    186.204  P6_3mc1'     186.2.1435  P6_3mc1'     1434  P6_3mc1'             P 6c  -2c  1'
1436    186.205  P6_3'm'c     186.3.1436  P6_3'm'c     1435  P6_3'm'c             P 6c' -2c
1437    186.206  P6_3'mc'     186.4.1437  P6_3'mc'     1436  P6_3'mc'             P 6c' -2c'
1438    186.207  P6_3m'c'     186.5.1438  P6_3m'c'     1437  P6_3m'c'             P 6c  -2c'
1422    186.208  P_c6_3mc     183.8.1422  P_2c6'mm'    1438  P6_3mc1'_c[P6mm]     P 6c -2c 1'c
1439    187.209  P-6m2        187.1.1439  P-6m2        1439  P-6m2                P -6 2
1440    187.210  P-6m21'      187.2.1440  P-6m21'      1440  P-6m21'              P -6 2  1'
1441    187.211  P-6'm'2      187.3.1441  P-6'm'2      1441  P-6'm'2              P -6' 2
1442    187.212  P-6'm2'      187.4.1442  P-6'm2'      1442  P-6'm2'              P -6' 2'
1443    187.213  P-6m'2'      187.5.1443  P-6m'2'      1443  P-6m'2'              P -6 2'
1444    187.214  P_c-6m2      187.6.1444  P_2c-6m2     1444  P-6m21'_c[P-6m2]     P -6 2 1'c
1446    188.215  P-6c2        188.1.1446  P-6c2        1445  P-6c2                P -6c 2
1447    188.216  P-6c21'      188.2.1447  P-6c21'      1446  P-6c21'              P -6c 2  1'
```

```
1448    188.217  P-6'c'2        188.3.1448  P-6'c'2        1447  P-6'c'2                    P -6c' 2
1449    188.218  P-6'c2'        188.4.1449  P-6'c2'        1448  P-6'c2'                    P -6c' 2'
1450    188.219  P-6c'2'        188.5.1450  P-6c'2'        1449  P-6c'2'                    P -6c  2'
1445    188.220  P_c-6c2        187.7.1445  P_2c-6'm'2     1450  P-6c21'_c[P-6m2]           P -6c 2 1'c
1451    189.221  P-62m          189.1.1451  P-62m          1451  P-62m                      P -6  -2
1452    189.222  P-62m1'        189.2.1452  P-62m1'        1452  P-62m1'                    P -6  -2  1'
1453    189.223  P-6'2'm        189.3.1453  P-6'2'm        1453  P-6'2'm                    P -6' -2
1454    189.224  P-6'2m'        189.4.1454  P-6'2m'        1454  P-6'2m'                    P -6' -2'
1455    189.225  P-62'm'        189.5.1455  P-62'm'        1455  P-62'm'                    P -6  -2'
1456    189.226  P_c-62m        189.6.1456  P_2c-62m       1456  P-62m1'_c[P-62m]           P -6 -2 1'c
1458    190.227  P-62c          190.1.1458  P-62c          1457  P-62c                      P -6c  -2c
1459    190.228  P-62c1'        190.2.1459  P-62c1'        1458  P-62c1'                    P -6c  -2c  1'
1460    190.229  P-6'2'c        190.3.1460  P-6'2'c        1459  P-6'2'c                    P -6c' -2c
1461    190.230  P-6'2c'        190.4.1461  P-6'2c'        1460  P-6'2c'                    P -6c' -2c'
1462    190.231  P-62'c'        190.5.1462  P-62'c'        1461  P-62'c'                    P -6c  -2c'
1457    190.232  P_c-62c        189.7.1457  P_2c-6'2m'     1462  P-62c1'_c[P-62m]           P -6c -2c 1'c
1463    191.233  P6/mmm         191.1.1463  P6/mmm         1463  P6/mmm                     -P 6 2
1464    191.234  P6/mmm1'       191.2.1464  P6/mmm1'       1464  P6/mmm1'                   -P 6 2 1'
1465    191.235  P6/m'mm        191.3.1465  P6/m'mm        1465  P6/m'mm                    P 6 2' -1'
1466    191.236  P6'/mm'm       191.4.1466  P6'/mm'm       1466  P6'/mm'm                   P 6' 2' -1'
1467    191.237  P6'/mmm'       191.5.1467  P6'/mmm'       1467  P6'/mmm'                   P 6' 2 -1'
1468    191.238  P6'/m'm'm      191.6.1468  P6'/m'm'm      1468  P6'/m'm'm                  -P 6' 2
1469    191.239  P6'/m'mm'      191.7.1469  P6'/m'mm'      1469  P6'/m'mm'                  -P 6' 2'
1470    191.240  P6/mm'm'       191.8.1470  P6/mm'm'       1470  P6/mm'm'                   -P 6 2'
1471    191.241  P6/m'm'm'      191.9.1471  P6/m'm'm'      1471  P6/m'm'm'                  P 6 2 -1'
1472    191.242  P_c6/mmm       191.10.1472 P_2c6/mmm      1472  P6/mmm1'_c[P6/mmm]         -P 6 2 1'c
1476    192.243  P6/mcc         192.1.1476  P6/mcc         1473  P6/mcc                     -P 6 2c
1477    192.244  P6/mcc1'       192.2.1477  P6/mcc1'       1474  P6/mcc1'                   -P 6 2c 1'
1478    192.245  P6/m'cc        192.3.1478  P6/m'cc        1475  P6/m'cc                    P 6 2c' -1'
1479    192.246  P6'/mc'c       192.4.1479  P6'/mc'c       1476  P6'/mc'c                   P 6' 2c' -1'
1480    192.247  P6'/mcc'       192.5.1480  P6'/mcc'       1477  P6'/mcc'                   P 6' 2c -1'
1481    192.248  P6'/m'c'c      192.6.1481  P6'/m'c'c      1478  P6'/m'c'c                  -P 6' 2c
1482    192.249  P6'/m'cc'      192.7.1482  P6'/m'cc'      1479  P6'/m'cc'                  -P 6' 2c'
1483    192.250  P6/mc'c'       192.8.1483  P6/mc'c'       1480  P6/mc'c'                   -P 6 2c'
1484    192.251  P6/m'c'c'      192.9.1484  P6/m'c'c'      1481  P6/m'c'c'                  P 6 2c -1'
1475    192.252  P_c6/mcc       191.13.1475 P_2c6/mm'm'    1482  P6/mcc1'_c[P6/mmm]         -P 6 2c 1'c
1485    193.253  P6_3/mcm       193.1.1485  P6_3/mcm       1483  P6_3/mcm                   -P 6c  2
1486    193.254  P6_3/mcm1'     193.2.1486  P6_3/mcm1'     1484  P6_3/mcm1'                 -P 6c  2    1'
1487    193.255  P6_3/m'cm      193.3.1487  P6_3/m'cm      1485  P6_3/m'cm                  P 6c  2'  -1'
1488    193.256  P6_3'/mc'm     193.4.1488  P6_3'/mc'm     1486  P6_3'/mc'm                 P 6c' 2'  -1'
1489    193.257  P6_3'/mcm'     193.5.1489  P6_3'/mcm'     1487  P6_3'/mcm'                 P 6c' 2   -1'
1490    193.258  P6_3'/m'c'm    193.6.1490  P6_3'/m'c'm    1488  P6_3'/m'c'm                -P 6c' 2
1491    193.259  P6_3'/m'cm'    193.7.1491  P6_3'/m'cm'    1489  P6_3'/m'cm'                -P 6c' 2'
1492    193.260  P6_3/mc'm'     193.8.1492  P6_3/mc'm'     1490  P6_3/mc'm'                 -P 6c  2'
1493    193.261  P6_3/m'c'm'    193.9.1493  P6_3/m'c'm'    1491  P6_3/m'c'm'                P 6c  2  -1'
1473    193.262  P_c6_3/mcm     191.11.1473 P_2c6'/mm'm    1492  P6_3/mcm1'_c[P6/mmm]       -P 6c 2 1'c
1494    194.263  P6_3/mmc       194.1.1494  P6_3/mmc       1493  P6_3/mmc                   -P 6c  2c
1495    194.264  P6_3/mmc1'     194.2.1495  P6_3/mmc1'     1494  P6_3/mmc1'                 -P 6c  2c   1'
1496    194.265  P6_3/m'mc      194.3.1496  P6_3/m'mc      1495  P6_3/m'mc                  P 6c  2c' -1'
1497    194.266  P6_3'/mm'c     194.4.1497  P6_3'/mm'c     1496  P6_3'/mm'c                 P 6c' 2c' -1'
1498    194.267  P6_3'/mmc'     194.5.1498  P6_3'/mmc'     1497  P6_3'/mmc'                 P 6c' 2c  -1'
1499    194.268  P6_3'/m'm'c    194.6.1499  P6_3'/m'm'c    1498  P6_3'/m'm'c                -P 6c' 2c
1500    194.269  P6_3'/m'mc'    194.7.1500  P6_3'/m'mc'    1499  P6_3'/m'mc'                -P 6c' 2c'
1501    194.270  P6_3/mm'c'     194.8.1501  P6_3/mm'c'     1500  P6_3/mm'c'                 -P 6c  2c'
1502    194.271  P6_3/m'm'c'    194.9.1502  P6_3/m'm'c'    1501  P6_3/m'm'c'                P 6c  2c  -1'
1474    194.272  P_c6_3/mmc     191.12.1474 P_2c6'/mmm'    1502  P6_3/mmc1'_c[P6/mmm]       -P 6c 2c 1'c
1503    195.1    P23            195.1.1503  P23            1503  P23                        P 2 2 3
1504    195.2    P231'          195.2.1504  P231'          1504  P231'                      P 2 2 3 1'
1510    195.3    P_I23          197.3.1510  I_P23          1505  P231'_I[I23]               P 2 2 3 1'I
1506    196.4    F23            196.1.1506  F23            1506  F23                        F 2 2 3
1507    196.5    F231'          196.2.1507  F231'          1507  F231'                      F 2 2 3 1'
1505    196.6    F_S23          195.3.1505  P_F23          1508  F231'_I[P23]               F 2 2 3 1'n
1508    197.7    I23            197.1.1508  I23            1509  I23                        I 2 2 3
1509    197.8    I231'          197.2.1509  I231'          1510  I231'                      I 2 2 3 1'
1511    198.9    P2_13          198.1.1511  P2_13          1511  P2_13                      P 2ac 2ab 3
1512    198.10   P2_131'        198.2.1512  P2_131'        1512  P2_131'                    P 2ac 2ab 3 1'
1515    198.11   P_I2_13        199.3.1515  I_P2_13        1513  P2_131'_I[I2_13]           P 2ac 2ab 3 1'I
1513    199.12   I2_13          199.1.1513  I2_13          1514  I2_13                      I 2b 2c 3
1514    199.13   I2_131'        199.2.1514  I2_131'        1515  I2_131'                    I 2b 2c 3 1'
1516    200.14   Pm-3           200.1.1516  Pm-3           1516  Pm-3                       -P 2 2 3
1517    200.15   Pm-31'         200.2.1517  Pm-31'         1517  Pm-31'                     -P 2 2 3 1'
1518    200.16   Pm'-3'         200.3.1518  Pm'-3'         1518  Pm'-3'                     P 2 2 3 -1'
1533    200.17   P_Im-3         204.4.1533  I_Pm-3         1519  Pm-31'_I[Im-3]             -P 2 2 3 1'I
1520    201.18   Pn-3           201.1.1520  Pn-3           1520  Pn-3                       -P 2ab 2bc 3
1521    201.19   Pn-31'         201.2.1521  Pn-31'         1521  Pn-31'                     -P 2ab 2bc 3  1'
1522    201.20   Pn'-3'         201.3.1522  Pn'-3'         1522  Pn'-3'                     P 2ab 2bc 3 -1'
1534    201.21   P_In-3         204.5.1534  I_Pm'-3'       1523  Pn-31'_I[Im-3]             -P 2ab 2bc 3 1'I
1524    202.22   Fm-3           202.1.1524  Fm-3           1524  Fm-3                       -F 2 2 3
1525    202.23   Fm-31'         202.2.1525  Fm-31'         1525  Fm-31'                     -F 2 2 3 1'
1526    202.24   Fm'-3'         202.3.1526  Fm'-3'         1526  Fm'-3'                     F 2 2 3 -1'
1519    202.25   F_Sm-3         200.4.1519  P_Fm-3         1527  Fm-31'_I[Pm-3]             -F 2 2 3 1'n
```

```
1527  203.26   Fd-3        203.1.1527  Fd-3        1528  Fd-3              -F 2uv 2vw 3
1528  203.27   Fd-31'      203.2.1528  Fd-31'      1529  Fd-31'            -F 2uv 2vw 3 1'
1529  203.28   Fd'-3'      203.3.1529  Fd'-3'      1530  Fd'-3'            F 2uv 2vw 3 -1'
1523  203.29   F_Sd-3      201.4.1523  P_Fn-3      1531  Fd-31'_I[Pn-3]    -F 2uv 2vw 3 1'n
1530  204.30   Im-3        204.1.1530  Im-3        1532  Im-3              -I 2 2 3
1531  204.31   Im-31'      204.2.1531  Im-31'      1533  Im-31'            -I 2 2 3 1'
1532  204.32   Im'-3'      204.3.1532  Im'-3'      1534  Im'-3'            I 2 2 3 -1'
1535  205.33   Pa-3        205.1.1535  Pa-3        1535  Pa-3              -P 2ac 2ab 3
1536  205.34   Pa-31'      205.2.1536  Pa-31'      1536  Pa-31'            -P 2ac 2ab 3 1'
1537  205.35   Pa'-3'      205.3.1537  Pa'-3'      1537  Pa'-3'            P 2ac 2ab 3 -1'
1541  205.36   P_Ia-3      206.4.1541  I_Pa-3'     1538  Pa-31'_I[Ia-3]    -P 2ac 2ab 3 1'I
1538  206.37   Ia-3        206.1.1538  Ia-3        1539  Ia-3              -I 2b 2c 3
1539  206.38   Ia-31'      206.2.1539  Ia-31'      1540  Ia-31'            -I 2b 2c 3 1'
1540  206.39   Ia'-3'      206.3.1540  Ia'-3'      1541  Ia'-3'            I 2b 2c 3 -1'
1542  207.40   P432        207.1.1542  P432        1542  P432              P 4 2 3
1543  207.41   P4321'      207.2.1543  P4321'      1543  P4321'            P 4 2 3 1'
1544  207.42   P4'32       207.3.1544  P4'32       1544  P4'32             P 4' 2 3
1559  207.43   P_I432      211.4.1559  I_P432      1545  P4321'_I[I432]    P 4 2 3 1'I
1546  208.44   P4_232      208.1.1546  P4_232      1546  P4_232            P 4n  2 3
1547  208.45   P4_2321'    208.2.1547  P4_2321'    1547  P4_2321'          P 4n  2 3 1'
1548  208.46   P4_2'32     208.3.1548  P4_2'32     1548  P4_2'32           P 4n' 2 3
1560  208.47   P_I4_232    211.5.1560  I_P4'32     1549  P4_2321'_I[I432]  P 4n 2 3 1'I
1550  209.48   F432        209.1.1550  F432        1550  F432              F 4 2 3
1551  209.49   F4321'      209.2.1551  F4321'      1551  F4321'            F 4 2 3 1'
1552  209.50   F4'32       209.3.1552  F4'32       1552  F4'32             F 4' 2 3
1545  209.51   F_S432      207.4.1545  P_F432      1553  F4321'_I[P432]    F 4 2 3 1'n
1553  210.52   F4_132      210.1.1553  F4_132      1554  F4_132            F 4d  2 3
1554  210.53   F4_1321'    210.2.1554  F4_1321'    1555  F4_1321'          F 4d  2 3 1'
1555  210.54   F4_1'32     210.3.1555  F4_1'32     1556  F4_1'32           F 4d' 2 3
1549  210.55   F_S4_132    208.4.1549  P_F4_232    1557  F4_1321'_I[P4_232] F 4d 2 3 1'n
1556  211.56   I432        211.1.1556  I432        1558  I432              I 4 2 3
1557  211.57   I4321'      211.2.1557  I4321'      1559  I4321'            I 4 2 3 1'
1558  211.58   I4'32       211.3.1558  I4'32       1560  I4'32             I 4' 2 3
1561  212.59   P4_332      212.1.1561  P4_332      1561  P4_332            P 4acd  2ab 3
1562  212.60   P4_3321'    212.2.1562  P4_3321'    1562  P4_3321'          P 4acd  2ab 3 1'
1563  212.61   P4_3'32     212.3.1563  P4_3'32     1563  P4_3'32           P 4acd' 2ab 3
1570  212.62   P_I4_332    214.4.1570  I_P4_132    1564  P4_3321'_I[I4_132] P 4acd 2ab 3 1'I
1564  213.63   P4_132      213.1.1564  P4_132      1565  P4_132            P 4bd  2ab 3
1565  213.64   P4_1321'    213.2.1565  P4_1321'    1566  P4_1321'          P 4bd  2ab 3 1'
1566  213.65   P4_1'32     213.3.1566  P4_1'32     1567  P4_1'32           P 4bd' 2ab 3
1571  213.66   P_I4_132    214.5.1571  I_P4_1'32'  1568  P4_1321'_I[I4_132] P 4bd 2ab 3 1'I
1567  214.67   I4_132      214.1.1567  I4_132      1569  I4_132            I 4bd  2c 3
1568  214.68   I4_1321'    214.2.1568  I4_1321'    1570  I4_1321'          I 4bd  2c 3 1'
1569  214.69   I4_1'32     214.3.1569  I4_1'32     1571  I4_1'32           I 4bd' 2c 3
1572  215.70   P-43m       215.1.1572  P-43m       1572  P-43m             P -4  2 3
1573  215.71   P-43m1'     215.2.1573  P-43m1'     1573  P-43m1'           P -4  2 3 1'
1574  215.72   P-4'3m'     215.3.1574  P-4'3m'     1574  P-4'3m'           P -4' 2 3
1583  215.73   P_I-43m     217.4.1583  I_P-43m     1575  P-43m1'_I[I-43m]  P -4 2 3 1'I
1577  216.74   F-43m       216.1.1577  F-43m       1576  F-43m             F -4  2 3
1578  216.75   F-43m1'     216.2.1578  F-43m1'     1577  F-43m1'           F -4  2 3 1'
1579  216.76   F-4'3m'     216.3.1579  F-4'3m'     1578  F-4'3m'           F -4' 2 3
1575  216.77   F_S-43m     215.4.1575  P_F-43m     1579  F-43m1'_I[P-43m]  F -4 2 3 1'n
1580  217.78   I-43m       217.1.1580  I-43m       1580  I-43m             I -4  2 3
1581  217.79   I-43m1'     217.2.1581  I-43m1'     1581  I-43m1'           I -4  2 3 1'
1582  217.80   I-4'3m'     217.3.1582  I-4'3m'     1582  I-4'3m'           I -4' 2 3
1585  218.81   P-43n       218.1.1585  P-43n       1583  P-43n             P -4n  2 3
1586  218.82   P-43n1'     218.2.1586  P-43n1'     1584  P-43n1'           P -4n  2 3 1'
1587  218.83   P-4'3n'     218.3.1587  P-4'3n'     1585  P-4'3n'           P -4n' 2 3
1584  218.84   P_I-43n     217.5.1584  I_P-4'3m'   1586  P-43n1'_I[I-43m]  P -4n 2 3 1'I
1588  219.85   F-43c       219.1.1588  F-43c       1587  F-43c             F -4c  2 3
1589  219.86   F-43c1'     219.2.1589  F-43c1'     1588  F-43c1'           F -4c  2 3 1'
1590  219.87   F-4'3c'     219.3.1590  F-4'3c'     1589  F-4'3c'           F -4c' 2 3
1576  219.88   F_S-43c     215.5.1576  P_F-4'3m'   1590  F-43c1'_I[P-43m]  F -4c 2 3 1'n
1591  220.89   I-43d       220.1.1591  I-43d       1591  I-43d             I -4bd  2c 3
1592  220.90   I-43d1'     220.2.1592  I-43d1'     1592  I-43d1'           I -4bd  2c 3 1'
1593  220.91   I-4'3d'     220.3.1593  I-4'3d'     1593  I-4'3d'           I -4bd' 2c 3
1594  221.92   Pm-3m       221.1.1594  Pm-3m       1594  Pm-3m             -P 4 2 3
1595  221.93   Pm-3m1'     221.2.1595  Pm-3m1'     1595  Pm-3m1'           -P 4 2 3 1'
1596  221.94   Pm'-3'm     221.3.1596  Pm'-3'm     1596  Pm'-3'm           P 4' 2 3 -1'
1597  221.95   Pm-3m'      221.4.1597  Pm-3m'      1597  Pm-3m'            -P 4' 2 3
1598  221.96   Pm'-3'm'    221.5.1598  Pm'-3'm'    1598  Pm'-3'm'          P 4 2 3 -1'
1643  221.97   P_Im-3m     229.6.1643  I_Pm-3m     1599  Pm-3m1'_I[Im-3m]  -P 4 2 3 1'I
1601  222.98   Pn-3n       222.1.1601  Pn-3n       1600  Pn-3n             -P 4a 2bc 3
1602  222.99   Pn-3n1'     222.2.1602  Pn-3n1'     1601  Pn-3n1'           -P 4a 2bc 3 1'
1603  222.100  Pn'-3'n     222.3.1603  Pn'-3'n     1602  Pn'-3'n           P 4a' 2bc 3 -1'
1604  222.101  Pn-3n'      222.4.1604  Pn-3n'      1603  Pn-3n'            -P 4a' 2bc 3
1605  222.102  Pn'-3'n'    222.5.1605  Pn'-3'n'    1604  Pn'-3'n'          P 4a 2bc 3 -1'
1646  222.103  P_In-3n     229.9.1646  I_Pm'-3'm'  1605  Pn-3n1'_I[Im-3m]  -P 4a 2bc 3 1'I
1606  223.104  Pm-3n       223.1.1606  Pm-3n       1606  Pm-3n             -P 4n  2 3
1607  223.105  Pm-3n1'     223.2.1607  Pm-3n1'     1607  Pm-3n1'           -P 4n  2 3 1'
1608  223.106  Pm'-3'n     223.3.1608  Pm'-3'n     1608  Pm'-3'n           P 4n' 2 3 -1'
```

```
1609   223.107  Pm-3n'      223.4.1609  Pm-3n'       1609  Pm-3n'              -P 4n' 2 3
1610   223.108  Pm'-3'n'    223.5.1610  Pm'-3'n'     1610  Pm'-3'n'            P 4n  2 3 -1'
1645   223.109  P_Im-3n     229.8.1645  I_Pm-3m'     1611  Pm-3n1'_I[Im-3m]    -P 4n 2 3 1'I
1611   224.110  Pn-3m       224.1.1611  Pn-3m        1612  Pn-3m               -P 4bc 2bc 3
1612   224.111  Pn-3m1'     224.2.1612  Pn-3m1'      1613  Pn-3m1'             -P 4bc  2bc 3  1'
1613   224.112  Pn'-3'm     224.3.1613  Pn'-3'm      1614  Pn'-3'm             P 4bc' 2bc 3 -1'
1614   224.113  Pn-3m'      224.4.1614  Pn-3m'       1615  Pn-3m'              -P 4bc' 2bc 3
1615   224.114  Pn'-3'm'    224.5.1615  Pn'-3'm'     1616  Pn'-3'm'            P 4bc  2bc 3 -1'
1644   224.115  P_In-3m     229.7.1644  I_Pm'-3'm    1617  Pn-3m1'_I[Im-3m]    -P 4bc 2bc 3 1'I
1618   225.116  Fm-3m       225.1.1618  Fm-3m        1618  Fm-3m               -F 4 2 3
1619   225.117  Fm-3m1'     225.2.1619  Fm-3m1'      1619  Fm-3m1'             -F 4 2 3 1'
1620   225.118  Fm'-3'm     225.3.1620  Fm'-3'm      1620  Fm'-3'm             F 4' 2 3 -1'
1621   225.119  Fm-3m'      225.4.1621  Fm-3m'       1621  Fm-3m'              -F 4' 2 3
1622   225.120  Fm'-3'm'    225.5.1622  Fm'-3'm'     1622  Fm'-3'm'            F 4 2 3 -1'
1599   225.121  F_Sm-3m     221.6.1599  P_Fm-3m      1623  Fm-3m1'_I[Pm-3m]    -F 4 2 3 1'n
1623   226.122  Fm-3c       226.1.1623  Fm-3c        1624  Fm-3c               -F 4c  2 3
1624   226.123  Fm-3c1'     226.2.1624  Fm-3c1'      1625  Fm-3c1'             -F 4c  2 3  1'
1625   226.124  Fm'-3'c     226.3.1625  Fm'-3'c      1626  Fm'-3'c             F 4c' 2 3 -1'
1626   226.125  Fm-3c'      226.4.1626  Fm-3c'       1627  Fm-3c'              -F 4c' 2 3
1627   226.126  Fm'-3'c'    226.5.1627  Fm'-3'c'     1628  Fm'-3'c'            F 4c  2 3 -1'
1600   226.127  F_Sm-3c     221.7.1600  P_Fm-3m'     1629  Fm-3c1'_I[Pm-3m]    -F 4c 2 3 1'n
1628   227.128  Fd-3m       227.1.1628  Fd-3m        1630  Fd-3m               -F 4vw  2vw 3
1629   227.129  Fd-3m1'     227.2.1629  Fd-3m1'      1631  Fd-3m1'             -F 4vw  2vw 3  1'
1630   227.130  Fd'-3'm     227.3.1630  Fd'-3'm      1632  Fd'-3'm             F 4vw' 2vw 3 -1'
1631   227.131  Fd-3m'      227.4.1631  Fd-3m'       1633  Fd-3m'              -F 4vw' 2vw 3
1632   227.132  Fd'-3'm'    227.5.1632  Fd'-3'm'     1634  Fd'-3'm'            F 4vw  2vw 3 -1'
1616   227.133  F_Sd-3m     224.6.1616  P_Fn-3m      1635  Fd-3m1'_I[Pn-3m]    -F 4vw 2vw 3 1'n
1633   228.134  Fd-3c       228.1.1633  Fd-3c        1636  Fd-3c               -F 4cvw  2vw 3
1634   228.135  Fd-3c1'     228.2.1634  Fd-3c1'      1637  Fd-3c1'             -F 4cvw  2vw 3  1'
1635   228.136  Fd'-3'c     228.3.1635  Fd'-3'c      1638  Fd'-3'c             F 4cvw' 2vw 3 -1'
1636   228.137  Fd-3c'      228.4.1636  Fd-3c'       1639  Fd-3c'              -F 4cvw' 2vw 3
1637   228.138  Fd'-3'c'    228.5.1637  Fd'-3'c'     1640  Fd'-3'c'            F 4cvw  2vw 3 -1'
1617   228.139  F_Sd-3c     224.7.1617  P_Fn-3m'     1641  Fd-3c1'_I[Pn-3m]    -F 4cvw 2vw 3 1'n
1638   229.140  Im-3m       229.1.1638  Im-3m        1642  Im-3m               -I 4 2 3
1639   229.141  Im-3m1'     229.2.1639  Im-3m1'      1643  Im-3m1'             -I 4 2 3 1'
1640   229.142  Im'-3'm     229.3.1640  Im'-3'm      1644  Im'-3m              I 4' 2 3 -1'
1641   229.143  Im-3m'      229.4.1641  Im-3m'       1645  Im-3m'              -I 4' 2 3
1642   229.144  Im'-3'm'    229.5.1642  Im'-3'm'     1646  Im'-3'm'            I 4 2 3 -1'
1647   230.145  Ia-3d       230.1.1647  Ia-3d        1647  Ia-3d               -I 4bd 2c 3
1648   230.146  Ia-3d1'     230.2.1648  Ia-3d1'      1648  Ia-3d1'             -I 4bd 2c 3  1'
1649   230.147  Ia'-3'd     230.3.1649  Ia'-3'd      1649  Ia'-3d              I 4bd' 2c 3 -1'
1650   230.148  Ia-3d'      230.4.1650  Ia-3d'       1650  Ia-3d'              -I 4bd' 2c 3
1651   230.149  Ia'-3'd'    230.5.1651  Ia'-3'd'     1651  Ia'-3d'             I 4bd 2c 3 -1'
```

## Appendix-II
## Procedure for installing and running the program MHALL

The program **MHALL** is distributed within the file `MHall.zip`. It contains the executable program `MHall.exe` for Windows (64 bits) and `MHall.x` for Linux. For working with the program, the following steps should be respected (this is for Windows, for Linux is similar):

1: Extract the files of `MHall.zip` in the directory of your choice

2: Open a Windows terminal (using cmd.exe) and go to the previous directory.

3: Type `MHall` in the terminal followed by the `<Enter>` key.

4: Introduce the Hall symbol of your choice (or a set of generators in Jones faithful notation separated by semicolons)

5: Type the `<Enter>` key

6: For exiting the program you should enter a void Hall symbol.

Notice that the database `magnetic_data.txt`, by Stokes & Campbell, is provided with the program. If you have already this database in another directory, you may define the environment variable `CRYSFML_DB` pointing to the directory where `magnetic_data.txt` is.

The program may be run also by entering the input information in the command line (do not forget to put the information within double quotes). For instance, we can run the program and redirect the standard output to the file `output.txt`:

MyPrompt> MHall "x,-y,z+1/4,-1;-x,-y,-z,1;x+1/2,y+1/2,z,1" > output.txt

MyPrompt> MHall "-F 2yw' -1'n" >> output.txt

This allows the preparation of a batch file for testing many examples at the fly. The standard output is sent to `output.txt`

### Examples without entering the input in the command line

*First example*: MSG P_C4_2/m using a strange setting (Hall symbol `-P 4n' 1u'`) followed by a change of basis putting back the operators in the standard setting.

```
  ---------------------------
   MHall: Testing Hall symbols
  ---------------------------
 => Enter the magnetic Hall symbol or a list of generators in Jones'faithful notation:
    -P 4n' 1u' : -a/4-b/4,a/4-b/4,c;-1/8,-1/8,0
 => Obtained generators: -y+1/2,x+1/2,z+1/2,-1;x+1/4,y,z,-1;-x,-y,-z,1
 => Followed by a change of basis: -a/4-b/4,a/4-b/4,c;-1/8,-1/8,0
 => Newly obtained generators: -y+1/2,x+1/2,z+1/2,-1;x+1/2,y+1/2,z,-1;-x,-y,-z,1
    General Space Group
    -------------------
               Op-Dimension:    4
            Space-Dimension:    3
               Multiplicity:   16
                    MagType:    4, Black-White:2
                    NumOps:    8
                    Centred:    2
```

```
       Num. Centring translation:    0
          Num. Anti-translations:    1
                  Crystal system: Tetragonal
   Crystallographic Point group: 4/m
                      Laue class: 4/m
              Space Group number:   84
           Shubnikov Group number:  717
                     Hall symbol: -P 4n' 1u' : -a/4-b/4,a/4-b/4,c;-1/8,-1/8,0
      Shubnikov Group BNS-symbol: P_C4_2/m
      Shubnikov Group BNS-label : 84.57
      Shubnikov Group  OG-symbol: P_P4_2/m
             Magnetic Point Group: 4/m1'
     To Standard Shubnikov Group: a,b,c;0,0,0
                  Generators List: -y+1/2,x+1/2,z+1/2,-1;x+1/2,y+1/2,z,-1;-x,-y,-z,1
                     Centre_coord: [ 0 0 0 ]
                Anti-Centre_coord: [ 1/4 1/4 0 ]

                Anti-translations:
                                   [ 1/2 1/2 0 ]

   Complete list of symmetry operators and symmetry symbols
   ========================================================
   SymmOp   1: x,y,z,1                          Symbol: 1
   SymmOp   2: -x,-y,z,1                        Symbol: 2 0,0,z
   SymmOp   3: -y,x,z+1/2,1                     Symbol: 4+ (0,0,1/2) 0,0,z
   SymmOp   4: y,-x,z+1/2,1                     Symbol: 4- (0,0,1/2) 0,0,z
   SymmOp   5: -y+1/2,x+1/2,z+1/2,-1            Symbol: 4+' (0,0,1/2) 0,1/2,z
   SymmOp   6: x+1/2,y+1/2,z,-1                 Symbol: t' (1/2,1/2,0)
   SymmOp   7: y+1/2,-x+1/2,z+1/2,-1            Symbol: 4-' (0,0,1/2) 1/2,0,z
   SymmOp   8: -x+1/2,-y+1/2,z,-1               Symbol: 2' 1/4,1/4,z
   SymmOp   9: -x,-y,-z,1                       Symbol: -1 0,0,0
   SymmOp  10: x,y,-z,1                         Symbol: m x,y,0
   SymmOp  11: y,-x,-z+1/2,1                    Symbol: -4+ 0,0,z; 0,0,1/4
   SymmOp  12: -y,x,-z+1/2,1                    Symbol: -4- 0,0,z; 0,0,1/4
   SymmOp  13: y+1/2,-x+1/2,-z+1/2,-1           Symbol: -4+' 1/2,0,z; 1/2,0,1/4
   SymmOp  14: -x+1/2,-y+1/2,-z,-1              Symbol: -1' 1/4,1/4,0
   SymmOp  15: -y+1/2,x+1/2,-z+1/2,-1           Symbol: -4-' 0,1/2,z; 0,1/2,1/4
   SymmOp  16: x+1/2,y+1/2,-z,-1                Symbol: n' (1/2,1/2,0) x,y,0

   => Total CPU_TIME for this calculation:       0.406 seconds
```

*Second example*: MSG `P4'` in a non-conventional setting (Hall symbol `X  4' 1u`) corresponding to the use a supercell `4a,4b,c` of the standard setting. Notice the high number of lattice centring vectors.

```
     ---------------------------
      MHall: Testing Hall symbols
     ---------------------------
   => Enter the magnetic Hall symbol or a list of generators in Jones' faithful notation: X 4' 1u
   => Obtained generators: -y,x,z,-1;x+1/4,y,z,1
      General Space Group
      -------------------
                   Op-Dimension:    4
                Space-Dimension:    3
                   Multiplicity:   64
                        MagType:   3, Black-White:1
                         NumOps:    4
                        Centred:    1
       Num. Centring translation:   15
          Num. Anti-translations:    0
                  Crystal system: Tetragonal
   Crystallographic Point group: 4
                      Laue class: 4/m
              Space Group number:   75
           Shubnikov Group number:  663
```

```
                    Hall symbol: X 4' 1u
    Shubnikov Group BNS-symbol: P4'
    Shubnikov Group BNS-label : 75.3
    Shubnikov Group  OG-symbol: P4'
          Magnetic Point Group: 4'
   To Standard Shubnikov Group: -a/4,-b/4,c;0,0,0
                Generators List: -y,x,z,-1;x+1/4,y,z,1
                    Centre_coord: none!

         Centring translations:
                                [ 1/4 0 0 ]
                                [ 1/2 0 0 ]
                                [ 3/4 0 0 ]
                                [ 0 3/4 0 ]
                                [ 0 1/4 0 ]
                                [ 1/4 3/4 0 ]
                                [ 0 1/2 0 ]
                                [ 3/4 3/4 0 ]
                                [ 1/4 1/4 0 ]
                                [ 3/4 1/4 0 ]
                                [ 1/2 3/4 0 ]
                                [ 1/4 1/2 0 ]
                                [ 3/4 1/2 0 ]
                                [ 1/2 1/4 0 ]
                                [ 1/2 1/2 0 ]

Complete list of symmetry operators and symmetry symbols
========================================================
SymmOp   1: x,y,z,1                        Symbol: 1
SymmOp   2: -x,-y,z,1                      Symbol: 2 0,0,z
SymmOp   3: -y,x,z,-1                      Symbol: 4+' 0,0,z
SymmOp   4: y,-x,z,-1                      Symbol: 4-' 0,0,z
SymmOp   5: x+1/4,y,z,1                    Symbol: t (1/4,0,0)
SymmOp   6: -x+1/4,-y,z,1                  Symbol: 2 1/8,0,z
SymmOp   7: -y+1/4,x,z,-1                  Symbol: 4+' 1/8,1/8,z
SymmOp   8: y+1/4,-x,z,-1                  Symbol: 4-' 1/8,-1/8,z
SymmOp   9: x+1/2,y,z,1                    Symbol: t (1/2,0,0)
SymmOp  10: -x+1/2,-y,z,1                  Symbol: 2 1/4,0,z
SymmOp  11: -y+1/2,x,z,-1                  Symbol: 4+' 1/4,1/4,z
SymmOp  12: y+1/2,-x,z,-1                  Symbol: 4-' 1/4,-1/4,z
SymmOp  13: x+3/4,y,z,1                    Symbol: t (3/4,0,0)
SymmOp  14: -x+3/4,-y,z,1                  Symbol: 2 3/8,0,z
SymmOp  15: -y+3/4,x,z,-1                  Symbol: 4+' 3/8,3/8,z
SymmOp  16: y+3/4,-x,z,-1                  Symbol: 4-' 3/8,-3/8,z
SymmOp  17: x,y+3/4,z,1                    Symbol: t (0,3/4,0)
SymmOp  18: -x,-y+3/4,z,1                  Symbol: 2 0,3/8,z
SymmOp  19: -y,x+3/4,z,-1                  Symbol: 4+' -3/8,3/8,z
SymmOp  20: y,-x+3/4,z,-1                  Symbol: 4-' 3/8,3/8,z
SymmOp  21: x,y+1/4,z,1                    Symbol: t (0,1/4,0)
SymmOp  22: -x,-y+1/4,z,1                  Symbol: 2 0,1/8,z
SymmOp  23: -y,x+1/4,z,-1                  Symbol: 4+' -1/8,1/8,z
SymmOp  24: y,-x+1/4,z,-1                  Symbol: 4-' 1/8,1/8,z
SymmOp  25: x+1/4,y+3/4,z,1                Symbol: t (1/4,3/4,0)
SymmOp  26: -x+1/4,-y+3/4,z,1              Symbol: 2 1/8,3/8,z
SymmOp  27: -y+1/4,x+3/4,z,-1              Symbol: 4+' -1/4,1/2,z
SymmOp  28: y+1/4,-x+3/4,z,-1              Symbol: 4-' 1/2,1/4,z
SymmOp  29: x,y+1/2,z,1                    Symbol: t (0,1/2,0)
SymmOp  30: -x,-y+1/2,z,1                  Symbol: 2 0,1/4,z
SymmOp  31: -y,x+1/2,z,-1                  Symbol: 4+' -1/4,1/4,z
SymmOp  32: y,-x+1/2,z,-1                  Symbol: 4-' 1/4,1/4,z
SymmOp  33: x+3/4,y+3/4,z,1                Symbol: t (3/4,3/4,0)
SymmOp  34: -x+3/4,-y+3/4,z,1              Symbol: 2 3/8,3/8,z
SymmOp  35: -y+3/4,x+3/4,z,-1              Symbol: 4+' 0,3/4,z
SymmOp  36: y+3/4,-x+3/4,z,-1              Symbol: 4-' 3/4,0,z
SymmOp  37: x+1/4,y+1/4,z,1                Symbol: t (1/4,1/4,0)
SymmOp  38: -x+1/4,-y+1/4,z,1              Symbol: 2 1/8,1/8,z
```

```
SymmOp  39: -y+1/4,x+1/4,z,-1              Symbol: 4+' 0,1/4,z
SymmOp  40: y+1/4,-x+1/4,z,-1              Symbol: 4-' 1/4,0,z
SymmOp  41: x+3/4,y+1/4,z,1                Symbol: t (3/4,1/4,0)
SymmOp  42: -x+3/4,-y+1/4,z,1              Symbol: 2 3/8,1/8,z
SymmOp  43: -y+3/4,x+1/4,z,-1              Symbol: 4+' 1/4,1/2,z
SymmOp  44: y+3/4,-x+1/4,z,-1              Symbol: 4-' 1/2,-1/4,z
SymmOp  45: x+1/2,y+3/4,z,1                Symbol: t (1/2,3/4,0)
SymmOp  46: -x+1/2,-y+3/4,z,1              Symbol: 2 1/4,3/8,z
SymmOp  47: -y+1/2,x+3/4,z,-1              Symbol: 4+' -1/8,5/8,z
SymmOp  48: y+1/2,-x+3/4,z,-1              Symbol: 4-' 5/8,1/8,z
SymmOp  49: x+1/4,y+1/2,z,1                Symbol: t (1/4,1/2,0)
SymmOp  50: -x+1/4,-y+1/2,z,1              Symbol: 2 1/8,1/4,z
SymmOp  51: -y+1/4,x+1/2,z,-1              Symbol: 4+' -1/8,3/8,z
SymmOp  52: y+1/4,-x+1/2,z,-1              Symbol: 4-' 3/8,1/8,z
SymmOp  53: x+3/4,y+1/2,z,1                Symbol: t (3/4,1/2,0)
SymmOp  54: -x+3/4,-y+1/2,z,1              Symbol: 2 3/8,1/4,z
SymmOp  55: -y+3/4,x+1/2,z,-1              Symbol: 4+' 1/8,5/8,z
SymmOp  56: y+3/4,-x+1/2,z,-1              Symbol: 4-' 5/8,-1/8,z
SymmOp  57: x+1/2,y+1/4,z,1                Symbol: t (1/2,1/4,0)
SymmOp  58: -x+1/2,-y+1/4,z,1              Symbol: 2 1/4,1/8,z
SymmOp  59: -y+1/2,x+1/4,z,-1              Symbol: 4+' 1/8,3/8,z
SymmOp  60: y+1/2,-x+1/4,z,-1              Symbol: 4-' 3/8,-1/8,z
SymmOp  61: x+1/2,y+1/2,z,1                Symbol: t (1/2,1/2,0)
SymmOp  62: -x+1/2,-y+1/2,z,1              Symbol: 2 1/4,1/4,z
SymmOp  63: -y+1/2,x+1/2,z,-1              Symbol: 4+' 0,1/2,z
SymmOp  64: y+1/2,-x+1/2,z,-1              Symbol: 4-' 1/2,0,z

=> Total CPU_TIME for this calculation:     0.391 seconds
```